\documentclass[apsprl,twocolumn,superscriptaddress,longbibliography]{revtex4-2}

\usepackage{color}
\usepackage{appendix}

\usepackage{graphicx}
\usepackage{dcolumn}
\usepackage{amsmath}
\usepackage{amssymb}
\usepackage{lipsum}

\usepackage[usenamesdvipsnames]{xcolor}
\usepackage[colorlinks=truelinkcolor=blue,urlcolor=blue,citecolor=blue]{hyperref}

\def\bra#1{\langle#1 |}
\def\ket#1{| #1\rangle}

\newcommand{\LE}{\mathrm{\scriptscriptstyle LE}}

\newcommand{\opadag}[1]{{\hat{a}^{\dagger}}_{#1}}
\newcommand{\opa}[1]{{\hat{a}^{\phantom \dagger}}_{#1}}
\newcommand{\opbdag}[1]{{\hat{b}^{\dagger}}_{#1}}
\newcommand{\opb}[1]{{\hat{b}^{\phantom \dagger}}_{#1}}

\newcommand{\nep}{\textrm{e}}

\usepackage{color}

\begin{document}

\title{Thouless pumping and topology}

\author{Roberta Citro}
\affiliation{Physics Department "E.R. Caianiello" and CNR-SPIN, c/o Universit\`a degli Studi di Salerno, Via Giovanni Paolo II, 132, I-84084 Fisciano (Sa), Italy}
\affiliation{INFN - Gruppo Collegato di Salerno, I-80184 Salerno, Italy}
\author{Monika Aidelsburger}
\affiliation{Faculty of Physics, Ludwig-Maximilians-Universit\"at M\"unchen, Schellingstr. 4, D-80799 Munich, Germany}
\affiliation{Munich Center for Quantum Science and Technology (MCQST), Schellingstr. 4, D-80799 Munich, Germany}

\begin{abstract}
Thouless pumping provides one of the simplest manifestations of topology in quantum systems, and has attracted a lot of recent interest, both theoretically and experimentally. Since the seminal works by Thouless and Niu in 1983 and 1984, it is argued that the quantization of the pumped charge is robust against weak disorder, but a clear characterization of the localization properties of the relevant states, and the breakdown of quantized transport in the presence of interaction or out of the adiabatic approximation, has been long debated. Thouless pumping is also the first example of a topological phase emerging in a periodically-driven system. Driven systems can exhibit exotic topological phases without any static analogue and have been the subject of many recent proposals both in fermionic and bosonic systems in diverse platforms ranging from cold atoms to photonics and condensed matter systems. In this respect, this review has a twofold purpose: On the one hand, it serves as a basis to understand the robustness of the topology of slowly-driven systems per se; On the other hand, it highlights the rich properties of topological pumps and their diverse range of applications, for instance, in systems with synthetic dimensions or for understanding higher-order topological phases. These examples underline the relevance of topological pumping for the fast growing field of topological quantum matter.
\end{abstract}

\maketitle

\section{Introduction}
\label{sec:intro}

A quantum pump is a device able to generate a particle current via slow and periodic modulation of at least two system parameters, in the absence of any external bias~\cite{altshuler_1999}. For this reason it is considered as one of the most intriguing effects in quantum mechanics. In a conductor a dc current is usually associated with a dissipative flow of electrons in response to an applied bias voltage. In systems of mesoscopic scale a dc current can be generated even at zero bias (e.g. in semiconductor nanostructures of nm size and tens of atoms) in the presence of slow periodic perturbations. In the adiabatic limit, when the applied perturbations are slow in comparison to the escape rate to external contacts, the electronic state of the quantum system is the same after a period, but a net charge has been transferred thanks to a squeezing of the wavefunction in the central region~\cite{zhou_1999}. Since the quantum state of the system remains coherent, this effect is known as quantum charge pumping. Although different in nature, quantum pumping shares some similarity with other fascinating phenomena such as persistent currents~\cite{altshuler_1999}
and superconductivity and allows exploring fundamental issues regarding the role of topology and symmetries, finding its maximum expression in the Thouless or {\it topological pump}, where transport is quantized.
Quantum pumping has received much attention in mesoscopic electronic systems, mainly due to its potential of reducing the dissipation of energy as wasteful heat, define a better current standard for metrological purpose~\cite{Niu_1990,Pekola_RMP_2013}, or even be used for quantum computing~\cite{Das_2006}. Recent experimental realizations of Thouless pumps have been observed in
photonics~\cite{kraus_2012,verbin_2015,ke_photonic_waveguide_2016,cerjan_2020}, magneto and electro-mechanical systems~\cite{grinberg_2020,electromechanical_waveguide}, and ultracold atoms~\cite{Lohse_2016,Nakajima_2016,Spielman_2016}, including pumps with interactions~\cite{tangpanitanon_2016,nonlinear-thouless-pump,esslinger_int_2022} and pumps in open systems~\cite{fedorova_2020,esslinger_2022}.

\section{Topological quantized transport in one-dimension}
\label{sec:quantized transport}

\begin{figure*}[ht!]
	\includegraphics{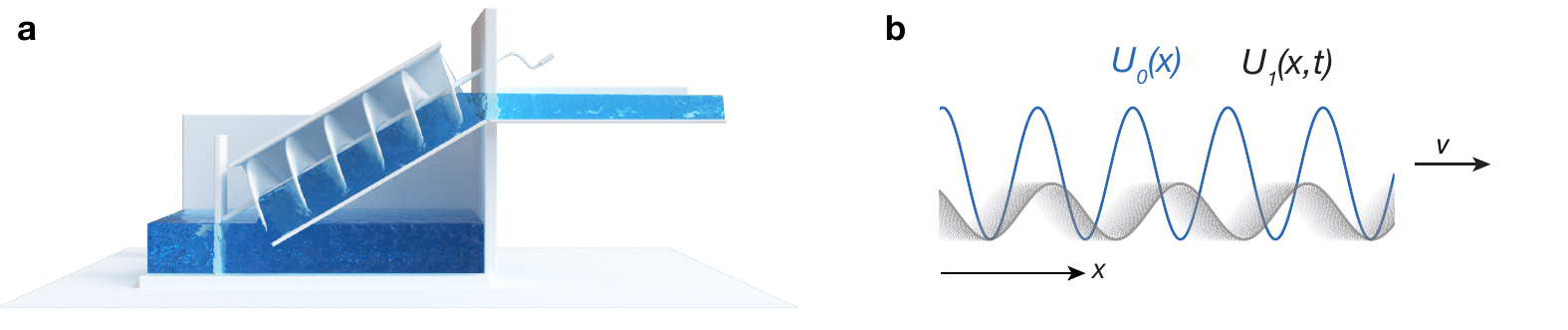}
\caption{\textbf a Illustration of a classical Archimede's screw (by C. Hohmann, MCQST). \textbf b Schematic drawing of two periodic potentials, $U_0(x)$ and $U_1(x,t)$, that share the same periodicity. One of the two potentials, $U_1(x,t)$, is moving with a small velocity $\text{v}$ with respect to the other in order to realize a quantum pump.}
\label{fig:thouless_pump}
\end{figure*}

Topological pumping entails the transport of charge, in the absence of a net external electric or magnetic field, through an adiabatic cyclic evolution of the underlying Hamiltonian, as originally proposed by Thouless~\cite{Thouless_1983}. In contrast to classical transport, the transported charge in a Thouless pump is quantized and purely determined by the topology of the pump cycle, making it robust against perturbations, such as interaction effects or disorder~\cite{Niu_1984}. Intuitively, the Thouless pump can be thought of as the quantum version of the famous Archimede’s screw (Fig.~\ref{fig:thouless_pump}a), where a directional motion of water is generated by slow and periodic movement of the handle. The key difference being that in the classical Archimede's screw the amount of water pumped per cycle can be continuously tuned by changing the tilt angle of the screw, while for the topological Thouless pump the amount of charge pumped per cycle cannot be continuously tuned, i.e., it is quantized according to the topology of the pump cycle.

Consider spinless electrons in 1D subject to a potential $U(x)$ with periodicity $a$, i.e., $U(x + a) = U(x)$. Provided the number of electrons $n a$ per period $a$ equals an integer $N$, the lowest $N$ bands of the energy spectrum are occupied while the higher bands are empty. Now let the potential slowly vary periodically in time, such that $U(x,t) = U_0(x) +U_1(x - \text{v} t)$, where $U_0$ and $U_1$ share the same periodicity $a$ and $\text{v}$ is some small velocity with $a/\text{v}=T$ (Fig.~\ref{fig:thouless_pump}b).
If the electrons follow the variation of the potential adiabatically, a quantized charge $Q$ is transferred per period $T$, where the quantization is determined by the topology of the energy bands that are occupied with electrons and the topology of the pump cycle. Each energy band including the time $t$ as a second dimension, hence realizing a (1+1)-D parameter space, is associated with an integer number, a topological invariant called \textit{Chern number}, and the combined response is given be the sum of all Chern numbers of the occupied energy bands (Appendix~\ref{sec:box1}), which can be viewed as a dynamical realization of the IQHE (Sec.~\ref{sec:dynamical_2DQH}).

Intuitively, the time evolution of the potential can be described in terms of a ``trajectory'' $\mathcal{C}$ of the system in a 2D plane, whose parameters are determined by the time-dependent potential $U(x,t)$.
If $\mathcal{C}$ encircles a degeneracy point of the Hamiltonian, the pumped charge is quantized (Sec.~\ref{sec:pump of charge}) and the response will remain quantized, as long as the trajectory encircles the degeneracy point. This illustrates the robustness of topological pumping to disorder and interactions, since small deformations of the trajectory cannot alter the pumped charge. However, the quantization of the pumped charge is only valid as long as the potential is varied adiabatically. Studies away from the adiabatic limit show that, despite its topological nature, this phenomenon is not generically robust to non-adiabatic effects. Indeed the mean value of the pumped charge shows a deviation from the topologically-quantized limit which is quadratic in the driving frequency for a sudden switch-on of the drive (Sec.~\ref{sec:beyond adiabatic}). Interestingly, even at fast driving one can realize an ideal pump under a family of protocols, which contain the adiabatic one as a limiting case~\cite{cheianov_2021}.

\section{Charge and spin pumping}
\label{sec:pump}

\subsection{Thouless pumping of charge}
\label{sec:pump of charge}

\begin{figure*}[ht!]
\includegraphics{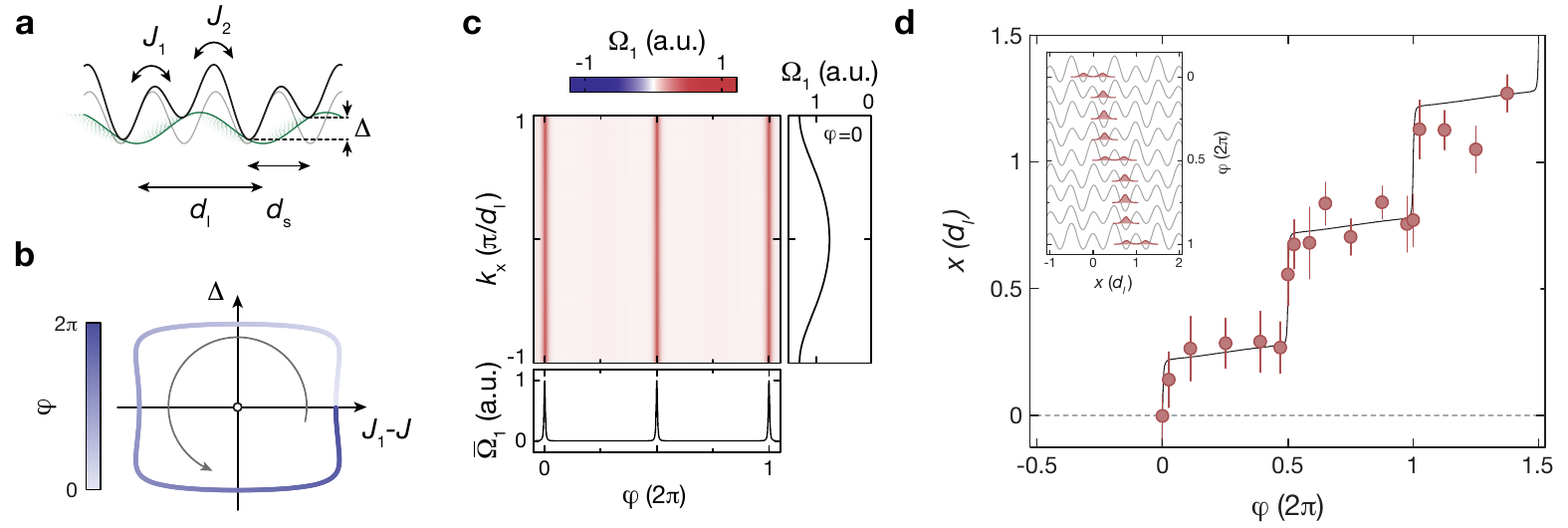}
\caption{Topological charge pumping in a superlattice (adapted from Ref.~\cite{Lohse_2016}). {\bf a} Superlattice potential formed by superimposing two lattices (gray and green) with constants $d_l$ and $d_s = d_l/2$. $J_{1,2}$ denote the tunnel couplings and $\Delta$ the energy offset between neighboring sites. {\bf b} Pumping cycle in the  $(J_1-J_2,\Delta)$ parameter space with winding numer $w=+1$, realized by varying the relative phase $\varphi$ of the superlattice. The closed path encircles the degeneracy point at the origin where $\Delta = 0$ and $J_1 = J_2$ resulting in quantized transport. {\bf c} Berry curvature $\Omega_1$ of the lowest band as a function of  $\varphi$ and quasimomentum $k_x$ for a lattice with $V_s = 10 E_r$ and $V_l = 5E_r$. The panel below shows the Berry curvature averaged over $k_x$ while the panel on the right shows $\Omega_1(k_x)$ for $\varphi = 0$. {\bf d} CoM motion during one pump cycle. The solid black line depicts the calculated CoM motion of a localized Wannier function. Inset: Evolution of the ground-state wave function illustrated for one double well.}
\label{fig:superlattice}
\end{figure*}

Topological charge pumping has remained out of reach in most electron-based condensed matter experiments because of challenges in realizing the adiabatic regime. Due to the versatility and control of synthetic quantum systems experimental demonstrations of adiabatic quantum pumps have been achieved with cold atoms~\cite{Lohse_2016,Nakajima_2016,Spielman_2016,minguzzi_topological_2021} and photons~\cite{kraus_2012,verbin_2015}.
Specifically, ultracold atoms in optical superlattices have emerged as an ideal platform for the implementation of quantized topological charge pumps~\cite{Lohse_2016,Nakajima_2016}. In these experiments, the Thouless potential, a sliding superlattice, is formed by
superimposing two lattices with different periodicities $d_l$ and $d_s = \alpha d_l$, $\alpha < 1$. This generates a potential
$V_s \sin^2 (\pi x/d_s + \pi/2)+V_l \sin^2 (\pi x/d_{l} - \varphi/2)$, where $V_s (V_l)$ denotes the depth of the short-(long-)wavelength lattice, respectively (Fig.~\ref{fig:superlattice}a). Changing the relative phase $\varphi$ between the two lattices results in a periodically-modulated superlattice potential. Varying the phase adiabatically by an amount of $2\pi$ realizes one pump cycle, where the long lattice has moved by $d_l$ with respect to the short lattice. Since the change is performed adiabatically, a particle initially in a Bloch eigenstate $|u_n(k_x,\varphi(0))\rangle$ of the Hamiltonian $\hat{H}(k_x,\varphi(t))$ follows the instantaneous eigenstate $|u_n(k_x,\varphi (t))\rangle$ and remains in an eigenstate at all times; here $n$ denotes the $n$th Bloch band and $k_x$ the quasimomentum. However, after one cycle the state has acquired a geometric phase proportional to $\partial_t \varphi$. This phase, know as Berry phase, is proportional to the integral over the Berry curvature $\Omega_n(k_x,\varphi)=i \left( \langle \partial_\varphi u_n|\partial_{k_x} u_n\rangle -\langle \partial_{k_x} u_n|\partial_\varphi u_n\rangle \right)$ of the occupied bands along the pump path (Fig.~\ref{fig:superlattice}c). Due to the non commutativity between position and momentum, this phase generates an anomalous velocity given by $\dot{x}_n=\Omega_n \partial_t \varphi$. The displacement of the cloud after one cycle can be obtained by integrating the anomalous velocity of the occupied Bloch states over one pump cycle. For a uniformly filled band the displacement is quantized according to the topological two-dimensioanl (2D) invariant, the so called {\it Chern number} $\nu_n$, which is an integer. It is defined as the integral over the closed surface in ($k_x$, $\varphi$) space:
\begin{equation}
\nu_n=\frac{1}{2\pi} \int_{BZ}\int_0^{2\pi}  \Omega_n(k_x, \varphi) \text{d}k_x \text{d}\varphi,
\label{eq:chern}
\end{equation}
\noindent where $BZ$ denotes the first Brillouin zone of the superlattice. During one pump cycle the center of mass (CoM) changes by $\nu_n d_l$ and is quantized in units of $d_l$. If more than one energy band is completely filled, the CoM response is given by the sum over all energy bands $\sum_n d_l \nu_n$.
Intriguingly, the induced motion can either occur in the same or opposite direction as the moving lattice, depending on the sign of $\nu_n$. This counterintuitive behavior has been demonstrated by preparing ultracold bosonic atoms in the first excited band of an optical superlattice, where quantized transport in the opposite direction of the moving lattice was found~\cite{Lohse_2016}.
\noindent Let us note that from the point of view of Eq.~(\ref{eq:chern}) topological pumping can be viewed as a dynamical version of the IQHE. In the latter case the conductance in the presence of a magnetic field in 2D can be related to an integral over the first Brillouin zone in the $(k_x,k_y)$ space~\cite{Hall_2020}. Indeed, the adiabatic variation of $\varphi$ is equivalent to threading a magnetic flux through a cylinder that generates an electric field in the orthogonal direction and leads to a quantized Hall conductance.
The $BZ$ spanned by $(k_x,\varphi)$ is then equivalent to the Brillouin zone spanned by $(k_x,k_y)$ in the corresponding 2D model (Sec.~\ref{sec:dynamical_2DQH}).
Since the displacement of the cloud is proportional to a topological invariant, it neither depends on the pumping speed, nor on the specific lattice parameters and is robust to small perturbations.
\\
Let us emphasize that quantized topological charge pumps require uniformly filled bands. In contrast, if a Bose-Einstein condensate (BEC) occupies just a single quasimomentum state, the system exhibits non-quantized charge pumping set by the \textit{local} geometric properties of the band~\cite{Spielman_2016}. This is known as {\it geometric charge pump}. Similar to topological charge pumps, there is an an overall displacement per pump cycle, which in this case is however not quantized.
Near perfect quantized pumping can be restored, however, by adding a linear tilt. Intuitively, the added potential assists in the uniform sampling of all momenta due to the Bloch oscillations induced by the tilt~\cite{assisted_Thouless_pump}.

The mechanism underlying the quantum pump can also be described on a microscopic level by considering the Wannier tunneling regime $V_s \ll V_l^2/(4 E_r)$, where $E_{r} = h^2/(8m d_s^2)$ denotes the recoil energy and $m$ the mass of an atom. In the tight-binding limit, the superlattice model with $d_l=2d_s$ and two sites per unit cell (Fig.~\ref{fig:superlattice}a) is described by the Rice-Mele Hamiltonian (RM)~\cite{ricemele_1982}:
\begin{eqnarray}
\hat{H}(\varphi) =&&-\sum_j [J_1 (\varphi) a^\dagger_{j} a_{j}+ J_2 (\varphi) a^\dagger_{j+1} b_{j}+ \text{h.c.}] \nonumber \\ \noindent 
&& +\frac{\Delta(\varphi)}{2}\sum_j (a^\dagger_{j} a_{j}- b^\dagger_{j}b_{j}),
\label{eq:rice-mele}
\end{eqnarray}
where $\opadag{j} (\opa{j})$ and $\opbdag{j} (\opb{j})$ are the creation (annihilation) operators acting on the left and right site of the $j$-th unit cell, $J_{1,2}$ denote the intra- and inter-unit cell hopping and $\Delta$ is the energy offset between neighbouring sites.
\\
The pumping mechanism can be understood from the point of view of a cycle in the parameters space $(J_1-J_2, \Delta)$, whose winding number is directly related to the Chern number of the pump~\cite{review_niu_2010}. For $J_1-J_2=\Delta=0$ the ground-state at half filling is gapless defining the degeneracy point in the center of the $(J_1-J_2, \Delta)$ parameter space (Fig.~\ref{fig:superlattice}b). The winding number $w$ is an integer that characterizes the chirality as well as the number of times the pump cycle winds around the degeneracy point. The winding number for the cycle shown in Fig.~\ref{fig:superlattice}b is $w=+1$, which is equal to the Chern number of the pump, and states that this cycle realizes quantized particle transport of one unit cell in the direction of the moving potential.

Intuitively, the mechanism of the quantized pump can be described as follows: Changing $\varphi$ is equivalent to changing the shape of the superlattice potential (Fig.~\ref{fig:superlattice}d). Starting at $\varphi=0$ ($\Delta=0$ and $J_1>J_2$) the ground-state at half filling for non-interacting fermions~\cite{Nakajima_2016} or equivalently hard-core bosons~\cite{Lohse_2016}, consists of localized particles in the symmetric-superposition between the two sites of the unit cell. For increasing $\varphi$ the long lattice is shifted to the right and the double wells are tilted ($\Delta>0$) in such a way that the atom tunnels to the lower-lying site on the right. The tilt is maximum at $\varphi=\frac{\pi}{2}$ and changes its sign afterwards until the lattice forms symmetric double wells again at $\varphi=\pi$, with a shift of one short lattice constant $d_s$ to the right. The atom that was on the lower site for large $\Delta$, delocalizes on the double well and moves by $d_l/2$ during the first half of the pumping cycle. After one full cycle the lattice configuration becomes identical to the initial one, but the atom has moved to the neighboring double well. Intriguingly, the lattice minima do not move in real space, hence, a classical particle would not move. In contrast, quantum mechanically the particle tunnels to the neighbouring sites during one pump cycle. Experimentally, the resulting motion of the atoms is observed by measuring the CoM position of the cloud (Fig.~\ref{fig:superlattice}d). The displacement occurs in steps and is indeed quantized~\cite{troyer_2013} in units of the long-lattice constant $d_l$ for $\nu=+1$.

Before topological Thouless pumps have been realized in cold atoms and photonics, a closely related quantized, however, non-topological, transport of particles in the absence of a bias voltage has been observed in quantum dots~\cite{Switkes_1999} and in 1D channels in the presence of acoustic waves~\cite{Talyanskii_1997}. In particular, the basic idea for the realization of an adiabatic quantum pump is that a dc current can be pumped through a quantum dot by periodically varying two independent parameters $X_1$ and $X_2$, e.g., a gate voltage or magnetic field. In this case one can relate the pumped current to the parametric derivatives of the scattering matrix $S(X_1,X_2)$ of the system~\cite{Brouwer_1998}. The charge pumped over one cycle is given by:
\begin{equation}
Q(m)=\frac{e}{\pi}\int_A dX_1 dX_2 \sum_\beta \sum_{\alpha\in m} \mathfrak{I} \frac{\partial S^\star_{\alpha \beta}}{\partial X_1}\frac{S_{\alpha \beta}}{\partial X_2},
\label{eq:scattering}
\end{equation}
where $m$ labels the contact, $X_1$ and $X_2$ are two external
parameters whose trace encloses the area $A$ in the
parameter space, $\alpha$  and $\beta$ label the conducting channels, and $\mathfrak{I}$ stands for the imaginary part. Although the physical description of open systems are dramatically
different from closed ones, the concept of a geometric phase can still be applied. The integrand
in Eq.~(\ref{eq:scattering}) can be thought as the Berry curvature~\cite{Zhou_2003} and the pumped charge is essentially the Abelian geometric phase of the scattering matrix, although non-topological in nature. Note, that the term \textit{geometric pump} is used in a different context here and its properties should not to be confused with the geometric pump introduced above, which was realized in Ref.~\cite{Spielman_2016}.

\subsection{Spin pumping}
\label{sec:pump of spin}

The idea of topological charge pumps has been generalized to spin pumps, which represent 1D dynamical versions
of 2D topological insulators. A quantum spin pump was realized experimentally with two-component ultracold bosonic atoms in an optical superlattice~\cite{Schweizer_2016}. To this aim, ultracold atoms in two different hyperfine states are prepared in a spin-dependent dynamically-controlled optical superlattice.
In the tight-binding limit, the dynamics of each component independently is well described by the RM model~\cite{ricemele_1982} [Eq.~(\ref{eq:rice-mele})]. If the two spin components do not interact with each other, their pumping motion is independent and a spin pump can be realized by a spin-dependent deformation of the potential, so that time-reversal symmetry is retained, while their Berry curvature is reversed. The two spin components are transported in opposite directions without any net charge transport, in contrast to the topological charge pump discussed above. In addition, the two spin components can be coupled by introducing on-site interactions~$U$ between the atoms.
For hardcore interactions and unit filling (two particles per unit-cell, one of each component), the bare tunneling is suppressed and the system can be described by a 1D spin-chain model:

\begin{eqnarray}
		\hat{\mathcal{H}} = && -\frac{1}{4}\sum_{m}\bigl(J_\text{ex}+(-1)^{m}\delta J_\text{ex}\bigr)\left(\hat{S}^+_{m}\hat{S}^-_{m+1}+\text{h.c.}\right)\\ \nonumber
		&&+ \frac{\Delta}{2}\sum_{m} (-1)^{m} \hat{S}^z_m,
		\label{eq:spinmodel}
\end{eqnarray}

where $\hat{S}^{\pm},\hat{S}^z$ are the spin operators, $\Delta$ is a spin-dependent tilt and  $\frac{1}{2}\left(J_\text{ex}\pm \delta J_\text{ex}\right) \simeq \left(J\pm \delta J\right)^2/U$ represents an alternating superexchange coupling.
In the limit of isolated double wells $\delta J_\text{ex}\approx J_\text{ex}$ and by applying a global gradient to a spin-independent superlattice the staggered tilts of the RM model can be locally reproduced.
In this situation, a cycle in the parameters ($\delta J$, $\Delta$) of the RM model corresponds to a modulation of ($J_\text{ex}$, $\Delta$) in the interacting 1D spin chain.

Direct evidence for the spin separation and spin transport can be obtained by measuring the CoM position of the two spin components using in-situ absorption images.
As a function of the pump parameter, on finds that the two components separate in real space.
Additional insights about the underlying mechanism of the spin pump was revealed by measuring spin currents between the left and the right site of a double well, e.g., it was shown that the integrated current does not depend on the specific pump parameters, as expected~\cite{Schweizer_2016}.

Interestingly, spin transport is described by a spin Chern number, which in the non-interacting case can be related to a $Z_2$ topological invariant~\cite{spin-chern-number,bernevig_book}. In its simplest the spin pump can be understood as to uncoupled Thouless pumps. A system with non-trivial $Z_2$-invariant can be realized with time-reversal invariant spin orbit interactions~\cite{Shindou_2005,Zhou_2014}.
When breaking time-reversal symmetry, the topological properties of the quantum spin Hall system remain but spin-Chern numbers are required for the description~\cite{Zhou_2014}. The interplay between glide symmetry and a coupling between different spin components in a quantum spin pumps was studied in Ref.~\cite{chen_spin_pumping_glide_symmetry}.
Moreover, for spin pumps with highly degenerate many-body ground states, a fractional transport is predicted~\cite{Meidan_2011}.
Away from the hard-core constraints for bosonic atoms, the effect of a finite interaction can be taken into account via a bosonization approach~\cite{Citro_RMP}. In this limit  the topological classification of the spin pump still remains valid, with one important difference: the topological excitations are solitons and antisolitons, which carry a spin 1/2.

\section{Beyond the adiabatic approximation}
\label{sec:beyond adiabatic}
As we have discussed above, having a topological nature, the quantization of the transported charge in a quantum pump shows robustness to various factors, such as  interactions or disorder~\cite{Niu_1984} with deviations that depend on the pumping protocol. Similarly, non-adiabatic effects have always been believed to be unimportant, i.e. exponentially small in the driving frequency $\omega$~\cite{Niu_1990,Shih_1994}
in analogy with the IQHE, where the Hall plateaus show corrections that are exponentially small in the longitudinal electric field~\cite{Klitzing_review}.
Theoretically, this follows from the fact that the quantized Chern number expression for the Hall conductivity, usually obtained through a Kubo formula
in linear response, is valid at all orders in perturbation theory~\cite{Avron_1999}.

However, deviations from this behavior have been recently found in Thouless pumping
out of the perfect adiabatic limit $\omega\to 0$~\cite{privitera_2018}. Consider a closed, clean, non-interacting system --- the driven RM model --- in the thermodynamic limit. The system starts from the initial ground-state Slater determinant
and the driving is switched on suddenly.
A careful Floquet analysis reveals that the charge pumped after many cycles shows a deviation from perfect quantization that is always \textit{polynomial} in the driving frequency $\omega$, contradicting the expected topological robustness. This quadratic deviation is present also after a finite number of pumping cycles, even if apparently hidden under a highly oscillatory non-analytic behaviour~\cite{Avron_1999} in $\omega$. An exponentially small deviation would only be obtained, if the system is prepared in a specific Floquet state, which requires a suitable switch-on of the driving. Recently, it was demonstrated that even at a fast driving frequency one can realize an ideal pump under a family of protocols which contains the adiabatic one as a limiting case~\cite{cheianov_2021}. Note, that the question about the breakdown of quantized pumping for increasing pumping speeds has also been addressed experimentally in Refs~\cite{Nakajima_2016,minguzzi_topological_2021,esslinger_int_2022}. These studies, however, consider frequencies $\omega$ that are rather far from the adiabatic limit. The fundamental question about robustness of pumping due to non-adiabatic effects discussed here cannot easily be studied in experiments, since small deviations of a few percent are extremely challenging to resolve due to other experimental imperfections.

Given the time-periodicity of the Hamiltonian in a Thouless pump, with period $T=2\pi/\omega$, one can employ a Floquet analysis ~\cite{Ferrari_1998,Avron_1999,Kitagawa_2010,Shih_1994,Russomanno_2011} (Appendix~\ref{sec:box2}).
 In a periodic boundary condition (PBC) ring geometry, the total current operator $\hat{J}(t)$ is obtained as a derivative of $\hat{H}(t)$ with respect to a flux $\Phi$ threading the ring, $\hat{J}=\partial_{\kappa} \hat{H}/\hbar$, where $\kappa = \frac{2\pi}{L} \frac{\Phi}{\Phi_0}$, $L$ is the length of the system and $\Phi_0$ the magnetic flux quantum.
As a consequence, the charge pumped in one period $T$ by a single Floquet state $|\psi_{\alpha}(t)\rangle$ is
$Q_{\alpha}(T) = \frac{1}{L} \int_{0}^{T} \mathrm{d}t \, \bra{\psi_{\alpha}(t)} \hat{J}(t) \ket{\psi_{\alpha}(t)}=\frac{T}{\hbar L} \partial_{\kappa} \varepsilon_{\alpha}$.
For a  translationally-invariant system with constant $a$, each completely filled Floquet-Bloch band with quasienergy dispersion $\varepsilon_{\alpha,k}$ contributes to the charge pumped as
\begin{equation} \label{eqn:Q}
Q_{\alpha}(T) = \frac{1}{\omega} \int_{-\frac{\pi}{a}}^{+\frac{\pi}{a}} \!\! \mathrm{d} k \, \frac{\partial \varepsilon_{\alpha,k}}{\partial k} \, ,
\end{equation}
where the $\kappa$-derivative has been replaced with a $k$-derivative, since $\varepsilon_{\alpha,k}$ depends on $k+\kappa$.
Thus, if $\varepsilon_{\alpha,k}$ wraps around the Floquet Brillouin zone (FBZ) in a continuous way as a function of $k$, $Q_{\alpha}(T)$ is equivalent to the {\em winding number} of the band, i.e.,
the number $n$ of times $\varepsilon_{\alpha,k}$ goes around the FBZ,
$\varepsilon_{\alpha,+\frac{\pi}{a}} - \varepsilon_{\alpha,-\frac{\pi}{a}} = n\, \hbar \omega $,
and $Q_{\alpha}(T)$ is therefore quantized{: $Q_{\alpha}(T)=n$}.
This result applies in the adiabatic limit, i.e., $\omega\to 0$: If $|\Psi_{\alpha}(t)\rangle$ is a Slater determinant made up of the instantaneous Hamiltonian Bloch eigenstates
$e^{ikx} u_{\alpha,k}(x,t)$,
the adiabatic theorem guarantees that such a state returns onto itself after a period $T$
by acquiring a geometric (Berry) phase
$\gamma_{\alpha,k} =  \int_{0}^{T} \mathrm{d}t \, i\bra{u_{\alpha,k}} \partial_t u_{\alpha,k} \rangle$
and a dynamical one $\theta_{\alpha,k} = \int_{0}^{T} \mathrm{d}t \, E_{\alpha,k}(t)$.
Thus $|\Psi_{\alpha}(t)\rangle$ is a Floquet state with quasienergy $\varepsilon_{\alpha,k}^0 = \hbar (-\gamma_{\alpha,k} + \theta_{\alpha,k})/T$.
Substituting this into~{Eq.~\eqref{eqn:Q}} only the geometric phase {survives, leading to the} Thouless' formula~\cite{Thouless_1983}
\begin{equation}
Q_{\alpha}(T) =  \int_{-\frac{\pi}{a}}^{+\frac{\pi}{a}} \! \frac{\mathrm{d}k}{2\pi} \int_{0}^{T} \!\! \mathrm{d}t \,
i(\bra{\partial_{k}u_{\alpha,k}} \partial_{t} u_{\alpha,k}\rangle - {\mathrm{c.c.}}) .
\end{equation}
This formula permits to identify the pumped charge with a Chern number~\cite{Avron_1999}, nicely recovering the result of Sec.~\ref{sec:pump of charge}.

\begin{figure*}[ht!]
\includegraphics{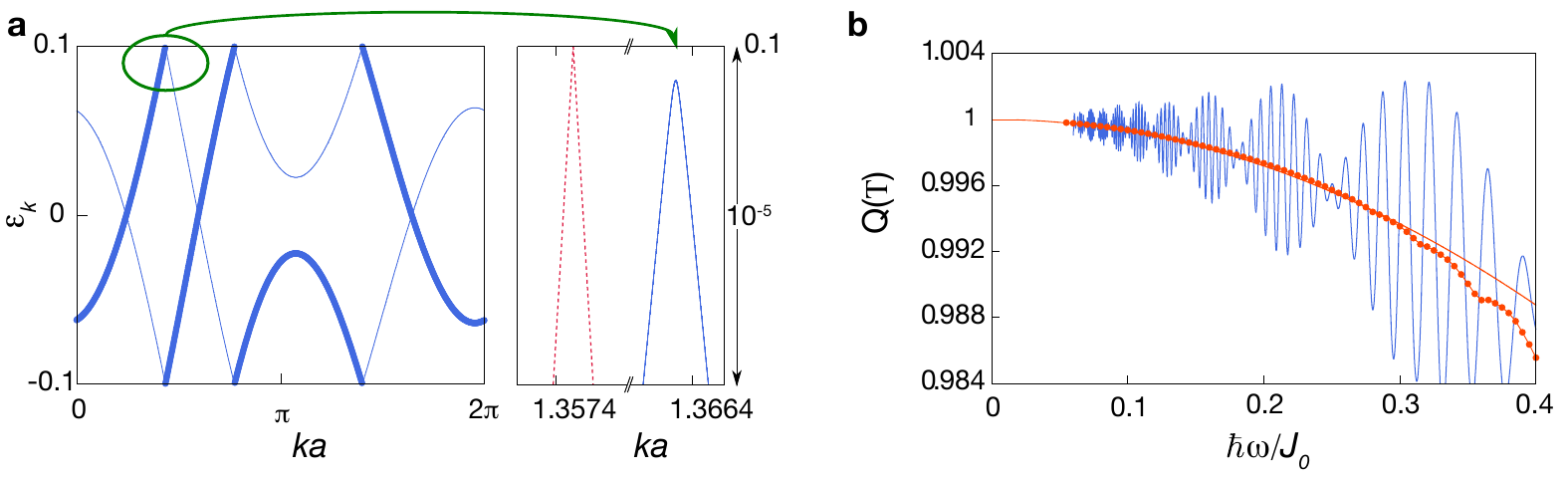}
\caption{{\bf a} Left: Exemplary quasienergy spectrum of the RM Model.
The thick band is the {lowest-energy Floquet band $\varepsilon_{\LE,k}$}. Right: (solid line) zoom of the previous figure close to the upper border of the FBZ around $ka = 1.3664$;
the dashed line denotes the quasienergies in the adiabatic limit  $\varepsilon_{\alpha,k}^{0}$.
The gap is of order $10^{-6}$;
{\bf b} The charge pumped after the first period, $Q(T)$, as a function of the frequency $\omega$, for the RM model with a suddenly switched-on driving (smooth blue line). The red dotted line is the corresponding diagonal ensemble value and $J_0=(J_1+J_2)/2$
(adapted from Ref.~\cite{privitera_2018}).}
\label{fig:spectrum_rm_deviation}
\end{figure*}

Away from the adiabatic limit $\omega\to 0$,
deviations from the perfect quantization strongly depend on how the system is brought away from the lowest-energy Floquet band. In fact, in the experiment one would be able to prepare an
initial state coinciding with the lowest-energy Floquet
band, such that the deviation from perfect quantization would be
exponentially small. This comes from the fact that, as noticed in Ref.~\cite{Avron_1999}, the quasienergy spectrum contains some crossings giving a non-vanishing winding number. Generically, that crossings turn into avoided crossings with opening of gaps for any finite $T$ --- in the present case (Fig.~\ref{fig:spectrum_rm_deviation}a) at the border of the FBZ ---
implying a deviation from perfect quantization of the pumped charge for the Floquet band under consideration~\cite{Avron_1999}.
Due to the presence of an avoided crossing an exponentially small $1/\omega$ gap appears~\cite{Lindner_2017}.
This implies that the  pumped charge deviates from an integer by terms proportional to the sum of the gaps for $\omega>0$.
This deviation is therefore exponentially small in $1/\omega$. However, this is more an artifact coming from the avoided crossings.

On the other hand one knows that, independently from the initial state, any local observable reaches
a periodic steady state with the same periodicity as the driving~\cite{Russomanno_2012}.
This asymptotic regime is described by the Floquet diagonal density matrix~\cite{Russomanno_2012,Lazarides_2014}. If $Q(mT)$ is the total charge pumped in the first $m$ periods starting from the initial ground state $|\Psi(0)\rangle$ of $\hat{H}(0)$,
the {\em asymptotic} charge pumped is given by the Floquet diagonal ensemble~\cite{Avron_1999}:

\begin{equation} \label{eq:ChargeDiag}
Q_{d} \equiv \lim_{m\to\infty} \frac{Q(mT)}{m} =
\frac{1}{\hbar\omega} \sum_{\alpha} \int_{-\frac{\pi}{a}}^{+\frac{\pi}{a}}  \mathrm{d}k\,  n_{\alpha,k} \frac{\partial \varepsilon_{\alpha,k}}{\partial k} \, ,
\end{equation}
where $n_{\alpha,k}$
is the initial ground-state occupation of the Floquet-Bloch mode labeled by $(\alpha,k)$.
The occupations $n_{\alpha,k}$ can give rise to a stronger deviation from quantization than the gaps, as numerically shown in Ref.~\cite{privitera_2018}.
Starting from the lowest-energy Floquet band $\varepsilon_{{\LE},k}$, and the associated occupations
$n_{{\LE},k}$,
one can develop a perturbation theory in $\omega$ for the Floquet modes, along the lines of Ref.~\cite{Rigolin_2008}, to show that:
\begin{equation} \label{npert2:eqn}
n_{{\LE},k} = 1- f(k, \varphi) \omega^2  + \; ...
\end{equation}
\noindent where $f$ is a function of the pump trajectory parametrized by the pump parameter $\varphi$
of the RM model defined in Eq.~(\ref{eq:rice-mele}), leading
to quadratic corrections to $Q_{d}$ (Fig.~\ref{fig:spectrum_rm_deviation}b).
Indeed one finds that the deviation from quantization
increases as $r^2$, where $r$ is the dimensionless radius of the closed trajectory in the $(\Delta,J_1-J_2)$-parameter space.

Fig.~\ref{fig:spectrum_rm_deviation}b shows the charge pumped after a single cycle, $Q(T)$, as a function of $\omega$: we see that $Q(T)$ exhibits remarkable beating-like oscillations,
on top of the overall quadratic decrease of $Q_{d}$, which become faster and faster as $\omega\to 0$.
Indeed, this  behaviour is compatible with the presence of non-analyticities, possibly of the kind of $\sin(c/\omega)$, where $c$ is a constant~\cite{Avron_1999}. Thus, within a Floquet framework, the Thouless pump is in general not robust to non-adiabatic effects despite its topological nature.

\section{Topological pumping with disorder}
\label{sec:dispump}
A key feature of topological states of matter is their robustness to disorder.
The robustness of topological pumps to disorder has been studied theoretically for different distributions ranging from true-random to quasi-periodic~\cite{wauters_2019,hayward_effect_2020,shi_topological_2020,ippoliti_dimensional_2020,marra_topologically_2020,wang_robustness_2019,qin_quantum_2016}. As discussed in Chapter~\ref{sec:beyond adiabatic}, since the Hamiltonian is time periodic, one can exploit the Floquet representation~\cite{Grifoni_1998} of the evolution operator.
For non-interacting fermions, it suffices to know the single-particle (SP) Floquet states $\ket{\psi_\alpha(t)}$ and their occupation number $n_\alpha $ to explicitly calculate the diagonal ensemble pumped charge~\cite{Avron_1999,Russomanno_2012,privitera_2018}, $Q_d$ [Eq.~(\ref{eq:ChargeDiag})].

\begin{figure}[ht!]
\includegraphics{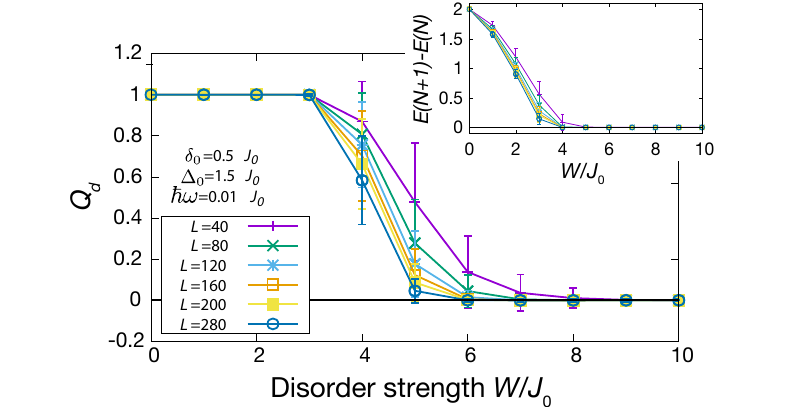}
\caption{Disorder average of $Q_d$ plotted against disorder strength $W$.
The transition between the quantized regime and the trivial one $Q_d=0$ is clearly linked to the closing of the minimum energy gap due to the disorder, shown in the inset. Here, $L$ is the system size, $J_0=(J_1+J_2)/2$ the mean tunnel coupling, $J_1-J_2=2\delta_0\cos\varphi$ the dimerization and $\Delta(\varphi)=2\Delta_0\sin\varphi$ the staggered offset in the RM model defined in Eq.~(\ref{eq:rice-mele}). Inset: minimal many-body gap as a function of disorder strength $W$. (figure adapted from Ref.~\cite{wauters_2019}).}
\label{fig:dis-pump}
\end{figure}

Let us now consider on-site disorder of the form $\hat{H}_{\text{dis}}= \sum_i W \xi_i \opbdag{i} \opb{i}$, where $W$ denotes the strength of the disorder and $\xi_i$ are uniformly distributed random numbers $\xi_i \in [-1/2,1/2]$.
Figure~\ref{fig:dis-pump} shows the disorder-averaged charge pumped per cycle as a function of the disorder strength $W$.
While topological pumping persists for sufficiently small $W\lesssim 3J_0$, it breaks down in
the regime of large $W\gtrsim 8 J_0$, where $Q_d=0$.
The intermediate region $W/J_0\approx 4$ exhibits large sample-to-sample fluctuations. The inset shows a correlation between the drop of $Q_d$
and the closing of the minimal many-body instantaneous gap $\Lambda_N \equiv \min_{t\in [0,T]} [E_{N+1}(t)-E_N(t)]$, where $E_N(t)$
is the $N$-particle ground state energy at time $t$.

The puzzling question is how a disordered 1D system that shows Anderson-localized instantaneous
energy eigenstates and a pure-point spectrum~\cite{Abrahams_1979}, can transport charge.
It has been found that topological pumping only takes place, if there is a significant fraction of {\it delocalized} SP Floquet states~\cite{wauters_2019}, which seems to be in contradiction with the adiabatic limit, where all states are necessarily Anderson localized.
Indeed it was found that the dynamics is adiabatic only at the many-body level. The driving mixes localized SP states, which results in extended Flqouet modes~\cite{Agarwal_2017,Hatami_2016}. This can be clearly shown by looking at the real-space inverse participation ratio (IPR)~\cite{wauters_2019}. For a finite system $\textrm{IPR}_\alpha \in [L^{-1},1]$, where $\textrm{IPR}_\alpha\sim L^{-1}$ signals a completely delocalized (plane-wave-like) state,
while $\textrm{IPR}_\alpha=1$ corresponds to a perfect localization on a single site. As shown in Ref.~\cite{wauters_2019} the IPR shows a localization/delocalization transition at crossover disorder strength $W^* \sim L^{-1/\beta}$ separating the two regimes, vanishing in the thermodynamic limit.
Delocalization renders quantized pumping robust, until
extended Floquet states with opposite winding coalesce
for large disorder. Even though the physics of quantum
pumping in clean systems is the same as the 2D IQHE, this
analogy is not trivial in the presence of disorder.
The competition between (quasi-periodic) disorder and topology has recently been studied in cold atoms~\cite{nakajima_2021} and photonics~\cite{cerjan_2020}. Both studies identify a clear connection between the closing of the gap in the instantaneous energy spectrum and a breakdown of quantized pumping, which can be understood as the result of Landau-Zener transitions induced by the periodic modulation. Interestingly, disorder can also induce topological transport in an otherwise topologically trivial regime, as recently investigated in a cold-atom experiment with quasi-periodic disorder~\cite{nakajima_2021}. This establishes exciting connections to 2D topological Anderson insulators~\cite{li_topological_2009,groth_theory_2009} and anomalous Floquet Anderson insulators~\cite{titum_anomalous_2016}.

\section{Topological pumps with interactions}
\label{sec:disinter}

The interplay between interactions and topology in 1D charge pumps gives rise to a rich variety of topological many-body phenomena and has been studied
both for fermionic~\cite{citro_2003,requist_2017,nakagawa_2018,bertok_splitting_2022,stenzel_quantum_2019} and bosonic atoms~\cite{Berg_2011,qian_quantum_2011,grusdt_realization_2014,zeng_fractional_2016,greschner_topological_2020}. The formalism presented above was based on a description using the instantaneous single-particle eigenstates of the time-dependent Hamiltonian. We start this chapter by presenting the generalized formalism for many-body systems introduced by Thouless and Niu~\cite{Niu_1984}, before discussing a few selected examples in more detail.

\subsection{Generalized many-body formalism}
Originally, it was shown by Thouless that quantization is unaffected by weak interactions under fairly broad assumptions~\cite{Niu_1984}. Starting from the Thouless and Niu paper~\cite{Niu_1984}, one can argue that the charge pump is robust to disorder and interaction as long as the system remains in its ground state during the pump cycle. The demonstration is based on the concept of twisted boundary conditions for the many-particle wavefunction:
\begin{eqnarray}
&& \Phi(x_1, \ldots,x_{i+L}, \ldots, x_N)= \nonumber \\ && \hspace{2cm} e^{i K L}\Phi(x_1, \ldots,x_{i}, \ldots, x_N),
\label{eq:pbc}\end{eqnarray}
where $L$ is the size of the system.
The corresponding Hamiltonian $\hat{H}(K,t)$ in the presence of a slowly time varying potential together with the boundary
condition (\ref{eq:pbc}) describes a 1D system
placed on a ring of length $L$ threaded by a magnetic
flux $2\pi K L/\Phi_0$~\cite{kohn_1964}, where $\Phi_0$ is the magnetic flux quantum. Thus, the current operator is given by
$\partial \hat{H}(K,t)/\partial K$ and one obtains:
\begin{equation}
j(K)=\frac{\partial \epsilon(K)}{\partial K}-\Omega_{K,t},
\end{equation}
where $\Omega_{K,t}$ is the Berry curvature in the many-body manifold $\Omega_{K,t}=
i \left( \langle \partial_K \tilde{\Phi}_0|\partial_{t} \tilde{\Phi}_0\rangle -\langle \partial_{t} \tilde{\Phi}_0|\partial_K \tilde{\Phi}_0\rangle \right)$, where $|\tilde{\Phi}_0\rangle$ is the many-body ground state.
The key step is achieved by realizing that if the Fermi energy lies
in a gap, then the current $j(K)$ should be insensitive to the
boundary condition specified by Eq.~(\ref{eq:pbc})~\cite{Niu_1984,Thouless_1982}.
Consequently one can take the thermodynamic limit and average $j(K)$ over different boundary conditions. Let us note that $K$ and $K+2\pi/L$ describe the same boundary condition in Eq.~(\ref{eq:pbc}). Therefore
the parameter space for $K$ and $t$ is a torus in 2D and the particle transport is given by the Chern number of the occupied band.
According to the previous discussion, it is quantized and can vary only in a discontinuous way.
This is a general outcome of topological invariance. However, deviations from the exact quantization can be observed for intermediate interactions, where numerical simulations have shown that interactions can lead to a breakdown of the quantized pumping by closing the many-body gap~\cite{nakagawa_2018,Lohse_2016}.

\begin{figure*}[ht!]
\includegraphics{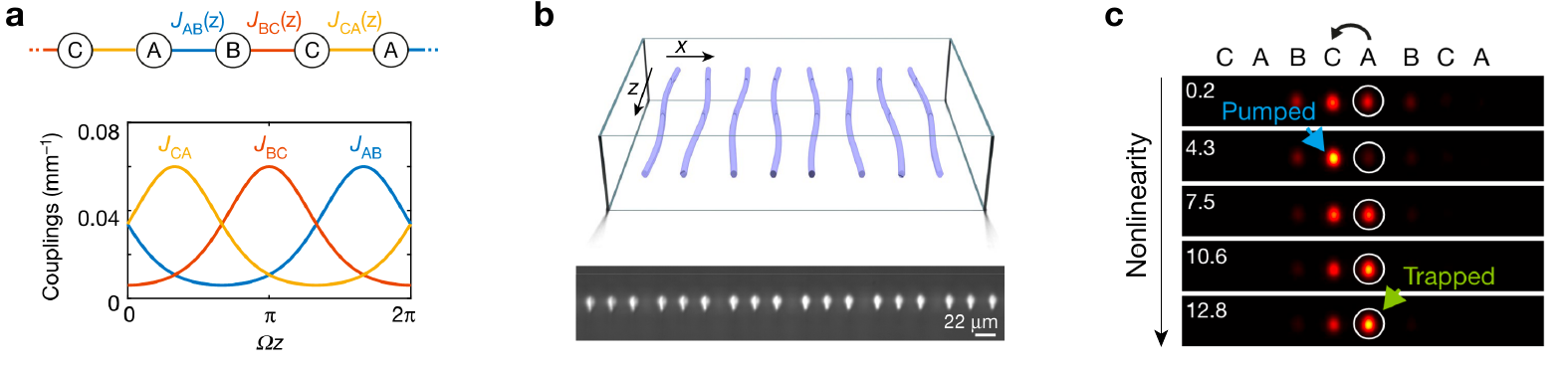}
\caption{{\bf Photonic implementation of a nonlinear topological Thouless pump} (adapted from Ref.~\cite{nonlinear-thouless-pump}). \textbf{a} Schematic illustration of the three-site model (A,B,C) with periodically-modulated couplings $J_n(z)$ between neighboring sites ($n=\{\text{AC}, \text{BC}, \text{CA}\}$) with driving frequency $\Omega$. \textbf{b} Top: Schematic illustration of the 1D waveguide with in-plane modulation. Bottom: Micrograph of the output facet with six unit cells. \textbf{c} Intensity pattern at the output facet after 1/3 of a full period. White circles denote the excited waveguide. Topological pumping (blue arrow) occurs up to a threshold nonlinearity (green arrow).}
\label{fig:nonlinear-pump}
\end{figure*}

\subsection{Nonlinear Thouless pumping}

One of the first experimental realizations of mean-field-type interacting topological pumps was realized in an optical waveguide array with Kerr non-linearity~\cite{nonlinear-thouless-pump}. In these experiments the propagation of monochromatic light is described by a nonlinear Schr\"odinger equation that is equivalent to an attractive Gross-Pitaevskii equation describing interacting bosonic atoms in the mean-field limit. In the experiment an off-diagonal implementation of the Aubry-André-Harper (AAH) model~\cite{aah_model} with three sites
per unit cell labeled as A, B and C is realized  (Fig.~\ref{fig:nonlinear-pump}a). The
corresponding tight-binding Hamiltonian is characterized by uniform on-site potentials and real off-diagonal nearest-neighbor couplings $J_n(z)$ that are periodic functions along the propagation direction $z$, which plays the role of time $t$.
In the linear (non-interacting) regime, one pumping cycle (of the lowest band) can be understood intuitively using the following simplified description. Suppose there is always only a single coupling switched on, such that the intensity couples completely from one site to the next. Starting with a single occupation on site A, after switching on $J_{CA}$, the
occupation shifts to site C. Subsequently, coupling $J_{BC}$ is turned on and finally $J_{AB}$, so that the wavefunction is pumped by one unit cell. The periodic modulation of the couplings is realized by changing the distance between the waveguides (Fig.~\ref{fig:nonlinear-pump}b), which in turn modifies the spatial overlap of neighboring waveguide modes~\cite{nonlinear-thouless-pump}.

In the linear regime quantized pumping is observed for an initial state with uniform occupation of the lowest band. This distribution however strongly depends on the strength of the nonlinearity. Experimentally this is controlled by changing the amount of power that is injected into the system. Surprisingly, quantized pumping is also observed for large nonlinearity~\cite{nonlinear-thouless-pump}, despite non-uniform band occupation (Fig.~\ref{fig:nonlinear-pump}c). This is explained by stable soliton solutions that exist at each instant in time and after one pump cycle are identical, up to translation invariance. Transport is dictated by the Chern number of the band the soliton bifurcates from. However, this constitutes a new mechanism since it does not rely on uniform band occupation, in contrast to Thouless pumping of fermions or hard-core bosons discussed in Sec.~\ref{sec:pump of charge}. These results highlight that nonlinearity or mean-field interactions can induce quantized transport and topological behavior in regimes where the linear limit is topologically trivial. Above a certain threshold nonlinearity matter transfer is completely arrested.

Quantized pumping of solitons was derived more formally by expressing the soliton solutions in the basis of linear Wannier orbitals~\cite{mostaan_quantized_2022,jurgensen_chern_vs_soliton_2022,konotop_2022}. While in the moderate nonlinearity regime, it is sufficient to consider a single band, excited bands need to be included for larger nonlinearities. The topology of these excited bands then determines the direction
and magnitude of the average velocity of the solitons. The motion remains quantized, admits fractional values and can even be fully arrested~\cite{konotop_2022}. Fractionally quantized motion of solitons has been observed recently in 1D coupled waveguide arrays with a five-site unit cell~\cite{jurgensen_quantized_2022}. Beyond optical waveguide arrays,
nonlinear Thouless pumping has also been theoretically proposed in a topological low-dimensional discrete nematic liquid crystal array that realizes a 1D topological Su-Schrieffer-Heeger (SSH) model~\cite{nematic_liquid_crystals}.

\subsection{Topological charge pumping in the interacting Rice-Mele model}
In the limit of hardcore bosons, the interacting 1D charge pump, as reported in Ref.~\cite{Lohse_2016}, has a simple interpretation. It can be mapped onto non-interacting spinless fermions. Hence, for a completely filled band, all quasimomenta are homogeneously populated resulting in quantized transport, as discussed above. This mapping, however, does not hold for finite Hubbard interactions $U$ described by $\hat{H}_{\text{int}}=\frac{U}{2}\sum_i \hat{n}_i(\hat{n}_1-1)$, where $\hat{n}_i=\opadag{i}\opa{i}$ is the bosonic number operator. While it was shown that in the strongly-interacting regime topological charge pumps remain quantized as long as the many-body gap does not close along the cycle~\cite{nakagawa_2018,Lohse_interaction_2018}, a breakdown of quantization occurs as the system enters the superfluid region.

To investigate charge pumps in interacting systems it is convenient to consider the
evolution of the many-body polarization $P(t)$ of a state $\ket{\Psi(t)}$
for the time-periodic pump Hamiltonian $\hat{H}(t+T)=\hat{H}(t)$. In general, the total transported charge $\Delta Q$ can be related to the polarization $P$ (Appendix~\ref{sec:box3}) via the current density $J(t)$:
\begin{equation}
  \label{eq:int-current}
  \Delta Q  =  \int _ {0} ^ {\Delta t } \text{d} t J ( t ) = \frac{1}{a}\int_0^{\Delta t}\text{d}t\,  \partial_t P(t).
\end{equation}
For charge pumps in the adiabatic limit, the polarization is also cyclic in $T$, with $[P(T) \text{mod}\ qa = P(0)]$, where $q$ denotes the charge of the particle (for cold atoms $q=1$) and $a$ is the size of the unit cell.
For $\Delta t = T$ the transported charge $\Delta Q(T)$ then corresponds to the winding number of
the many-body polarization, which implies quantization.

\begin{figure*}[ht!]
  \includegraphics{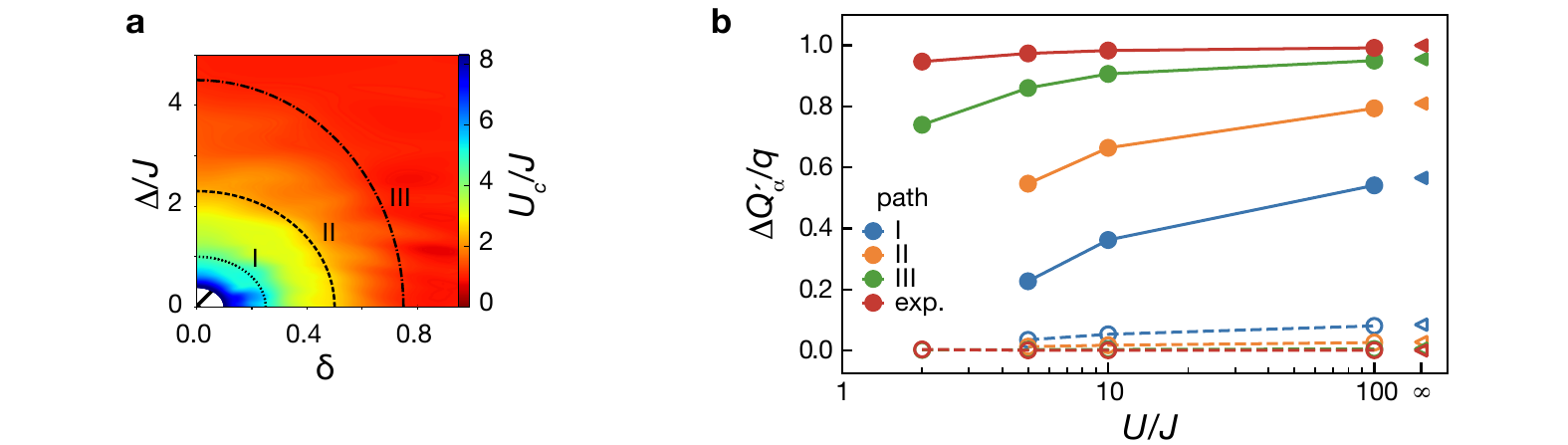}
  \caption{{\bf Transported charge in an interacting RM pump}: \textbf{a} Phase diagram of the interacting bosonic RM model: The color indicates the critical Hubbard interaction strength $U_c$ as a function of the RM model parameters as defined in Eq.~(\ref{eq:rice-mele}) with mean tunneling $J=(J_1+J_2)/2$ and dimerization $\delta=(J_1-J_2)/(2J)$. The lines show the paths of the pump cycles used for the calculations in \textbf{b}, where the dimerization is varied as $\delta(\varphi)=\delta_0\cos\varphi$ and the staggered potential as $\Delta(\varphi)=\Delta_0\sin\varphi$.
 \textbf{b} Theoretical value of the transported charge $\Delta Q'_\alpha/q$ after one pump cycle using the weighted single-particle Berry curvature [Eq.~(\ref{eq:qprime})].
Data shown in red corresponds to the path used in the bosonic charge-pump experiment of Ref.~\cite{Lohse_2016}. Full (empty)
symbols indicate $\Delta Q'_\alpha$ for the lower (upper) band (adapted from Ref.~\cite{Lohse_interaction_2018}).}
\label{fig:delta-q}
\end{figure*}

It is interesting to address the question, for which parameters a single-particle interpretation of the quantized response in the strongly-interacting Mott-insulating regime remains valid. To this end one can consider the momentum-weighted single-particle Berry curvature, defined as $\Omega^W_\alpha(k, \varphi) = \Omega_\alpha(k, \varphi)n_\alpha(k, \varphi)$, where $n_\alpha(k, \varphi)$ denotes the momentum distribution of the interacting system expressed in the single-particle basis and $\varphi$ is the pump parameter. In analogy to the single-particle formalism the transported charge would then be given by the integral of the weighted Berry curvature $\Omega^W_\alpha(k, \varphi)$
\begin{equation}
\label{eq:qprime}
\Delta Q'_\alpha = q\int^{\pi/a}_{-\pi/a} \text{d}k \, \int_0^{2\pi} \text{d}\varphi\, \Omega^W_\alpha(k,\varphi).
\end{equation}
$\Delta Q_\alpha'$ will, in general, deviate from the exactly quantized $\Delta Q_\alpha$ since $n_\alpha (k, \varphi)$ is not flat. The connection between quantized transport in interacting pumps an the single-particle interpretation has been studied numerically in Ref.~\cite{Lohse_interaction_2018} using matrix-product-state (MPS) methods in an infinite-system size formulation (iDMRG)~\cite{schollwocks_2011}. As shown in Fig.~\ref{fig:delta-q} a clear deviation of $\Delta Q'_\alpha/q$ is found for most pump trajectories. This is expected, since in general the interacting ground-state wavefunction is far from a product state and therefore, applying a single-particle picture breaks down. Interestingly, for the experimental trajectory used in Ref.~\cite{Lohse_2016} a single-particle interpretation remains valid even down to very small interaction strength. The reason is that for this particular parametrization of the pump, the atoms remain essentially localized to individual sites or double-wells during the entire evolution along the pump cycle. Recently, breakdown of quantized particle transport was studied experimentally for an interacting fermionic RM model in a dynamical superlattice~\cite{esslinger_int_2022}. Moreover, interaction-induced topological pumping has been predicted in generalized RM models~\cite{lin_interaction-induced_2020,Yoshihito_interaction_induced_pump_2022} as well as for quasi steady-states that exist in a Floquet-prethermal regime~\cite{Lindner_2017,esin_universal_2022}.

Beyond the many-particle framework discussed above, the few-body problem can be treated within a generalized Wannier-state formalism~\cite{multiparticle_wannier}. The general understanding is that the interparticle interaction breaks the translational symmetry of individual particles and the Wannier states cannot be constructed in the usual way. Despite this, it is possible to introduce multiparticle Wannier
states for interacting systems~\cite{multiparticle_wannier} which provide an orthogonal basis for constructing effective Hamiltonians. The shift of multiparticle Wannier state relates to the Chern number of the multiparticle Bloch band allowing for the Thouless pumping of bound states when two or more particles move unidirectionally as a whole.
In general, it's possible to perform topological pumping of cluster and kink excitations which can be related to spin flips by duality transformations~\cite{amico_2020}. This offers a paradigm for multiparticle pumping.

\section{Synthetic dimensions and higher-dimensional systems}

\subsection{1D charge pumps as dynamical realizations of 2D quantum Hall physics}
\label{sec:dynamical_2DQH}

There is a deep connection between 1D topological charge pumps and the 2D IQHE~\cite{ando_1975_theory,Laughlin_1981}, where time plays the role of a second real-space dimension. To illustrate this connection better, let us consider a 1D topological charge pump described by the generalized RM model
\begin{eqnarray}
    \hat{H}(\varphi)= 
   && -\sum_m [ J_m(\varphi) + \delta J_m (\varphi) ] a^\dagger_{m+1} a_{m} + \textrm{h.c.}  \nonumber\\ 
   &&+ \sum_m \Delta_m(\varphi)a^\dagger_{m}a_{m}. 
\label{eq:RMpump}
\end{eqnarray}
This Hamiltonian is periodic in the site index $m$ and the pump parameter $\varphi \in [0,2\pi[$. Therefore the eigenstates of the pump are parametrized by $k_x,\varphi$. For implementations based on cold atoms in bichromatic superlattices~\cite{Lohse_2016,Nakajima_2016}, the local site-dependent potential can be expressed as $\Delta_m(\varphi)=-\Delta \cos(2\pi \alpha m - \varphi)$, where $\alpha=d_s/d_l$ is the ratio of the two lattice constants. Assuming $\delta J_m=0$ it is apparent that Eq.~(\ref{eq:RMpump}) corresponds to the 1D Harper equation, i.e, the 1D eigenvalue equation for states with well defined transverse quasimomentum $k_y$ of the 2D Harper-Hofstadter model~\cite{harper_general_1955,azbel_1964,hofstadter_energy_1976} that describes the dynamics of charged particles on a square lattice with homogeneous magnetic field. Hence, we can identify the phase $\varphi$ with the transverse quasimomentum $k_y$ and the topological charge pump can be interpreted as a dynamical version of the IQHE. The Brillouin zone spanned by $k_x,\varphi$ in 1D is then equivalent to the Brillouin zone spanned by $k_x,k_y$ in the corresponding 2D model. Note, that in the context of localization, Hamiltonian~(\ref{eq:RMpump}) is known as the Aubry-André model~\cite{aubry_1980}, if $\alpha$ is irrational.

The quantization of the transverse Hall conductance in the 2D IQHE is one of the hallmark phenomena of topological condensed matter systems. The plateau values are uniquely defined by the topological invariant characterizing the electron bands, i.e., the (first) Chern number. Similarly, quantized charge transport in 1D Thouless pumps is determined by the first Chern number of the RM model, as we have seen in Sec.~\ref{sec:pump of charge}, and can be interpreted as the dynamical version of the quantum Hall effect. In general, a 1D pump Hamiltonian can be extended to a 2D Hamiltonian via dimensional extension~\cite{kraus_topological_2012,kraus_four-dimensional_2013}
\begin{equation}
    \hat{H}_{2D}=\frac{1}{2\pi}\int_0^{2\pi}\hat{H}(\varphi)\textrm{d}\varphi.
    \label{eq:dimext}
\end{equation}
For $\delta J_m=0$ this yields a square-lattice tight-binding Hamiltonian with uniform flux $2\pi\alpha$
\begin{eqnarray}
    \hat{H}_{\text{HH}}=&& -\sum_{m,n} ( J_x \hat{a}^\dagger_{m+1,n}\hat{a}_{m,n} \nonumber \\
    &&+ \frac{\Delta}{4} \text{e}^{i 2\pi\alpha m}\hat{a}^\dagger_{m,n+1}\hat{a}_{m,n} + \textrm{h.c.}),
\end{eqnarray}
where $\alpha$ corresponds to the spatial periodicity of the potential, which for the superlattice potential introduced in Sec.~\ref{sec:pump of charge} is determined by the value $\alpha=d_l/d_s$. Intriguingly, this paves the way towards studying a variety of different topological pumps, where the Chern number of the lowest band can in principle take arbitrary integer or fractional integer values~\cite{Marra_2015}, which can be experimentally realized by simply adjusting the ratio of the lattice constants $\alpha=d_l/d_s$. This can lead to counterintuitive cases of charge pumping, where the atoms move faster than the sliding lattice~\cite{wei_anomalous_2015}.

\begin{figure*}[ht!]
\includegraphics{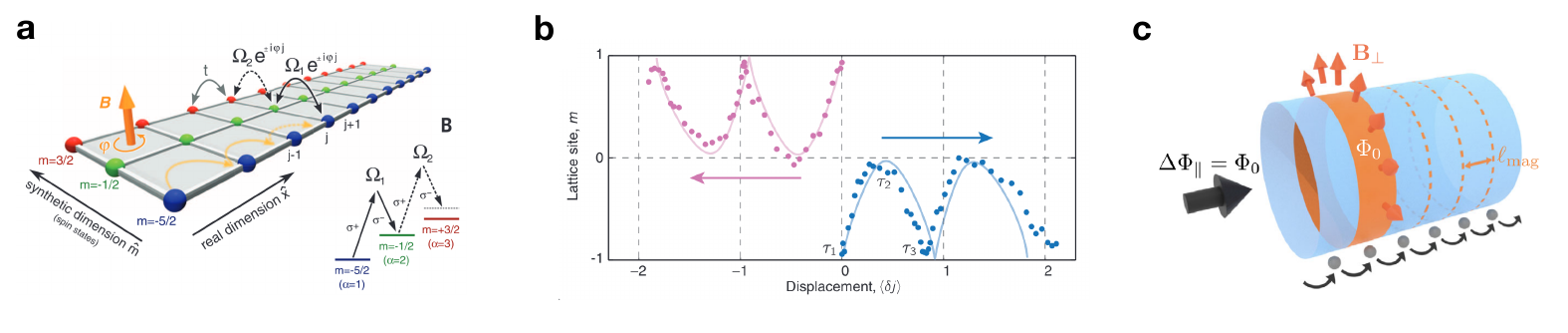}
\caption{\textbf{a} Schematic of synthetic dimensions realized using internal states of fermionic $^{173}$Yb atoms that are coupled by complex two-photon Raman transitions (adapted from Ref.~\cite{mancini_observation_2015}). \textbf{b} Observation of skipping orbits in a synthetic quantum Hall ribbon with $^{87}$Rb atoms (adapted from Ref.~\cite{stuhl_visualizing_2015}). \textbf{c} Schematic drawing of Laughlin's charge pump (adapted from Ref.~\cite{fabre_laughlins_2022}).
}\label{fig:Syn_dim}
\end{figure*}

\subsection{The concept of synthetic dimensions}
The pump parameter can more generally also be seen as a so-called \textit{synthetic dimension}, where conventional real-space degrees of freedom are replaced by any other degrees of freedom that are available in an experimental platform. This concept is very general and has found application in a number of experiments ranging from cold atoms~\cite{celi_synthetic_2014,mancini_observation_2015,stuhl_visualizing_2015,livi_synthetic_2016,kolkowitz_spinorbit-coupled_2017,an_direct_2017} to photonics~\cite{lustig_photonic_2019,ozawa_topological_2019}, where synthetic dimensions have been realized, e.g., in the form of internal levels in an atom~\cite{mancini_observation_2015,stuhl_visualizing_2015} as illustrated in Fig.~\ref{fig:Syn_dim}a, discrete momentum-space degrees of freedom~\cite{an_direct_2017,meier_2018}, or different eigenmodes in photonic waveguide arrays~\cite{lustig_photonic_2019}. This opens the door to a variety of different implementations of 2D quantum Hall physics in engineered, well-controlled quantum systems.

The first cold-atom experiments have realized 2D synthetic quantum Hall ribbons by replacing one spatial direction with internal atomic states that are coupled by additional laser beams. Due to the finite number of internal states, this setting naturally realizes open boundary conditions and facilitated, e.g., the observation of chiral edge modes~\cite{mancini_observation_2015,stuhl_visualizing_2015}, as shown in Fig.~\ref{fig:Syn_dim}b. The experimental realization of Laughlin's charge pump, however, requires periodic boundary conditions and an adiabatic change of the axial magnetic flux. In his gedankenexperiment Laughlin considered a quantum Hall state on a cylinder as shown in Fig.~\ref{fig:Syn_dim}c, where the perpendicular magnetic field is pointing radially outwards. When the additional axial magnetic flux is varied adiabatically in time, a quantized axial transport is induced between the two ends of the cylinder. In contrast to real-space lattices, synthetic dimensions facilitate the realization of periodic boundary conditions~\cite{han_band_2019,liang_coherence_2021,li_bose-einstein_2022}. Moreover, by controlling the phases of the laser-induced spin-orbit couplings an additional varying axial magnetic flux was recently realized in synthetic quantum Hall cylinders using Dy atoms, which facilitated the first experimental realization of Laughlin's charge pump~\cite{fabre_laughlins_2022}.

\subsection{2D topological pumps -- exploring 4D quantum Hall physics}
\label{sec:pump_4D}
Synthetic parameter spaces provide an excited path towards studying higher-dimensional systems that cannot be accessed with experiments in three real-space dimension, such as 4D quantum Hall physics~\cite{price_four-dimensional_2015,ozawa_synthetic_2016,lu_topological_2018,kolodrubetz_measuring_2016}.
Using similar arguments as described above, the concept of dimensional extension offers a direct path towards studying higher-dimensional topological systems~\cite{kraus_four-dimensional_2013}.

\begin{figure*}[ht!]
\includegraphics[width=\textwidth]{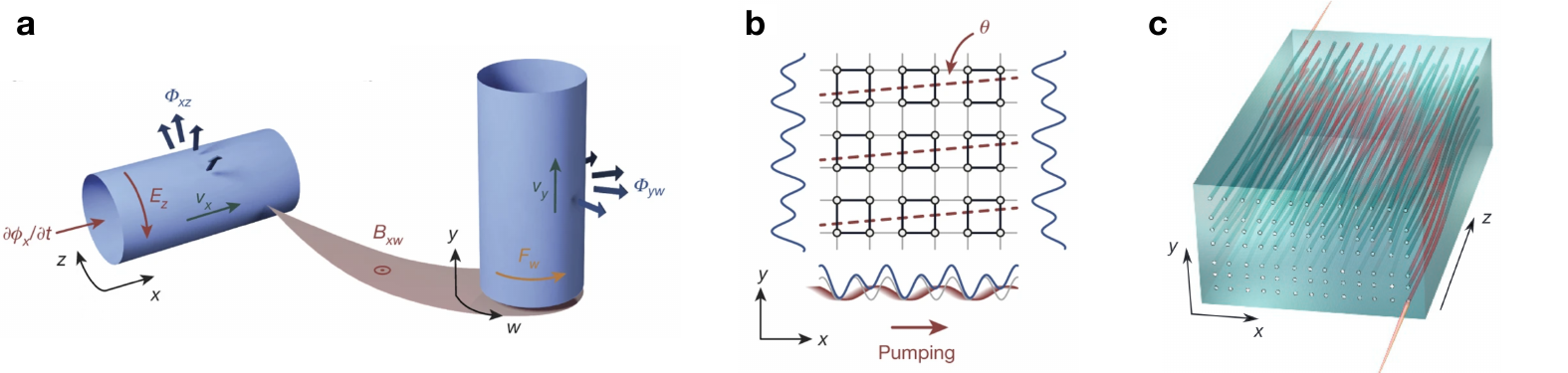}
\caption{\textbf{a} Schematic of a 4D quantum Hall system, which can be composed of two 2D quantum Hall systems in the $x-z$ and $y-w$ planes (adapted from Ref.~\cite{lohse2018}). \textbf{b} Schematic drawing of the experimental realization of the 2D pump using optical superlattices (adapted from Ref.~\cite{lohse2018}). \textbf{c} Topological edge transport in a coupled waveguide array realizing a 2D topological charge pump (adapted from Ref.~\cite{zilberberg_photonic_2018}).
}\label{fig:4dqh}
\end{figure*}

Indeed the quantized nonlinear response in 4D arises in the presence of two perturbing external fields, an electric field and a magnetic field that couples the two 2D quantum Hall systems. The resulting response is then orthogonal to both perturbing fields and is proportional to the so-called second Chern number. The corresponding 2D topological pump was realized with cold atoms in a 2D superlattice~\cite{lohse2018} with phases $\varphi_x$ and $\varphi_y$, where the phase $\varphi_y$ depends on the position along $x$. The pump parameter or the electric field is implemented by varying $\varphi_x$, while the magnetic perturbation is realized by the spatial dependence of $\varphi_y$ which couples transport in the two 1D charge pumps. This induces a nonlinear response along $y$, which is determined by the second Chern number $\nu_2= \frac{1}{4\pi^2} \oint_{BZ} \Omega^x \Omega^y dk_x dk_y d\varphi_x d\varphi_y$,
where BZ denotes the 4D Brillouin zone and $\Omega^{x}(k_x,\varphi_x)$ and $\Omega^{x}(k_y,\varphi_y)$ are the respective Berry curvatures along $x$ and $y$.

At the same time a similar implementation has been reported with coupled photonic waveguide arrays, where topological edge pumping in a 2D array is supported by a non-zero second Chern number~\cite{zilberberg_photonic_2018}. The second Chern number has further been measured in an artificial parameter space, which was realized by a cyclic coupling of four internal levels of bosonic $^{87}$Rb atoms~\cite{sugawa_second_2018}. Beyond cold atoms, higher-dimensional pumping has been achieved by replicating a 2D-Chern insulator in one spatial and one temporal dimension using a 1D metamaterial composed of magnetically coupled mechanical resonators~\cite{grinberg_2020}.

\section{Higher-Order Symmetry-Protected Topological Phases}

Intriguingly, topological charge pumps can also be used to characterize the topological properties of higher-order topological insulators (HOTIs) and higher-order symmetry-protected topological phases (HOSPTs)~\cite{Benalcazar2017, Benalcazar2017a}, where an $n$-dimensional bulk with topology of order $m$ can exhibit ($n-m$)-dimensional boundary modes. Such systems have been realized, for instance, in solids and classical meta-materials~\cite{Benalcazar2020, Noguchi2021,Peterson2018, Imhof2018, SerraGarcia2018, Bao2019, Mittal2019, Ni2019, Ni2020, Xue2018, Dutt2020}. In order to characterize their topological properties, efforts have been made to define higher-order Thouless pumps~\cite{Benalcazar2020, Benalcazar2017a, Kang2019, Petrides2020, Kang2021,wienand_2021}. The concept of higher-order topological pumps developed in Ref.~\cite{wienand_2021} based on the specific example of interacting bosonic $C_4\times\mathbb{Z}_2$-symmetric HOSPTs can be summarized as follows: By introducing corner-periodic boundary conditions (CPBC), Resta's argument on polarization~\cite{Resta1998} can be extended to higher-order systems. CPBCs are realized by adding links that connect the corners of the underlying 2D system. Based on these CPBCs a tuple of Zak (Berry) phases can be defined that act as topological invariants for HOSPTs and govern charge transport through the corners during Thouless pumping in a direction-dependent manner.

The generalization of the Zak (Berry) phases for higher-order systems~\cite{araki_berry_2020,wienand_2021} relies on the idea of magnetic flux insertion through the two ``super-cells'', which are delimited by the edge and the corner-connecting links and meet at the corner labeled by the index $i$. This process is associated with an induced electric field pointing along a diagonal and can be formally described by gauge transformations $\hat{U}_i$, $i\in \{1,2,3,4\}$, applied \emph{only} to the corner-parts of a Hamiltonian $\hat{H}^{\rm{C}}$:
\begin{equation}
  \hat{H}_{i}^{\rm{C}}(\theta) = \hat{U}^{\dagger}_i(\theta) \hat{H}^{\rm{C}} \hat{U}_{i}(\theta) ,
  \label{eq:gauge_hamil}
\end{equation}
with $\hat{U}_{i}(\theta) = e^{i\,\theta \hat{n}_{c_{i}}}$ and
$\hat{n}_{c_{i}}$ is the particle number operator at the $i$-th corner.
Separately for each gauge choice $\hat{U}_i$, one can define a higher-order Zak (Berry) phase~$\gamma_i$ as the geometric phase picked up by the ground state wavefunction $\ket{\psi_i(\theta)}$ of $\hat{H}_i(\theta)$, when changing $\theta$ from $0$ to $2 \pi$~\cite{Berry_1984,zak_1989}:
\begin{equation}
    \gamma_i = \oint_{0}^{2 \pi} d\mathbf{\theta} ~ \bra{\psi_i(\theta)} i \partial_\theta \ket{\psi_i(\theta)}.
    \label{eq:berryphase}
\end{equation}
We note that, like the 1D Zak phase~\cite{zak_1989}, the higher-order version $\gamma_i$ is gauge dependent, while the difference $\Delta\gamma_i$ is gauge-invariant.
The final step is to relate the higher-order Zak (Berry) phase to charge transport.
Indeed the adiabatic flux insertion in Eq.~\eqref{eq:gauge_hamil} can be directly related to the current passing diagonally through a corner, $\hat{J}_{i} = \partial_\theta \hat{H}_i (\theta)|_{\theta=0}$ for $i=1,...,4$. Integrating these currents along an adiabatic path that connects two HOSPTs yields the change of the charge $\Delta q_{c_i}$ in corner $i$,  i.e.
\begin{equation}
    \Delta q_{c_i} = -\frac{\Delta \gamma_i}{2 \pi}.
    \label{eqDqciResult}
\end{equation}
Since $\gamma_i$ is quantized by $C_4 \times \mathbb{Z}_2$-symmetry, it follows that the corner charge $\Delta q_{c_i}$ is also quantized and represents an intrinsic topological invariant. The total amount of charge $\Delta Q_{c_i} = \oint dq_{c_i}$ transported during one full pumping cycle, or equivalently, the amount of charge piling up at the corners in a system with OBC, can be measured by four Chern numbers $\nu_i$ with $i\in\{1,2,3,4\}$ which are obtained as winding numbers of the higher-order Zak (Berry) phase.
Since the Zak (Berry) phase is defined mod $2 \pi$, it follows directly that the Chern numbers, $\nu_i$, and the associated bulk charge transport is integer quantized.

\section{Outlook and  conclusions}

Quantum pumps are transport mechanisms where a dc current results from a cyclic evolution of
the potential~\cite{altshuler_1999}. Starting from the Thouless pump,
we have discussed the topological origin of the pumping process,
when considering the motion of quantum particles in spatially and temporally periodic potentials~\cite{Thouless_1983}. Initially, the periodic evolution that drives these pumps has mostly been assumed to be adiabatic and the transported particles were considered noninteracting. Within this review we have discussed quantum pumping away from these limits. Indeed, interactions and disorder can lead to a breakdown of particle quantization and nonadiabatic phenomena may affect the occupation of the lowest band in a detrimental way. At the same time, however, disorder and interaction effects can induce topological transport in regimes that are otherwise trivial and the concept of synthetic dimensions enables studies of higher-dimensional topological systems highlighting the rich properties of topological pumps. \\
Different generalizations of the concept of Thouless pumping have recently appeared. One of this is the concept of {\it non-Abelian} Thouless pumping~\cite{brosco_2021}. It  yields a displacement across the lattice but it also generates a holonomic transformation among the different bands, extending the known results of Resta and Vanderbilt~\cite{resta_1994}. It hints at the possibility to detect the signatures of the Wilczek-Zee connection in solids; furthermore, it points at the possibility to investigate the role of non-Abelian holonomies also in quantum Hall experiments.
Moreover, in a standard Thouless pump the drive is usually imparted from outside, as was the case in the experimental systems studied so far. However, recently, an emergent mechanism for geometric pumping in a quantum gas coupled to
an optical resonator has been reported, where a particle
current has been observed without applying a periodic drive~\cite{esslinger_2022}. The
pumping potential experienced by the atoms is
formed by the self-consistent cavity field interfering with the static laser field driving the atoms~\cite{esslinger_2022}. Last but not least an exciting new direction has been opened with the realization of nonlinear Thouless pumps in optical waveguide arrays~\cite{nonlinear-thouless-pump,jurgensen_quantized_2022} that combines topology and the dynamics of solitons. \\
Despite its longstanding discovery, Thouless pumps still deserve special interest and the recent developments within cold atoms and photonics raises interesting questions that call for future investigations. Natural extensions would be realising a dissipative topological time crystal or dissipative $Z_2$ spin pumps. In addition, the Wilczek-Zee connection in solids comes within reach through a Thouless pumping mechanism.

\section*{Acknowledgements}
The authors thank all of their collaborators on the topic of topological pumping and related work, with whom they have had many stimulating interactions and discussions. M.~A. acknowledges funding from the Deutsche Forschungsgemeinschaft (DFG, German Research Foundation) via Research Unit FOR 2414 under project number 277974659, and under Germany’s Excellence Strategy – EXC-2111 – 390814868. R.C. acknowledges the project Quantox of QuantERA ERA-NET Cofund in Quantum Technologies (Grant Agreement N. 731473).

\bibliography{sample}

\begin{thebibliography}{149}%
\makeatletter
\providecommand \@ifxundefined [1]{%
 \@ifx{#1\undefined}
}%
\providecommand \@ifnum [1]{%
 \ifnum #1\expandafter \@firstoftwo
 \else \expandafter \@secondoftwo
 \fi
}%
\providecommand \@ifx [1]{%
 \ifx #1\expandafter \@firstoftwo
 \else \expandafter \@secondoftwo
 \fi
}%
\providecommand \natexlab [1]{#1}%
\providecommand \enquote  [1]{``#1''}%
\providecommand \bibnamefont  [1]{#1}%
\providecommand \bibfnamefont [1]{#1}%
\providecommand \citenamefont [1]{#1}%
\providecommand \href@noop [0]{\@secondoftwo}%
\providecommand \href [0]{\begingroup \@sanitize@url \@href}%
\providecommand \@href[1]{\@@startlink{#1}\@@href}%
\providecommand \@@href[1]{\endgroup#1\@@endlink}%
\providecommand \@sanitize@url [0]{\catcode `\\12\catcode `\$12\catcode
  `\&12\catcode `\#12\catcode `\^12\catcode `\_12\catcode `\%12\relax}%
\providecommand \@@startlink[1]{}%
\providecommand \@@endlink[0]{}%
\providecommand \url  [0]{\begingroup\@sanitize@url \@url }%
\providecommand \@url [1]{\endgroup\@href {#1}{\urlprefix }}%
\providecommand \urlprefix  [0]{URL }%
\providecommand \Eprint [0]{\href }%
\providecommand \doibase [0]{https://doi.org/}%
\providecommand \selectlanguage [0]{\@gobble}%
\providecommand \bibinfo  [0]{\@secondoftwo}%
\providecommand \bibfield  [0]{\@secondoftwo}%
\providecommand \translation [1]{[#1]}%
\providecommand \BibitemOpen [0]{}%
\providecommand \bibitemStop [0]{}%
\providecommand \bibitemNoStop [0]{.\EOS\space}%
\providecommand \EOS [0]{\spacefactor3000\relax}%
\providecommand \BibitemShut  [1]{\csname bibitem#1\endcsname}%
\let\auto@bib@innerbib\@empty
\bibitem [{\citenamefont {Altshuler}\ and\ \citenamefont
  {Glazman}(1999)}]{altshuler_1999}%
  \BibitemOpen
  \bibfield  {author} {\bibinfo {author} {\bibfnamefont {B.}~\bibnamefont
  {Altshuler}}\ and\ \bibinfo {author} {\bibfnamefont {L.}~\bibnamefont
  {Glazman}},\ }\bibfield  {title} {\bibinfo {title} {Pumping electrons},\
  }\href {https://doi.org/10.1126/science.283.5409.18} {\bibfield  {journal}
  {\bibinfo  {journal} {Science}\ }\textbf {\bibinfo {volume} {283}},\ \bibinfo
  {pages} {1864} (\bibinfo {year} {1999})}\BibitemShut {NoStop}%
\bibitem [{\citenamefont {Zhou}\ \emph {et~al.}(1999)\citenamefont {Zhou},
  \citenamefont {Spivak},\ and\ \citenamefont {Altshuler}}]{zhou_1999}%
  \BibitemOpen
  \bibfield  {author} {\bibinfo {author} {\bibfnamefont {F.}~\bibnamefont
  {Zhou}}, \bibinfo {author} {\bibfnamefont {B.}~\bibnamefont {Spivak}},\ and\
  \bibinfo {author} {\bibfnamefont {B.}~\bibnamefont {Altshuler}},\ }\bibfield
  {title} {\bibinfo {title} {Mesoscopic mechanism of adiabatic charge
  transport},\ }\href {https://doi.org/10.1103/PhysRevLett.82.608} {\bibfield
  {journal} {\bibinfo  {journal} {Phys. Rev. Lett.}\ }\textbf {\bibinfo
  {volume} {82}},\ \bibinfo {pages} {608} (\bibinfo {year} {1999})}\BibitemShut
  {NoStop}%
\bibitem [{\citenamefont {Niu}(1990)}]{Niu_1990}%
  \BibitemOpen
  \bibfield  {author} {\bibinfo {author} {\bibfnamefont {Q.}~\bibnamefont
  {Niu}},\ }\bibfield  {title} {\bibinfo {title} {Towards a quantum pump of
  electric charges},\ }\href {https://doi.org/10.1103/PhysRevLett.64.1812}
  {\bibfield  {journal} {\bibinfo  {journal} {Phys. Rev. Lett.}\ }\textbf
  {\bibinfo {volume} {64}},\ \bibinfo {pages} {1812} (\bibinfo {year}
  {1990})}\BibitemShut {NoStop}%
\bibitem [{\citenamefont {Pekola}\ \emph {et~al.}(2013)\citenamefont {Pekola},
  \citenamefont {Saira}, \citenamefont {Maisi}, \citenamefont {Kemppinen},
  \citenamefont {M\"ott\"onen}, \citenamefont {Pashkin},\ and\ \citenamefont
  {Averin}}]{Pekola_RMP_2013}%
  \BibitemOpen
  \bibfield  {author} {\bibinfo {author} {\bibfnamefont {J.~P.}\ \bibnamefont
  {Pekola}}, \bibinfo {author} {\bibfnamefont {O.-P.}\ \bibnamefont {Saira}},
  \bibinfo {author} {\bibfnamefont {V.~F.}\ \bibnamefont {Maisi}}, \bibinfo
  {author} {\bibfnamefont {A.}~\bibnamefont {Kemppinen}}, \bibinfo {author}
  {\bibfnamefont {M.}~\bibnamefont {M\"ott\"onen}}, \bibinfo {author}
  {\bibfnamefont {Y.~A.}\ \bibnamefont {Pashkin}},\ and\ \bibinfo {author}
  {\bibfnamefont {D.~V.}\ \bibnamefont {Averin}},\ }\bibfield  {title}
  {\bibinfo {title} {Single-electron current sources: Toward a refined
  definition of the ampere},\ }\href
  {https://doi.org/10.1103/RevModPhys.85.1421} {\bibfield  {journal} {\bibinfo
  {journal} {Rev. Mod. Phys.}\ }\textbf {\bibinfo {volume} {85}},\ \bibinfo
  {pages} {1421} (\bibinfo {year} {2013})}\BibitemShut {NoStop}%
\bibitem [{\citenamefont {Das}\ \emph {et~al.}(2006)\citenamefont {Das},
  \citenamefont {Kim},\ and\ \citenamefont {Mizel}}]{Das_2006}%
  \BibitemOpen
  \bibfield  {author} {\bibinfo {author} {\bibfnamefont {K.~K.}\ \bibnamefont
  {Das}}, \bibinfo {author} {\bibfnamefont {S.}~\bibnamefont {Kim}},\ and\
  \bibinfo {author} {\bibfnamefont {A.}~\bibnamefont {Mizel}},\ }\bibfield
  {title} {\bibinfo {title} {Controlled flow of spin-entangled electrons via
  adiabatic quantum pumping},\ }\href
  {https://doi.org/10.1103/PhysRevLett.97.096602} {\bibfield  {journal}
  {\bibinfo  {journal} {Phys. Rev. Lett.}\ }\textbf {\bibinfo {volume} {97}},\
  \bibinfo {pages} {096602} (\bibinfo {year} {2006})}\BibitemShut {NoStop}%
\bibitem [{\citenamefont {Kraus}\ \emph
  {et~al.}(2012{\natexlab{a}})\citenamefont {Kraus}, \citenamefont {Lahini},
  \citenamefont {Ringel}, \citenamefont {Verbin},\ and\ \citenamefont
  {Zilberberg}}]{kraus_2012}%
  \BibitemOpen
  \bibfield  {author} {\bibinfo {author} {\bibfnamefont {Y.~E.}\ \bibnamefont
  {Kraus}}, \bibinfo {author} {\bibfnamefont {Y.}~\bibnamefont {Lahini}},
  \bibinfo {author} {\bibfnamefont {Z.}~\bibnamefont {Ringel}}, \bibinfo
  {author} {\bibfnamefont {M.}~\bibnamefont {Verbin}},\ and\ \bibinfo {author}
  {\bibfnamefont {O.}~\bibnamefont {Zilberberg}},\ }\bibfield  {title}
  {\bibinfo {title} {Topological states and adiabatic pumping in
  quasicrystals},\ }\href {https://doi.org/10.1103/PhysRevLett.109.106402}
  {\bibfield  {journal} {\bibinfo  {journal} {Phys. Rev. Lett.}\ }\textbf
  {\bibinfo {volume} {109}},\ \bibinfo {pages} {106402} (\bibinfo {year}
  {2012}{\natexlab{a}})}\BibitemShut {NoStop}%
\bibitem [{\citenamefont {Verbin}\ \emph {et~al.}(2015)\citenamefont {Verbin},
  \citenamefont {Zilberberg}, \citenamefont {Lahini}, \citenamefont {Kraus},\
  and\ \citenamefont {Silberberg}}]{verbin_2015}%
  \BibitemOpen
  \bibfield  {author} {\bibinfo {author} {\bibfnamefont {M.}~\bibnamefont
  {Verbin}}, \bibinfo {author} {\bibfnamefont {O.}~\bibnamefont {Zilberberg}},
  \bibinfo {author} {\bibfnamefont {Y.}~\bibnamefont {Lahini}}, \bibinfo
  {author} {\bibfnamefont {Y.~E.}\ \bibnamefont {Kraus}},\ and\ \bibinfo
  {author} {\bibfnamefont {Y.}~\bibnamefont {Silberberg}},\ }\bibfield  {title}
  {\bibinfo {title} {Topological pumping over a photonic fibonacci
  quasicrystal},\ }\href {https://doi.org/10.1103/PhysRevB.91.064201}
  {\bibfield  {journal} {\bibinfo  {journal} {Phys. Rev. B}\ }\textbf {\bibinfo
  {volume} {91}},\ \bibinfo {pages} {064201} (\bibinfo {year}
  {2015})}\BibitemShut {NoStop}%
\bibitem [{\citenamefont {Ke}\ \emph {et~al.}(2016)\citenamefont {Ke},
  \citenamefont {Qin}, \citenamefont {Mei}, \citenamefont {Zhong},
  \citenamefont {Kivshar},\ and\ \citenamefont
  {Lee}}]{ke_photonic_waveguide_2016}%
  \BibitemOpen
  \bibfield  {author} {\bibinfo {author} {\bibfnamefont {Y.}~\bibnamefont
  {Ke}}, \bibinfo {author} {\bibfnamefont {X.}~\bibnamefont {Qin}}, \bibinfo
  {author} {\bibfnamefont {F.}~\bibnamefont {Mei}}, \bibinfo {author}
  {\bibfnamefont {H.}~\bibnamefont {Zhong}}, \bibinfo {author} {\bibfnamefont
  {Y.~S.}\ \bibnamefont {Kivshar}},\ and\ \bibinfo {author} {\bibfnamefont
  {C.}~\bibnamefont {Lee}},\ }\bibfield  {title} {\bibinfo {title} {Topological
  phase transitions and thouless pumping of light in photonic waveguide
  arrays},\ }\href {https://doi.org/10.1002/lpor.201600119} {\bibfield
  {journal} {\bibinfo  {journal} {Laser \& Photonics Reviews}\ }\textbf
  {\bibinfo {volume} {10}},\ \bibinfo {pages} {995} (\bibinfo {year}
  {2016})}\BibitemShut {NoStop}%
\bibitem [{\citenamefont {Cerjan}\ \emph {et~al.}(2020)\citenamefont {Cerjan},
  \citenamefont {Wang}, \citenamefont {Huang}, \citenamefont {Chen},\ and\
  \citenamefont {Rechtsman}}]{cerjan_2020}%
  \BibitemOpen
  \bibfield  {author} {\bibinfo {author} {\bibfnamefont {A.}~\bibnamefont
  {Cerjan}}, \bibinfo {author} {\bibfnamefont {M.}~\bibnamefont {Wang}},
  \bibinfo {author} {\bibfnamefont {S.}~\bibnamefont {Huang}}, \bibinfo
  {author} {\bibfnamefont {K.~P.}\ \bibnamefont {Chen}},\ and\ \bibinfo
  {author} {\bibfnamefont {M.~C.}\ \bibnamefont {Rechtsman}},\ }\bibfield
  {title} {\bibinfo {title} {Thouless pumping in disordered photonic systems},\
  }\href {https://doi.org/10.1038/s41377-020-00408-2} {\bibfield  {journal}
  {\bibinfo  {journal} {Light: Science \& Applications}\ }\textbf {\bibinfo
  {volume} {9}},\ \bibinfo {pages} {1} (\bibinfo {year} {2020})}\BibitemShut
  {NoStop}%
\bibitem [{\citenamefont {Grinberg}\ \emph {et~al.}(2020)\citenamefont
  {Grinberg}, \citenamefont {Lin}, \citenamefont {Harris}, \citenamefont
  {Benalcazar}, \citenamefont {Peterson}, \citenamefont {Hughes},\ and\
  \citenamefont {Bahl}}]{grinberg_2020}%
  \BibitemOpen
  \bibfield  {author} {\bibinfo {author} {\bibfnamefont {I.~H.}\ \bibnamefont
  {Grinberg}}, \bibinfo {author} {\bibfnamefont {M.}~\bibnamefont {Lin}},
  \bibinfo {author} {\bibfnamefont {C.}~\bibnamefont {Harris}}, \bibinfo
  {author} {\bibfnamefont {W.~A.}\ \bibnamefont {Benalcazar}}, \bibinfo
  {author} {\bibfnamefont {C.~W.}\ \bibnamefont {Peterson}}, \bibinfo {author}
  {\bibfnamefont {T.~L.}\ \bibnamefont {Hughes}},\ and\ \bibinfo {author}
  {\bibfnamefont {G.}~\bibnamefont {Bahl}},\ }\bibfield  {title} {\bibinfo
  {title} {Robust temporal pumping in a magneto-mechanical topological
  insulator},\ }\href {https://doi.org/10.1038/s41467-020-14804-0} {\bibfield
  {journal} {\bibinfo  {journal} {Nat. Commun.}\ }\textbf {\bibinfo {volume}
  {11}},\ \bibinfo {pages} {1} (\bibinfo {year} {2020})}\BibitemShut {NoStop}%
\bibitem [{\citenamefont {Xia}\ \emph {et~al.}(2021)\citenamefont {Xia},
  \citenamefont {Riva}, \citenamefont {Rosa}, \citenamefont {Cazzulani},
  \citenamefont {Erturk}, \citenamefont {Braghin},\ and\ \citenamefont
  {Ruzzene}}]{electromechanical_waveguide}%
  \BibitemOpen
  \bibfield  {author} {\bibinfo {author} {\bibfnamefont {Y.}~\bibnamefont
  {Xia}}, \bibinfo {author} {\bibfnamefont {E.}~\bibnamefont {Riva}}, \bibinfo
  {author} {\bibfnamefont {M.~I.~N.}\ \bibnamefont {Rosa}}, \bibinfo {author}
  {\bibfnamefont {G.}~\bibnamefont {Cazzulani}}, \bibinfo {author}
  {\bibfnamefont {A.}~\bibnamefont {Erturk}}, \bibinfo {author} {\bibfnamefont
  {F.}~\bibnamefont {Braghin}},\ and\ \bibinfo {author} {\bibfnamefont
  {M.}~\bibnamefont {Ruzzene}},\ }\bibfield  {title} {\bibinfo {title}
  {Experimental observation of temporal pumping in electromechanical
  waveguides},\ }\href {https://doi.org/10.1103/PhysRevLett.126.095501}
  {\bibfield  {journal} {\bibinfo  {journal} {Phys. Rev. Lett.}\ }\textbf
  {\bibinfo {volume} {126}},\ \bibinfo {pages} {095501} (\bibinfo {year}
  {2021})}\BibitemShut {NoStop}%
\bibitem [{\citenamefont {Lohse}\ \emph {et~al.}(2016)\citenamefont {Lohse},
  \citenamefont {Schweizer}, \citenamefont {Zilberberg}, \citenamefont
  {Aidelsburger},\ and\ \citenamefont {Bloch}}]{Lohse_2016}%
  \BibitemOpen
  \bibfield  {author} {\bibinfo {author} {\bibfnamefont {M.}~\bibnamefont
  {Lohse}}, \bibinfo {author} {\bibfnamefont {C.}~\bibnamefont {Schweizer}},
  \bibinfo {author} {\bibfnamefont {O.}~\bibnamefont {Zilberberg}}, \bibinfo
  {author} {\bibfnamefont {M.}~\bibnamefont {Aidelsburger}},\ and\ \bibinfo
  {author} {\bibfnamefont {I.}~\bibnamefont {Bloch}},\ }\bibfield  {title}
  {\bibinfo {title} {A thouless quantum pump with ultracold bosonic atoms in an
  optical superlattice},\ }\href {https://doi.org/10.1038/nphys3584} {\bibfield
   {journal} {\bibinfo  {journal} {Nature Physics}\ }\textbf {\bibinfo {volume}
  {12}},\ \bibinfo {pages} {350} (\bibinfo {year} {2016})}\BibitemShut
  {NoStop}%
\bibitem [{\citenamefont {Nakajima}\ \emph {et~al.}(2016)\citenamefont
  {Nakajima}, \citenamefont {Tomita}, \citenamefont {Taie}, \citenamefont
  {Ichinose}, \citenamefont {Ozawa}, \citenamefont {Wang}, \citenamefont
  {Troyer},\ and\ \citenamefont {Takahashi}}]{Nakajima_2016}%
  \BibitemOpen
  \bibfield  {author} {\bibinfo {author} {\bibfnamefont {S.}~\bibnamefont
  {Nakajima}}, \bibinfo {author} {\bibfnamefont {T.}~\bibnamefont {Tomita}},
  \bibinfo {author} {\bibfnamefont {S.}~\bibnamefont {Taie}}, \bibinfo {author}
  {\bibfnamefont {T.}~\bibnamefont {Ichinose}}, \bibinfo {author}
  {\bibfnamefont {H.}~\bibnamefont {Ozawa}}, \bibinfo {author} {\bibfnamefont
  {L.}~\bibnamefont {Wang}}, \bibinfo {author} {\bibfnamefont {M.}~\bibnamefont
  {Troyer}},\ and\ \bibinfo {author} {\bibfnamefont {Y.}~\bibnamefont
  {Takahashi}},\ }\bibfield  {title} {\bibinfo {title} {Topological thouless
  pumping of ultracold fermions},\ }\href {https://doi.org/10.1038/nphys3622}
  {\bibfield  {journal} {\bibinfo  {journal} {Nat. Phys.}\ }\textbf {\bibinfo
  {volume} {12}},\ \bibinfo {pages} {296} (\bibinfo {year} {2016})}\BibitemShut
  {NoStop}%
\bibitem [{\citenamefont {Lu}\ \emph {et~al.}(2016)\citenamefont {Lu},
  \citenamefont {Schemmer}, \citenamefont {Aycock}, \citenamefont {Genkina},
  \citenamefont {Sugawa},\ and\ \citenamefont {Spielman}}]{Spielman_2016}%
  \BibitemOpen
  \bibfield  {author} {\bibinfo {author} {\bibfnamefont {H.-I.}\ \bibnamefont
  {Lu}}, \bibinfo {author} {\bibfnamefont {M.}~\bibnamefont {Schemmer}},
  \bibinfo {author} {\bibfnamefont {L.~M.}\ \bibnamefont {Aycock}}, \bibinfo
  {author} {\bibfnamefont {D.}~\bibnamefont {Genkina}}, \bibinfo {author}
  {\bibfnamefont {S.}~\bibnamefont {Sugawa}},\ and\ \bibinfo {author}
  {\bibfnamefont {I.~B.}\ \bibnamefont {Spielman}},\ }\bibfield  {title}
  {\bibinfo {title} {Geometrical pumping with a bose-einstein condensate},\
  }\href {https://doi.org/10.1103/PhysRevLett.116.200402} {\bibfield  {journal}
  {\bibinfo  {journal} {Phys. Rev. Lett.}\ }\textbf {\bibinfo {volume} {116}},\
  \bibinfo {pages} {200402} (\bibinfo {year} {2016})}\BibitemShut {NoStop}%
\bibitem [{\citenamefont {Tangpanitanon}\ \emph {et~al.}(2016)\citenamefont
  {Tangpanitanon}, \citenamefont {Bastidas}, \citenamefont {Al-Assam},
  \citenamefont {Roushan}, \citenamefont {Jaksch},\ and\ \citenamefont
  {Angelakis}}]{tangpanitanon_2016}%
  \BibitemOpen
  \bibfield  {author} {\bibinfo {author} {\bibfnamefont {J.}~\bibnamefont
  {Tangpanitanon}}, \bibinfo {author} {\bibfnamefont {V.~M.}\ \bibnamefont
  {Bastidas}}, \bibinfo {author} {\bibfnamefont {S.}~\bibnamefont {Al-Assam}},
  \bibinfo {author} {\bibfnamefont {P.}~\bibnamefont {Roushan}}, \bibinfo
  {author} {\bibfnamefont {D.}~\bibnamefont {Jaksch}},\ and\ \bibinfo {author}
  {\bibfnamefont {D.~G.}\ \bibnamefont {Angelakis}},\ }\bibfield  {title}
  {\bibinfo {title} {Topological pumping of photons in nonlinear resonator
  arrays},\ }\href {https://doi.org/10.1103/PhysRevLett.117.213603} {\bibfield
  {journal} {\bibinfo  {journal} {Physical Review Letters}\ }\textbf {\bibinfo
  {volume} {117}},\ \bibinfo {pages} {213603} (\bibinfo {year}
  {2016})}\BibitemShut {NoStop}%
\bibitem [{\citenamefont {J{\"u}rgensen}\ \emph {et~al.}(2021)\citenamefont
  {J{\"u}rgensen}, \citenamefont {Mukherjee},\ and\ \citenamefont
  {Rechtsman}}]{nonlinear-thouless-pump}%
  \BibitemOpen
  \bibfield  {author} {\bibinfo {author} {\bibfnamefont {M.}~\bibnamefont
  {J{\"u}rgensen}}, \bibinfo {author} {\bibfnamefont {S.}~\bibnamefont
  {Mukherjee}},\ and\ \bibinfo {author} {\bibfnamefont {M.~C.}\ \bibnamefont
  {Rechtsman}},\ }\bibfield  {title} {\bibinfo {title} {Quantized nonlinear
  thouless pumping},\ }\href {https://doi.org/10.1038/s41586-021-03688-9}
  {\bibfield  {journal} {\bibinfo  {journal} {Nature}\ }\textbf {\bibinfo
  {volume} {596}},\ \bibinfo {pages} {63} (\bibinfo {year} {2021})}\BibitemShut
  {NoStop}%
\bibitem [{\citenamefont {Walter}\ \emph {et~al.}(2022)\citenamefont {Walter},
  \citenamefont {Zhu}, \citenamefont {G{\"a}chter}, \citenamefont {Minguzzi},
  \citenamefont {Roschinski}, \citenamefont {Sandholzer}, \citenamefont
  {Viebahn},\ and\ \citenamefont {Esslinger}}]{esslinger_int_2022}%
  \BibitemOpen
  \bibfield  {author} {\bibinfo {author} {\bibfnamefont {A.-S.}\ \bibnamefont
  {Walter}}, \bibinfo {author} {\bibfnamefont {Z.}~\bibnamefont {Zhu}},
  \bibinfo {author} {\bibfnamefont {M.}~\bibnamefont {G{\"a}chter}}, \bibinfo
  {author} {\bibfnamefont {J.}~\bibnamefont {Minguzzi}}, \bibinfo {author}
  {\bibfnamefont {S.}~\bibnamefont {Roschinski}}, \bibinfo {author}
  {\bibfnamefont {K.}~\bibnamefont {Sandholzer}}, \bibinfo {author}
  {\bibfnamefont {K.}~\bibnamefont {Viebahn}},\ and\ \bibinfo {author}
  {\bibfnamefont {T.}~\bibnamefont {Esslinger}},\ }\bibfield  {title} {\bibinfo
  {title} {Breakdown of quantisation in a {H}ubbard-{T}houless pump},\
  }\bibfield  {journal} {\bibinfo  {journal} {arXiv preprint}\ }\href
  {https://doi.org/10.48550/arXiv.2204.06561} {10.48550/arXiv.2204.06561}
  (\bibinfo {year} {2022})\BibitemShut {NoStop}%
\bibitem [{\citenamefont {Fedorova}\ \emph {et~al.}(2020)\citenamefont
  {Fedorova}, \citenamefont {Qiu}, \citenamefont {Linden},\ and\ \citenamefont
  {Kroha}}]{fedorova_2020}%
  \BibitemOpen
  \bibfield  {author} {\bibinfo {author} {\bibfnamefont {Z.}~\bibnamefont
  {Fedorova}}, \bibinfo {author} {\bibfnamefont {H.}~\bibnamefont {Qiu}},
  \bibinfo {author} {\bibfnamefont {S.}~\bibnamefont {Linden}},\ and\ \bibinfo
  {author} {\bibfnamefont {J.}~\bibnamefont {Kroha}},\ }\bibfield  {title}
  {\bibinfo {title} {Observation of topological transport quantization by
  dissipation in fast {T}houless pumps},\ }\href
  {https://doi.org/10.1038/s41467-020-17510-z} {\bibfield  {journal} {\bibinfo
  {journal} {Nat. Commun.}\ }\textbf {\bibinfo {volume} {11}},\ \bibinfo
  {pages} {1} (\bibinfo {year} {2020})}\BibitemShut {NoStop}%
\bibitem [{\citenamefont {Dreon}\ \emph {et~al.}(2021)\citenamefont {Dreon},
  \citenamefont {Baumg{\"a}rtner}, \citenamefont {Li}, \citenamefont
  {Hertlein}, \citenamefont {Esslinger},\ and\ \citenamefont
  {Donner}}]{esslinger_2022}%
  \BibitemOpen
  \bibfield  {author} {\bibinfo {author} {\bibfnamefont {D.}~\bibnamefont
  {Dreon}}, \bibinfo {author} {\bibfnamefont {A.}~\bibnamefont
  {Baumg{\"a}rtner}}, \bibinfo {author} {\bibfnamefont {X.}~\bibnamefont {Li}},
  \bibinfo {author} {\bibfnamefont {S.}~\bibnamefont {Hertlein}}, \bibinfo
  {author} {\bibfnamefont {T.}~\bibnamefont {Esslinger}},\ and\ \bibinfo
  {author} {\bibfnamefont {T.}~\bibnamefont {Donner}},\ }\bibfield  {title}
  {\bibinfo {title} {Self-oscillating geometric pump in a dissipative
  atom-cavity system},\ }\bibfield  {journal} {\bibinfo  {journal} {arXiv
  preprint}\ }\href {https://doi.org/10.48550/arXiv.2112.11502}
  {10.48550/arXiv.2112.11502} (\bibinfo {year} {2021})\BibitemShut {NoStop}%
\bibitem [{\citenamefont {Thouless}(1983)}]{Thouless_1983}%
  \BibitemOpen
  \bibfield  {author} {\bibinfo {author} {\bibfnamefont {D.~J.}\ \bibnamefont
  {Thouless}},\ }\bibfield  {title} {\bibinfo {title} {Quantization of particle
  transport},\ }\href {https://doi.org/10.1103/PhysRevB.27.6083} {\bibfield
  {journal} {\bibinfo  {journal} {Phys. Rev. B}\ }\textbf {\bibinfo {volume}
  {27}},\ \bibinfo {pages} {6083} (\bibinfo {year} {1983})}\BibitemShut
  {NoStop}%
\bibitem [{\citenamefont {Niu}\ and\ \citenamefont
  {Thouless}(1984)}]{Niu_1984}%
  \BibitemOpen
  \bibfield  {author} {\bibinfo {author} {\bibfnamefont {Q.}~\bibnamefont
  {Niu}}\ and\ \bibinfo {author} {\bibfnamefont {D.~J.}\ \bibnamefont
  {Thouless}},\ }\bibfield  {title} {\bibinfo {title} {Quantised adiabatic
  charge transport in the presence of substrate disorder and many-body
  interaction},\ }\href {https://doi.org/10.1088/0305-4470/17/12/016}
  {\bibfield  {journal} {\bibinfo  {journal} {Journal of Physics A:
  Mathematical and General}\ }\textbf {\bibinfo {volume} {17}},\ \bibinfo
  {pages} {2453} (\bibinfo {year} {1984})}\BibitemShut {NoStop}%
\bibitem [{\citenamefont {Malikis}\ and\ \citenamefont
  {Cheianov}(2021)}]{cheianov_2021}%
  \BibitemOpen
  \bibfield  {author} {\bibinfo {author} {\bibfnamefont {S.}~\bibnamefont
  {Malikis}}\ and\ \bibinfo {author} {\bibfnamefont {V.}~\bibnamefont
  {Cheianov}},\ }\bibfield  {title} {\bibinfo {title} {An ideal rapid-cycle
  {T}houless pump},\ }\bibfield  {journal} {\bibinfo  {journal} {arXiv
  preprint}\ }\href {https://doi.org/10.48550/arXiv.2104.02751}
  {10.48550/arXiv.2104.02751} (\bibinfo {year} {2021})\BibitemShut {NoStop}%
\bibitem [{\citenamefont {Minguzzi}\ \emph {et~al.}(2021)\citenamefont
  {Minguzzi}, \citenamefont {Zhu}, \citenamefont {Sandholzer}, \citenamefont
  {Walter}, \citenamefont {Viebahn},\ and\ \citenamefont
  {Esslinger}}]{minguzzi_topological_2021}%
  \BibitemOpen
  \bibfield  {author} {\bibinfo {author} {\bibfnamefont {J.}~\bibnamefont
  {Minguzzi}}, \bibinfo {author} {\bibfnamefont {Z.}~\bibnamefont {Zhu}},
  \bibinfo {author} {\bibfnamefont {K.}~\bibnamefont {Sandholzer}}, \bibinfo
  {author} {\bibfnamefont {A.-S.}\ \bibnamefont {Walter}}, \bibinfo {author}
  {\bibfnamefont {K.}~\bibnamefont {Viebahn}},\ and\ \bibinfo {author}
  {\bibfnamefont {T.}~\bibnamefont {Esslinger}},\ }\bibfield  {title} {\bibinfo
  {title} {Topological pumping in a {Floquet}-{Bloch} band},\ }\bibfield
  {journal} {\bibinfo  {journal} {arXiv preprint}\ }\href
  {https://doi.org/10.48550/arXiv.2112.12788} {10.48550/arXiv.2112.12788}
  (\bibinfo {year} {2021})\BibitemShut {NoStop}%
\bibitem [{\citenamefont {Klais}\ \emph {et~al.}(2020)\citenamefont {Klais},
  \citenamefont {Chakraborty}, \citenamefont {Kim},\ and\ \citenamefont
  {al}}]{Hall_2020}%
  \BibitemOpen
  \bibfield  {author} {\bibinfo {author} {\bibfnamefont {v.~K.}\ \bibnamefont
  {Klais}}, \bibinfo {author} {\bibfnamefont {T.}~\bibnamefont {Chakraborty}},
  \bibinfo {author} {\bibfnamefont {P.}~\bibnamefont {Kim}},\ and\ \bibinfo
  {author} {\bibnamefont {al}},\ }\bibfield  {title} {\bibinfo {title} {40
  years of the quantum hall effect},\ }\href
  {https://doi.org/10.1038/s42254-020-0209-1} {\bibfield  {journal} {\bibinfo
  {journal} {Nat. Rev. Phys.}\ }\textbf {\bibinfo {volume} {2}},\ \bibinfo
  {pages} {397} (\bibinfo {year} {2020})}\BibitemShut {NoStop}%
\bibitem [{\citenamefont {Ke}\ \emph {et~al.}(2020)\citenamefont {Ke},
  \citenamefont {Hu}, \citenamefont {Zhu}, \citenamefont {Gong}, \citenamefont
  {Kivshar},\ and\ \citenamefont {Lee}}]{assisted_Thouless_pump}%
  \BibitemOpen
  \bibfield  {author} {\bibinfo {author} {\bibfnamefont {Y.}~\bibnamefont
  {Ke}}, \bibinfo {author} {\bibfnamefont {S.}~\bibnamefont {Hu}}, \bibinfo
  {author} {\bibfnamefont {B.}~\bibnamefont {Zhu}}, \bibinfo {author}
  {\bibfnamefont {J.}~\bibnamefont {Gong}}, \bibinfo {author} {\bibfnamefont
  {Y.}~\bibnamefont {Kivshar}},\ and\ \bibinfo {author} {\bibfnamefont
  {C.}~\bibnamefont {Lee}},\ }\bibfield  {title} {\bibinfo {title} {Topological
  pumping assisted by bloch oscillations},\ }\href
  {https://doi.org/10.1103/PhysRevResearch.2.033143} {\bibfield  {journal}
  {\bibinfo  {journal} {Phys. Rev. Research}\ }\textbf {\bibinfo {volume}
  {2}},\ \bibinfo {pages} {033143} (\bibinfo {year} {2020})}\BibitemShut
  {NoStop}%
\bibitem [{\citenamefont {Rice}\ and\ \citenamefont
  {Mele}(1982)}]{ricemele_1982}%
  \BibitemOpen
  \bibfield  {author} {\bibinfo {author} {\bibfnamefont {M.~J.}\ \bibnamefont
  {Rice}}\ and\ \bibinfo {author} {\bibfnamefont {E.~J.}\ \bibnamefont
  {Mele}},\ }\bibfield  {title} {\bibinfo {title} {Elementary excitations of a
  linearly conjugated diatomic polymer},\ }\href
  {https://doi.org/10.1103/PhysRevLett.49.1455} {\bibfield  {journal} {\bibinfo
   {journal} {Phys. Rev. Lett.}\ }\textbf {\bibinfo {volume} {49}},\ \bibinfo
  {pages} {1455} (\bibinfo {year} {1982})}\BibitemShut {NoStop}%
\bibitem [{\citenamefont {Xiao}\ \emph {et~al.}(2010)\citenamefont {Xiao},
  \citenamefont {Chang},\ and\ \citenamefont {Niu}}]{review_niu_2010}%
  \BibitemOpen
  \bibfield  {author} {\bibinfo {author} {\bibfnamefont {D.}~\bibnamefont
  {Xiao}}, \bibinfo {author} {\bibfnamefont {M.-C.}\ \bibnamefont {Chang}},\
  and\ \bibinfo {author} {\bibfnamefont {Q.}~\bibnamefont {Niu}},\ }\bibfield
  {title} {\bibinfo {title} {Berry phase effects on electronic properties},\
  }\href {https://doi.org/10.1103/RevModPhys.82.1959} {\bibfield  {journal}
  {\bibinfo  {journal} {Rev. Mod. Phys.}\ }\textbf {\bibinfo {volume} {82}},\
  \bibinfo {pages} {1959} (\bibinfo {year} {2010})}\BibitemShut {NoStop}%
\bibitem [{\citenamefont {Wang}\ \emph {et~al.}(2013)\citenamefont {Wang},
  \citenamefont {Troyer},\ and\ \citenamefont {Dai}}]{troyer_2013}%
  \BibitemOpen
  \bibfield  {author} {\bibinfo {author} {\bibfnamefont {L.}~\bibnamefont
  {Wang}}, \bibinfo {author} {\bibfnamefont {M.}~\bibnamefont {Troyer}},\ and\
  \bibinfo {author} {\bibfnamefont {X.}~\bibnamefont {Dai}},\ }\bibfield
  {title} {\bibinfo {title} {Topological charge pumping in a one-dimensional
  optical lattice},\ }\href {https://doi.org/10.1103/PhysRevLett.111.026802}
  {\bibfield  {journal} {\bibinfo  {journal} {Phys. Rev. Lett.}\ }\textbf
  {\bibinfo {volume} {111}},\ \bibinfo {pages} {026802} (\bibinfo {year}
  {2013})}\BibitemShut {NoStop}%
\bibitem [{\citenamefont {Switkes}\ \emph {et~al.}(1999)\citenamefont
  {Switkes}, \citenamefont {Marcus}, \citenamefont {Campman},\ and\
  \citenamefont {Gossard}}]{Switkes_1999}%
  \BibitemOpen
  \bibfield  {author} {\bibinfo {author} {\bibfnamefont {M.}~\bibnamefont
  {Switkes}}, \bibinfo {author} {\bibfnamefont {C.~M.}\ \bibnamefont {Marcus}},
  \bibinfo {author} {\bibfnamefont {K.}~\bibnamefont {Campman}},\ and\ \bibinfo
  {author} {\bibfnamefont {A.~C.}\ \bibnamefont {Gossard}},\ }\bibfield
  {title} {\bibinfo {title} {An adiabatic quantum electron pump},\ }\href
  {https://doi.org/10.1126/science.283.5409.1905} {\bibfield  {journal}
  {\bibinfo  {journal} {Science}\ }\textbf {\bibinfo {volume} {283}},\ \bibinfo
  {pages} {1905} (\bibinfo {year} {1999})}\BibitemShut {NoStop}%
\bibitem [{\citenamefont {Talyanskii}\ \emph {et~al.}(1997)\citenamefont
  {Talyanskii}, \citenamefont {Shilton}, \citenamefont {Pepper}, \citenamefont
  {Smith}, \citenamefont {Ford}, \citenamefont {Linfield}, \citenamefont
  {Ritchie},\ and\ \citenamefont {Jones}}]{Talyanskii_1997}%
  \BibitemOpen
  \bibfield  {author} {\bibinfo {author} {\bibfnamefont {V.~I.}\ \bibnamefont
  {Talyanskii}}, \bibinfo {author} {\bibfnamefont {J.~M.}\ \bibnamefont
  {Shilton}}, \bibinfo {author} {\bibfnamefont {M.}~\bibnamefont {Pepper}},
  \bibinfo {author} {\bibfnamefont {C.~G.}\ \bibnamefont {Smith}}, \bibinfo
  {author} {\bibfnamefont {C.~J.~B.}\ \bibnamefont {Ford}}, \bibinfo {author}
  {\bibfnamefont {E.~H.}\ \bibnamefont {Linfield}}, \bibinfo {author}
  {\bibfnamefont {D.~A.}\ \bibnamefont {Ritchie}},\ and\ \bibinfo {author}
  {\bibfnamefont {G.~A.~C.}\ \bibnamefont {Jones}},\ }\bibfield  {title}
  {\bibinfo {title} {Single-electron transport in a one-dimensional channel by
  high-frequency surface acoustic waves},\ }\href
  {https://doi.org/10.1103/PhysRevB.56.15180} {\bibfield  {journal} {\bibinfo
  {journal} {Phys. Rev. B}\ }\textbf {\bibinfo {volume} {56}},\ \bibinfo
  {pages} {15180} (\bibinfo {year} {1997})}\BibitemShut {NoStop}%
\bibitem [{\citenamefont {Brouwer}(1998)}]{Brouwer_1998}%
  \BibitemOpen
  \bibfield  {author} {\bibinfo {author} {\bibfnamefont {P.~W.}\ \bibnamefont
  {Brouwer}},\ }\bibfield  {title} {\bibinfo {title} {Scattering approach to
  parametric pumping},\ }\href {https://doi.org/10.1103/PhysRevB.58.R10135}
  {\bibfield  {journal} {\bibinfo  {journal} {Phys. Rev. B}\ }\textbf {\bibinfo
  {volume} {58}},\ \bibinfo {pages} {R10135} (\bibinfo {year}
  {1998})}\BibitemShut {NoStop}%
\bibitem [{\citenamefont {Zhou}\ \emph {et~al.}(2003)\citenamefont {Zhou},
  \citenamefont {Cho},\ and\ \citenamefont {McKenzie}}]{Zhou_2003}%
  \BibitemOpen
  \bibfield  {author} {\bibinfo {author} {\bibfnamefont {H.-Q.}\ \bibnamefont
  {Zhou}}, \bibinfo {author} {\bibfnamefont {S.~Y.}\ \bibnamefont {Cho}},\ and\
  \bibinfo {author} {\bibfnamefont {R.~H.}\ \bibnamefont {McKenzie}},\
  }\bibfield  {title} {\bibinfo {title} {Gauge fields, geometric phases, and
  quantum adiabatic pumps},\ }\href
  {https://doi.org/10.1103/PhysRevLett.91.186803} {\bibfield  {journal}
  {\bibinfo  {journal} {Phys. Rev. Lett.}\ }\textbf {\bibinfo {volume} {91}},\
  \bibinfo {pages} {186803} (\bibinfo {year} {2003})}\BibitemShut {NoStop}%
\bibitem [{\citenamefont {Schweizer}\ \emph {et~al.}(2016)\citenamefont
  {Schweizer}, \citenamefont {Lohse}, \citenamefont {Citro},\ and\
  \citenamefont {Bloch}}]{Schweizer_2016}%
  \BibitemOpen
  \bibfield  {author} {\bibinfo {author} {\bibfnamefont {C.}~\bibnamefont
  {Schweizer}}, \bibinfo {author} {\bibfnamefont {M.}~\bibnamefont {Lohse}},
  \bibinfo {author} {\bibfnamefont {R.}~\bibnamefont {Citro}},\ and\ \bibinfo
  {author} {\bibfnamefont {I.}~\bibnamefont {Bloch}},\ }\bibfield  {title}
  {\bibinfo {title} {Spin pumping and measurement of spin currents in optical
  superlattices},\ }\href {https://doi.org/10.1103/PhysRevLett.117.170405}
  {\bibfield  {journal} {\bibinfo  {journal} {Phys. Rev. Lett.}\ }\textbf
  {\bibinfo {volume} {117}},\ \bibinfo {pages} {170405} (\bibinfo {year}
  {2016})}\BibitemShut {NoStop}%
\bibitem [{\citenamefont {Sheng}\ \emph {et~al.}(2006)\citenamefont {Sheng},
  \citenamefont {Weng}, \citenamefont {Sheng},\ and\ \citenamefont
  {Haldane}}]{spin-chern-number}%
  \BibitemOpen
  \bibfield  {author} {\bibinfo {author} {\bibfnamefont {D.~N.}\ \bibnamefont
  {Sheng}}, \bibinfo {author} {\bibfnamefont {Z.~Y.}\ \bibnamefont {Weng}},
  \bibinfo {author} {\bibfnamefont {L.}~\bibnamefont {Sheng}},\ and\ \bibinfo
  {author} {\bibfnamefont {F.~D.~M.}\ \bibnamefont {Haldane}},\ }\bibfield
  {title} {\bibinfo {title} {Quantum spin-hall effect and topologically
  invariant chern numbers},\ }\href
  {https://doi.org/10.1103/PhysRevLett.97.036808} {\bibfield  {journal}
  {\bibinfo  {journal} {Phys. Rev. Lett.}\ }\textbf {\bibinfo {volume} {97}},\
  \bibinfo {pages} {036808} (\bibinfo {year} {2006})}\BibitemShut {NoStop}%
\bibitem [{\citenamefont {Bernevig}(2013)}]{bernevig_book}%
  \BibitemOpen
  \bibfield  {author} {\bibinfo {author} {\bibfnamefont {B.~A.}\ \bibnamefont
  {Bernevig}},\ }\bibfield  {title} {\bibinfo {title} {Topological insulators
  and topological superconductors},\ }in\ \href@noop {} {\emph {\bibinfo
  {booktitle} {Topological Insulators and Topological Superconductors}}}\
  (\bibinfo  {publisher} {Princeton university press},\ \bibinfo {year}
  {2013})\BibitemShut {NoStop}%
\bibitem [{\citenamefont {Shindou}(2005)}]{Shindou_2005}%
  \BibitemOpen
  \bibfield  {author} {\bibinfo {author} {\bibfnamefont {R.}~\bibnamefont
  {Shindou}},\ }\bibfield  {title} {\bibinfo {title} {Quantum spin pump in
  $s=1/2$ antiferromagnetic chains –holonomy of phase operators in
  sine-gordon theory–},\ }\href {https://doi.org/10.1143/JPSJ.74.1214}
  {\bibfield  {journal} {\bibinfo  {journal} {J. Phys. Soc. Jpn.}\ }\textbf
  {\bibinfo {volume} {74}},\ \bibinfo {pages} {1214} (\bibinfo {year}
  {2005})}\BibitemShut {NoStop}%
\bibitem [{\citenamefont {Zhou}\ \emph {et~al.}(2014)\citenamefont {Zhou},
  \citenamefont {Zhang}, \citenamefont {Sheng}, \citenamefont {Shen},
  \citenamefont {Sheng},\ and\ \citenamefont {Xing}}]{Zhou_2014}%
  \BibitemOpen
  \bibfield  {author} {\bibinfo {author} {\bibfnamefont {C.~Q.}\ \bibnamefont
  {Zhou}}, \bibinfo {author} {\bibfnamefont {Y.~F.}\ \bibnamefont {Zhang}},
  \bibinfo {author} {\bibfnamefont {L.}~\bibnamefont {Sheng}}, \bibinfo
  {author} {\bibfnamefont {R.}~\bibnamefont {Shen}}, \bibinfo {author}
  {\bibfnamefont {D.~N.}\ \bibnamefont {Sheng}},\ and\ \bibinfo {author}
  {\bibfnamefont {D.~Y.}\ \bibnamefont {Xing}},\ }\bibfield  {title} {\bibinfo
  {title} {Proposal for a topological spin chern pump},\ }\href
  {https://doi.org/10.1103/PhysRevB.90.085133} {\bibfield  {journal} {\bibinfo
  {journal} {Phys. Rev. B}\ }\textbf {\bibinfo {volume} {90}},\ \bibinfo
  {pages} {085133} (\bibinfo {year} {2014})}\BibitemShut {NoStop}%
\bibitem [{\citenamefont {Chen}\ \emph {et~al.}(2020)\citenamefont {Chen},
  \citenamefont {Cai},\ and\ \citenamefont
  {Zhang}}]{chen_spin_pumping_glide_symmetry}%
  \BibitemOpen
  \bibfield  {author} {\bibinfo {author} {\bibfnamefont {Q.}~\bibnamefont
  {Chen}}, \bibinfo {author} {\bibfnamefont {J.}~\bibnamefont {Cai}},\ and\
  \bibinfo {author} {\bibfnamefont {S.}~\bibnamefont {Zhang}},\ }\bibfield
  {title} {\bibinfo {title} {Topological quantum pumping in spin-dependent
  superlattices with glide symmetry},\ }\href
  {https://doi.org/10.1103/PhysRevA.101.043614} {\bibfield  {journal} {\bibinfo
   {journal} {Phys. Rev. A}\ }\textbf {\bibinfo {volume} {101}},\ \bibinfo
  {pages} {043614} (\bibinfo {year} {2020})}\BibitemShut {NoStop}%
\bibitem [{\citenamefont {Meidan}\ \emph {et~al.}(2011)\citenamefont {Meidan},
  \citenamefont {Micklitz},\ and\ \citenamefont {Brouwer}}]{Meidan_2011}%
  \BibitemOpen
  \bibfield  {author} {\bibinfo {author} {\bibfnamefont {D.}~\bibnamefont
  {Meidan}}, \bibinfo {author} {\bibfnamefont {T.}~\bibnamefont {Micklitz}},\
  and\ \bibinfo {author} {\bibfnamefont {P.~W.}\ \bibnamefont {Brouwer}},\
  }\bibfield  {title} {\bibinfo {title} {Topological classification of
  interaction-driven spin pumps},\ }\href
  {https://doi.org/10.1103/PhysRevB.84.075325} {\bibfield  {journal} {\bibinfo
  {journal} {Phys. Rev. B}\ }\textbf {\bibinfo {volume} {84}},\ \bibinfo
  {pages} {075325} (\bibinfo {year} {2011})}\BibitemShut {NoStop}%
\bibitem [{\citenamefont {Cazalilla}\ \emph {et~al.}(2011)\citenamefont
  {Cazalilla}, \citenamefont {Citro}, \citenamefont {Giamarchi}, \citenamefont
  {Orignac},\ and\ \citenamefont {Rigol}}]{Citro_RMP}%
  \BibitemOpen
  \bibfield  {author} {\bibinfo {author} {\bibfnamefont {M.~A.}\ \bibnamefont
  {Cazalilla}}, \bibinfo {author} {\bibfnamefont {R.}~\bibnamefont {Citro}},
  \bibinfo {author} {\bibfnamefont {T.}~\bibnamefont {Giamarchi}}, \bibinfo
  {author} {\bibfnamefont {E.}~\bibnamefont {Orignac}},\ and\ \bibinfo {author}
  {\bibfnamefont {M.}~\bibnamefont {Rigol}},\ }\bibfield  {title} {\bibinfo
  {title} {One dimensional bosons: From condensed matter systems to ultracold
  gases},\ }\href {https://doi.org/10.1103/RevModPhys.83.1405} {\bibfield
  {journal} {\bibinfo  {journal} {Rev. Mod. Phys.}\ }\textbf {\bibinfo {volume}
  {83}},\ \bibinfo {pages} {1405} (\bibinfo {year} {2011})}\BibitemShut
  {NoStop}%
\bibitem [{\citenamefont {Shih}\ and\ \citenamefont {Niu}(1994)}]{Shih_1994}%
  \BibitemOpen
  \bibfield  {author} {\bibinfo {author} {\bibfnamefont {W.-K.}\ \bibnamefont
  {Shih}}\ and\ \bibinfo {author} {\bibfnamefont {Q.}~\bibnamefont {Niu}},\
  }\bibfield  {title} {\bibinfo {title} {Nonadiabatic particle transport in a
  one-dimensional electron system},\ }\href
  {https://doi.org/10.1103/PhysRevB.50.11902} {\bibfield  {journal} {\bibinfo
  {journal} {Phys. Rev. B}\ }\textbf {\bibinfo {volume} {50}},\ \bibinfo
  {pages} {11902} (\bibinfo {year} {1994})}\BibitemShut {NoStop}%
\bibitem [{\citenamefont {von Klitzing}(1986)}]{Klitzing_review}%
  \BibitemOpen
  \bibfield  {author} {\bibinfo {author} {\bibfnamefont {K.}~\bibnamefont {von
  Klitzing}},\ }\bibfield  {title} {\bibinfo {title} {The quantized hall
  effect},\ }\href {https://doi.org/10.1103/RevModPhys.58.519} {\bibfield
  {journal} {\bibinfo  {journal} {Rev. Mod. Phys.}\ }\textbf {\bibinfo {volume}
  {58}},\ \bibinfo {pages} {519} (\bibinfo {year} {1986})}\BibitemShut
  {NoStop}%
\bibitem [{\citenamefont {Avron}\ and\ \citenamefont
  {Kons}(1999)}]{Avron_1999}%
  \BibitemOpen
  \bibfield  {author} {\bibinfo {author} {\bibfnamefont {J.~E.}\ \bibnamefont
  {Avron}}\ and\ \bibinfo {author} {\bibfnamefont {Z.}~\bibnamefont {Kons}},\
  }\bibfield  {title} {\bibinfo {title} {Quantum response at finite fields and
  breakdown of chern numbers},\ }\href
  {https://doi.org/10.1088/0305-4470/32/33/308} {\bibfield  {journal} {\bibinfo
   {journal} {Journal of Physics A: Mathematical and General}\ }\textbf
  {\bibinfo {volume} {32}},\ \bibinfo {pages} {6097} (\bibinfo {year}
  {1999})}\BibitemShut {NoStop}%
\bibitem [{\citenamefont {Privitera}\ \emph {et~al.}(2018)\citenamefont
  {Privitera}, \citenamefont {Russomanno}, \citenamefont {Citro},\ and\
  \citenamefont {Santoro}}]{privitera_2018}%
  \BibitemOpen
  \bibfield  {author} {\bibinfo {author} {\bibfnamefont {L.}~\bibnamefont
  {Privitera}}, \bibinfo {author} {\bibfnamefont {A.}~\bibnamefont
  {Russomanno}}, \bibinfo {author} {\bibfnamefont {R.}~\bibnamefont {Citro}},\
  and\ \bibinfo {author} {\bibfnamefont {G.~E.}\ \bibnamefont {Santoro}},\
  }\bibfield  {title} {\bibinfo {title} {Nonadiabatic breaking of topological
  pumping},\ }\href {https://doi.org/10.1103/PhysRevLett.120.106601} {\bibfield
   {journal} {\bibinfo  {journal} {Phys. Rev. Lett.}\ }\textbf {\bibinfo
  {volume} {120}},\ \bibinfo {pages} {106601} (\bibinfo {year}
  {2018})}\BibitemShut {NoStop}%
\bibitem [{\citenamefont {Ferrari}(1998)}]{Ferrari_1998}%
  \BibitemOpen
  \bibfield  {author} {\bibinfo {author} {\bibfnamefont {R.}~\bibnamefont
  {Ferrari}},\ }\bibfield  {title} {\bibinfo {title} {Floquet energies and
  quantum hall effect in a periodic potential},\ }\href
  {https://doi.org/10.1142/S0217979298000600} {\bibfield  {journal} {\bibinfo
  {journal} {Int. J. Mod. Phys. B}\ }\textbf {\bibinfo {volume} {12}},\
  \bibinfo {pages} {1105} (\bibinfo {year} {1998})}\BibitemShut {NoStop}%
\bibitem [{\citenamefont {Kitagawa}\ \emph {et~al.}(2010)\citenamefont
  {Kitagawa}, \citenamefont {Berg}, \citenamefont {Rudner},\ and\ \citenamefont
  {Demler}}]{Kitagawa_2010}%
  \BibitemOpen
  \bibfield  {author} {\bibinfo {author} {\bibfnamefont {T.}~\bibnamefont
  {Kitagawa}}, \bibinfo {author} {\bibfnamefont {E.}~\bibnamefont {Berg}},
  \bibinfo {author} {\bibfnamefont {M.}~\bibnamefont {Rudner}},\ and\ \bibinfo
  {author} {\bibfnamefont {E.}~\bibnamefont {Demler}},\ }\bibfield  {title}
  {\bibinfo {title} {Topological characterization of periodically driven
  quantum systems},\ }\href {https://doi.org/10.1103/PhysRevB.82.235114}
  {\bibfield  {journal} {\bibinfo  {journal} {Phys. Rev. B}\ }\textbf {\bibinfo
  {volume} {82}},\ \bibinfo {pages} {235114} (\bibinfo {year}
  {2010})}\BibitemShut {NoStop}%
\bibitem [{\citenamefont {Russomanno}\ \emph {et~al.}(2011)\citenamefont
  {Russomanno}, \citenamefont {Pugnetti}, \citenamefont {Brosco},\ and\
  \citenamefont {Fazio}}]{Russomanno_2011}%
  \BibitemOpen
  \bibfield  {author} {\bibinfo {author} {\bibfnamefont {A.}~\bibnamefont
  {Russomanno}}, \bibinfo {author} {\bibfnamefont {S.}~\bibnamefont
  {Pugnetti}}, \bibinfo {author} {\bibfnamefont {V.}~\bibnamefont {Brosco}},\
  and\ \bibinfo {author} {\bibfnamefont {R.}~\bibnamefont {Fazio}},\ }\bibfield
   {title} {\bibinfo {title} {Floquet theory of cooper pair pumping},\ }\href
  {https://doi.org/10.1103/PhysRevB.83.214508} {\bibfield  {journal} {\bibinfo
  {journal} {Phys. Rev. B}\ }\textbf {\bibinfo {volume} {83}},\ \bibinfo
  {pages} {214508} (\bibinfo {year} {2011})}\BibitemShut {NoStop}%
\bibitem [{\citenamefont {Lindner}\ \emph {et~al.}(2017)\citenamefont
  {Lindner}, \citenamefont {Berg},\ and\ \citenamefont
  {Rudner}}]{Lindner_2017}%
  \BibitemOpen
  \bibfield  {author} {\bibinfo {author} {\bibfnamefont {N.~H.}\ \bibnamefont
  {Lindner}}, \bibinfo {author} {\bibfnamefont {E.}~\bibnamefont {Berg}},\ and\
  \bibinfo {author} {\bibfnamefont {M.~S.}\ \bibnamefont {Rudner}},\ }\bibfield
   {title} {\bibinfo {title} {Universal chiral quasisteady states in
  periodically driven many-body systems},\ }\href
  {https://doi.org/10.1103/PhysRevX.7.011018} {\bibfield  {journal} {\bibinfo
  {journal} {Phys. Rev. X}\ }\textbf {\bibinfo {volume} {7}},\ \bibinfo {pages}
  {011018} (\bibinfo {year} {2017})}\BibitemShut {NoStop}%
\bibitem [{\citenamefont {Russomanno}\ \emph {et~al.}(2012)\citenamefont
  {Russomanno}, \citenamefont {Silva},\ and\ \citenamefont
  {Santoro}}]{Russomanno_2012}%
  \BibitemOpen
  \bibfield  {author} {\bibinfo {author} {\bibfnamefont {A.}~\bibnamefont
  {Russomanno}}, \bibinfo {author} {\bibfnamefont {A.}~\bibnamefont {Silva}},\
  and\ \bibinfo {author} {\bibfnamefont {G.~E.}\ \bibnamefont {Santoro}},\
  }\bibfield  {title} {\bibinfo {title} {Periodic steady regime and
  interference in a periodically driven quantum system},\ }\href
  {https://doi.org/10.1103/PhysRevLett.109.257201} {\bibfield  {journal}
  {\bibinfo  {journal} {Phys. Rev. Lett.}\ }\textbf {\bibinfo {volume} {109}},\
  \bibinfo {pages} {257201} (\bibinfo {year} {2012})}\BibitemShut {NoStop}%
\bibitem [{\citenamefont {Lazarides}\ \emph {et~al.}(2014)\citenamefont
  {Lazarides}, \citenamefont {Das},\ and\ \citenamefont
  {Moessner}}]{Lazarides_2014}%
  \BibitemOpen
  \bibfield  {author} {\bibinfo {author} {\bibfnamefont {A.}~\bibnamefont
  {Lazarides}}, \bibinfo {author} {\bibfnamefont {A.}~\bibnamefont {Das}},\
  and\ \bibinfo {author} {\bibfnamefont {R.}~\bibnamefont {Moessner}},\
  }\bibfield  {title} {\bibinfo {title} {Periodic thermodynamics of isolated
  quantum systems},\ }\href {https://doi.org/10.1103/PhysRevLett.112.150401}
  {\bibfield  {journal} {\bibinfo  {journal} {Phys. Rev. Lett.}\ }\textbf
  {\bibinfo {volume} {112}},\ \bibinfo {pages} {150401} (\bibinfo {year}
  {2014})}\BibitemShut {NoStop}%
\bibitem [{\citenamefont {Rigolin}\ \emph {et~al.}(2008)\citenamefont
  {Rigolin}, \citenamefont {Ortiz},\ and\ \citenamefont
  {Ponce}}]{Rigolin_2008}%
  \BibitemOpen
  \bibfield  {author} {\bibinfo {author} {\bibfnamefont {G.}~\bibnamefont
  {Rigolin}}, \bibinfo {author} {\bibfnamefont {G.}~\bibnamefont {Ortiz}},\
  and\ \bibinfo {author} {\bibfnamefont {V.~H.}\ \bibnamefont {Ponce}},\
  }\bibfield  {title} {\bibinfo {title} {Beyond the quantum adiabatic
  approximation: Adiabatic perturbation theory},\ }\href
  {https://doi.org/10.1103/PhysRevA.78.052508} {\bibfield  {journal} {\bibinfo
  {journal} {Phys. Rev. A}\ }\textbf {\bibinfo {volume} {78}},\ \bibinfo
  {pages} {052508} (\bibinfo {year} {2008})}\BibitemShut {NoStop}%
\bibitem [{\citenamefont {Wauters}\ \emph {et~al.}(2019)\citenamefont
  {Wauters}, \citenamefont {Russomanno}, \citenamefont {Citro}, \citenamefont
  {Santoro},\ and\ \citenamefont {Privitera}}]{wauters_2019}%
  \BibitemOpen
  \bibfield  {author} {\bibinfo {author} {\bibfnamefont {M.~M.}\ \bibnamefont
  {Wauters}}, \bibinfo {author} {\bibfnamefont {A.}~\bibnamefont {Russomanno}},
  \bibinfo {author} {\bibfnamefont {R.}~\bibnamefont {Citro}}, \bibinfo
  {author} {\bibfnamefont {G.~E.}\ \bibnamefont {Santoro}},\ and\ \bibinfo
  {author} {\bibfnamefont {L.}~\bibnamefont {Privitera}},\ }\bibfield  {title}
  {\bibinfo {title} {Localization, topology, and quantized transport in
  disordered floquet systems},\ }\href
  {https://doi.org/10.1103/PhysRevLett.123.266601} {\bibfield  {journal}
  {\bibinfo  {journal} {Phys. Rev. Lett.}\ }\textbf {\bibinfo {volume} {123}},\
  \bibinfo {pages} {266601} (\bibinfo {year} {2019})}\BibitemShut {NoStop}%
\bibitem [{\citenamefont {Hayward}\ \emph {et~al.}(2021)\citenamefont
  {Hayward}, \citenamefont {Bertok}, \citenamefont {Schneider},\ and\
  \citenamefont {Heidrich-Meisner}}]{hayward_effect_2020}%
  \BibitemOpen
  \bibfield  {author} {\bibinfo {author} {\bibfnamefont {A.~L.~C.}\
  \bibnamefont {Hayward}}, \bibinfo {author} {\bibfnamefont {E.}~\bibnamefont
  {Bertok}}, \bibinfo {author} {\bibfnamefont {U.}~\bibnamefont {Schneider}},\
  and\ \bibinfo {author} {\bibfnamefont {F.}~\bibnamefont {Heidrich-Meisner}},\
  }\bibfield  {title} {\bibinfo {title} {Effect of disorder on topological
  charge pumping in the {R}ice-{M}ele model},\ }\href
  {https://doi.org/10.1103/PhysRevA.103.043310} {\bibfield  {journal} {\bibinfo
   {journal} {Phys. Rev. A}\ }\textbf {\bibinfo {volume} {103}},\ \bibinfo
  {pages} {043310} (\bibinfo {year} {2021})}\BibitemShut {NoStop}%
\bibitem [{\citenamefont {Hu}\ \emph {et~al.}(2020)\citenamefont {Hu},
  \citenamefont {Ke},\ and\ \citenamefont {Lee}}]{shi_topological_2020}%
  \BibitemOpen
  \bibfield  {author} {\bibinfo {author} {\bibfnamefont {S.}~\bibnamefont
  {Hu}}, \bibinfo {author} {\bibfnamefont {Y.}~\bibnamefont {Ke}},\ and\
  \bibinfo {author} {\bibfnamefont {C.}~\bibnamefont {Lee}},\ }\bibfield
  {title} {\bibinfo {title} {Topological quantum transport and spatial
  entanglement distribution via a disordered bulk channel},\ }\href
  {https://doi.org/10.1103/PhysRevA.101.052323} {\bibfield  {journal} {\bibinfo
   {journal} {Phys. Rev. A}\ }\textbf {\bibinfo {volume} {101}},\ \bibinfo
  {pages} {052323} (\bibinfo {year} {2020})}\BibitemShut {NoStop}%
\bibitem [{\citenamefont {Ippoliti}\ and\ \citenamefont
  {Bhatt}(2020)}]{ippoliti_dimensional_2020}%
  \BibitemOpen
  \bibfield  {author} {\bibinfo {author} {\bibfnamefont {M.}~\bibnamefont
  {Ippoliti}}\ and\ \bibinfo {author} {\bibfnamefont {R.~N.}\ \bibnamefont
  {Bhatt}},\ }\bibfield  {title} {\bibinfo {title} {Dimensional crossover of
  the integer quantum hall plateau transition and disordered topological
  pumping},\ }\href {https://doi.org/10.1103/PhysRevLett.124.086602} {\bibfield
   {journal} {\bibinfo  {journal} {Phys. Rev. Lett.}\ }\textbf {\bibinfo
  {volume} {124}},\ \bibinfo {pages} {086602} (\bibinfo {year}
  {2020})}\BibitemShut {NoStop}%
\bibitem [{\citenamefont {Marra}\ and\ \citenamefont
  {Nitta}(2020)}]{marra_topologically_2020}%
  \BibitemOpen
  \bibfield  {author} {\bibinfo {author} {\bibfnamefont {P.}~\bibnamefont
  {Marra}}\ and\ \bibinfo {author} {\bibfnamefont {M.}~\bibnamefont {Nitta}},\
  }\bibfield  {title} {\bibinfo {title} {Topologically quantized current in
  quasiperiodic thouless pumps},\ }\href
  {https://doi.org/10.1103/PhysRevResearch.2.042035} {\bibfield  {journal}
  {\bibinfo  {journal} {Phys. Rev. Research}\ }\textbf {\bibinfo {volume}
  {2}},\ \bibinfo {pages} {042035} (\bibinfo {year} {2020})}\BibitemShut
  {NoStop}%
\bibitem [{\citenamefont {Wang}\ and\ \citenamefont
  {Song}(2019)}]{wang_robustness_2019}%
  \BibitemOpen
  \bibfield  {author} {\bibinfo {author} {\bibfnamefont {R.}~\bibnamefont
  {Wang}}\ and\ \bibinfo {author} {\bibfnamefont {Z.}~\bibnamefont {Song}},\
  }\bibfield  {title} {\bibinfo {title} {Robustness of the pumping charge to
  dynamic disorder},\ }\href {https://doi.org/10.1103/PhysRevB.100.184304}
  {\bibfield  {journal} {\bibinfo  {journal} {Phys. Rev. B}\ }\textbf {\bibinfo
  {volume} {100}},\ \bibinfo {pages} {184304} (\bibinfo {year}
  {2019})}\BibitemShut {NoStop}%
\bibitem [{\citenamefont {Qin}\ and\ \citenamefont
  {Guo}(2016)}]{qin_quantum_2016}%
  \BibitemOpen
  \bibfield  {author} {\bibinfo {author} {\bibfnamefont {J.}~\bibnamefont
  {Qin}}\ and\ \bibinfo {author} {\bibfnamefont {H.}~\bibnamefont {Guo}},\
  }\bibfield  {title} {\bibinfo {title} {Quantum pumping induced by disorder in
  one dimension},\ }\href {https://doi.org/10.1016/j.physleta.2016.05.014}
  {\bibfield  {journal} {\bibinfo  {journal} {Physics Letters A}\ }\textbf
  {\bibinfo {volume} {380}},\ \bibinfo {pages} {2317} (\bibinfo {year}
  {2016})}\BibitemShut {NoStop}%
\bibitem [{\citenamefont {Grifoni}\ and\ \citenamefont
  {Hänggi}(1998)}]{Grifoni_1998}%
  \BibitemOpen
  \bibfield  {author} {\bibinfo {author} {\bibfnamefont {M.}~\bibnamefont
  {Grifoni}}\ and\ \bibinfo {author} {\bibfnamefont {P.}~\bibnamefont
  {Hänggi}},\ }\bibfield  {title} {\bibinfo {title} {Driven quantum
  tunneling},\ }\href
  {https://doi.org/https://doi.org/10.1016/S0370-1573(98)00022-2} {\bibfield
  {journal} {\bibinfo  {journal} {Physics Reports}\ }\textbf {\bibinfo {volume}
  {304}},\ \bibinfo {pages} {229} (\bibinfo {year} {1998})}\BibitemShut
  {NoStop}%
\bibitem [{\citenamefont {Abrahams}\ \emph {et~al.}(1979)\citenamefont
  {Abrahams}, \citenamefont {Anderson}, \citenamefont {Licciardello},\ and\
  \citenamefont {Ramakrishnan}}]{Abrahams_1979}%
  \BibitemOpen
  \bibfield  {author} {\bibinfo {author} {\bibfnamefont {E.}~\bibnamefont
  {Abrahams}}, \bibinfo {author} {\bibfnamefont {P.~W.}\ \bibnamefont
  {Anderson}}, \bibinfo {author} {\bibfnamefont {D.~C.}\ \bibnamefont
  {Licciardello}},\ and\ \bibinfo {author} {\bibfnamefont {T.~V.}\ \bibnamefont
  {Ramakrishnan}},\ }\bibfield  {title} {\bibinfo {title} {Scaling theory of
  localization: Absence of quantum diffusion in two dimensions},\ }\href
  {https://doi.org/10.1103/PhysRevLett.42.673} {\bibfield  {journal} {\bibinfo
  {journal} {Phys. Rev. Lett.}\ }\textbf {\bibinfo {volume} {42}},\ \bibinfo
  {pages} {673} (\bibinfo {year} {1979})}\BibitemShut {NoStop}%
\bibitem [{\citenamefont {Agarwal}\ \emph {et~al.}(2017)\citenamefont
  {Agarwal}, \citenamefont {Ganeshan},\ and\ \citenamefont
  {Bhatt}}]{Agarwal_2017}%
  \BibitemOpen
  \bibfield  {author} {\bibinfo {author} {\bibfnamefont {K.}~\bibnamefont
  {Agarwal}}, \bibinfo {author} {\bibfnamefont {S.}~\bibnamefont {Ganeshan}},\
  and\ \bibinfo {author} {\bibfnamefont {R.~N.}\ \bibnamefont {Bhatt}},\
  }\bibfield  {title} {\bibinfo {title} {Localization and transport in a
  strongly driven anderson insulator},\ }\href
  {https://doi.org/10.1103/PhysRevB.96.014201} {\bibfield  {journal} {\bibinfo
  {journal} {Phys. Rev. B}\ }\textbf {\bibinfo {volume} {96}},\ \bibinfo
  {pages} {014201} (\bibinfo {year} {2017})}\BibitemShut {NoStop}%
\bibitem [{\citenamefont {Hatami}\ \emph {et~al.}(2016)\citenamefont {Hatami},
  \citenamefont {Danieli}, \citenamefont {Bodyfelt},\ and\ \citenamefont
  {Flach}}]{Hatami_2016}%
  \BibitemOpen
  \bibfield  {author} {\bibinfo {author} {\bibfnamefont {H.}~\bibnamefont
  {Hatami}}, \bibinfo {author} {\bibfnamefont {C.}~\bibnamefont {Danieli}},
  \bibinfo {author} {\bibfnamefont {J.~D.}\ \bibnamefont {Bodyfelt}},\ and\
  \bibinfo {author} {\bibfnamefont {S.}~\bibnamefont {Flach}},\ }\bibfield
  {title} {\bibinfo {title} {Quasiperiodic driving of anderson localized waves
  in one dimension},\ }\href {https://doi.org/10.1103/PhysRevE.93.062205}
  {\bibfield  {journal} {\bibinfo  {journal} {Phys. Rev. E}\ }\textbf {\bibinfo
  {volume} {93}},\ \bibinfo {pages} {062205} (\bibinfo {year}
  {2016})}\BibitemShut {NoStop}%
\bibitem [{\citenamefont {Nakajima}\ \emph {et~al.}(2021)\citenamefont
  {Nakajima}, \citenamefont {Takei}, \citenamefont {Sakuma}, \citenamefont
  {Kuno}, \citenamefont {Marra},\ and\ \citenamefont
  {Takahashi}}]{nakajima_2021}%
  \BibitemOpen
  \bibfield  {author} {\bibinfo {author} {\bibfnamefont {S.}~\bibnamefont
  {Nakajima}}, \bibinfo {author} {\bibfnamefont {N.}~\bibnamefont {Takei}},
  \bibinfo {author} {\bibfnamefont {K.}~\bibnamefont {Sakuma}}, \bibinfo
  {author} {\bibfnamefont {Y.}~\bibnamefont {Kuno}}, \bibinfo {author}
  {\bibfnamefont {P.}~\bibnamefont {Marra}},\ and\ \bibinfo {author}
  {\bibfnamefont {Y.}~\bibnamefont {Takahashi}},\ }\bibfield  {title} {\bibinfo
  {title} {Competition and interplay between topology and quasi-periodic
  disorder in thouless pumping of ultracold atoms},\ }\href
  {https://doi.org/10.1038/s41567-021-01229-9} {\bibfield  {journal} {\bibinfo
  {journal} {Nature Physics}\ }\textbf {\bibinfo {volume} {17}},\ \bibinfo
  {pages} {844} (\bibinfo {year} {2021})}\BibitemShut {NoStop}%
\bibitem [{\citenamefont {Li}\ \emph {et~al.}(2009)\citenamefont {Li},
  \citenamefont {Chu}, \citenamefont {Jain},\ and\ \citenamefont
  {Shen}}]{li_topological_2009}%
  \BibitemOpen
  \bibfield  {author} {\bibinfo {author} {\bibfnamefont {J.}~\bibnamefont
  {Li}}, \bibinfo {author} {\bibfnamefont {R.-L.}\ \bibnamefont {Chu}},
  \bibinfo {author} {\bibfnamefont {J.~K.}\ \bibnamefont {Jain}},\ and\
  \bibinfo {author} {\bibfnamefont {S.-Q.}\ \bibnamefont {Shen}},\ }\bibfield
  {title} {\bibinfo {title} {Topological {Anderson} {Insulator}},\ }\href
  {https://doi.org/10.1103/PhysRevLett.102.136806} {\bibfield  {journal}
  {\bibinfo  {journal} {Phys. Rev. Lett.}\ }\textbf {\bibinfo {volume} {102}},\
  \bibinfo {pages} {136806} (\bibinfo {year} {2009})}\BibitemShut {NoStop}%
\bibitem [{\citenamefont {Groth}\ \emph {et~al.}(2009)\citenamefont {Groth},
  \citenamefont {Wimmer}, \citenamefont {Akhmerov}, \citenamefont
  {Tworzydło},\ and\ \citenamefont {Beenakker}}]{groth_theory_2009}%
  \BibitemOpen
  \bibfield  {author} {\bibinfo {author} {\bibfnamefont {C.~W.}\ \bibnamefont
  {Groth}}, \bibinfo {author} {\bibfnamefont {M.}~\bibnamefont {Wimmer}},
  \bibinfo {author} {\bibfnamefont {A.~R.}\ \bibnamefont {Akhmerov}}, \bibinfo
  {author} {\bibfnamefont {J.}~\bibnamefont {Tworzydło}},\ and\ \bibinfo
  {author} {\bibfnamefont {C.~W.~J.}\ \bibnamefont {Beenakker}},\ }\bibfield
  {title} {\bibinfo {title} {Theory of the {Topological} {Anderson}
  {Insulator}},\ }\href {https://doi.org/10.1103/PhysRevLett.103.196805}
  {\bibfield  {journal} {\bibinfo  {journal} {Phys. Rev. Lett.}\ }\textbf
  {\bibinfo {volume} {103}},\ \bibinfo {pages} {196805} (\bibinfo {year}
  {2009})}\BibitemShut {NoStop}%
\bibitem [{\citenamefont {Titum}\ \emph {et~al.}(2016)\citenamefont {Titum},
  \citenamefont {Berg}, \citenamefont {Rudner}, \citenamefont {Refael},\ and\
  \citenamefont {Lindner}}]{titum_anomalous_2016}%
  \BibitemOpen
  \bibfield  {author} {\bibinfo {author} {\bibfnamefont {P.}~\bibnamefont
  {Titum}}, \bibinfo {author} {\bibfnamefont {E.}~\bibnamefont {Berg}},
  \bibinfo {author} {\bibfnamefont {M.~S.}\ \bibnamefont {Rudner}}, \bibinfo
  {author} {\bibfnamefont {G.}~\bibnamefont {Refael}},\ and\ \bibinfo {author}
  {\bibfnamefont {N.~H.}\ \bibnamefont {Lindner}},\ }\bibfield  {title}
  {\bibinfo {title} {Anomalous {Floquet}-{Anderson} {Insulator} as a
  {Nonadiabatic} {Quantized} {Charge} {Pump}},\ }\href
  {https://doi.org/10.1103/PhysRevX.6.021013} {\bibfield  {journal} {\bibinfo
  {journal} {Phys. Rev. X}\ }\textbf {\bibinfo {volume} {6}},\ \bibinfo {pages}
  {021013} (\bibinfo {year} {2016})}\BibitemShut {NoStop}%
\bibitem [{\citenamefont {Citro}\ \emph {et~al.}(2003)\citenamefont {Citro},
  \citenamefont {Andrei},\ and\ \citenamefont {Niu}}]{citro_2003}%
  \BibitemOpen
  \bibfield  {author} {\bibinfo {author} {\bibfnamefont {R.}~\bibnamefont
  {Citro}}, \bibinfo {author} {\bibfnamefont {N.}~\bibnamefont {Andrei}},\ and\
  \bibinfo {author} {\bibfnamefont {Q.}~\bibnamefont {Niu}},\ }\bibfield
  {title} {\bibinfo {title} {Pumping in an interacting quantum wire},\ }\href
  {https://doi.org/10.1103/PhysRevB.68.165312} {\bibfield  {journal} {\bibinfo
  {journal} {Phys. Rev. B}\ }\textbf {\bibinfo {volume} {68}},\ \bibinfo
  {pages} {165312} (\bibinfo {year} {2003})}\BibitemShut {NoStop}%
\bibitem [{\citenamefont {Requist}\ and\ \citenamefont
  {Gross}(2018)}]{requist_2017}%
  \BibitemOpen
  \bibfield  {author} {\bibinfo {author} {\bibfnamefont {R.}~\bibnamefont
  {Requist}}\ and\ \bibinfo {author} {\bibfnamefont {E.~K.~U.}\ \bibnamefont
  {Gross}},\ }\bibfield  {title} {\bibinfo {title} {Accurate formula for the
  macroscopic polarization of strongly correlated materials},\ }\href
  {https://doi.org/10.1021/acs.jpclett.8b03028} {\bibfield  {journal} {\bibinfo
   {journal} {J. Phys. Chem. Lett.}\ }\textbf {\bibinfo {volume} {9}},\
  \bibinfo {pages} {7045} (\bibinfo {year} {2018})}\BibitemShut {NoStop}%
\bibitem [{\citenamefont {Nakagawa}\ \emph {et~al.}(2018)\citenamefont
  {Nakagawa}, \citenamefont {Yoshida}, \citenamefont {Peters},\ and\
  \citenamefont {Kawakami}}]{nakagawa_2018}%
  \BibitemOpen
  \bibfield  {author} {\bibinfo {author} {\bibfnamefont {M.}~\bibnamefont
  {Nakagawa}}, \bibinfo {author} {\bibfnamefont {T.}~\bibnamefont {Yoshida}},
  \bibinfo {author} {\bibfnamefont {R.}~\bibnamefont {Peters}},\ and\ \bibinfo
  {author} {\bibfnamefont {N.}~\bibnamefont {Kawakami}},\ }\bibfield  {title}
  {\bibinfo {title} {Breakdown of topological thouless pumping in the strongly
  interacting regime},\ }\href {https://doi.org/10.1103/PhysRevB.98.115147}
  {\bibfield  {journal} {\bibinfo  {journal} {Phys. Rev. B}\ }\textbf {\bibinfo
  {volume} {98}},\ \bibinfo {pages} {115147} (\bibinfo {year}
  {2018})}\BibitemShut {NoStop}%
\bibitem [{\citenamefont {Bertok}\ \emph {et~al.}(2022)\citenamefont {Bertok},
  \citenamefont {Heidrich-Meisner},\ and\ \citenamefont
  {Aligia}}]{bertok_splitting_2022}%
  \BibitemOpen
  \bibfield  {author} {\bibinfo {author} {\bibfnamefont {E.}~\bibnamefont
  {Bertok}}, \bibinfo {author} {\bibfnamefont {F.}~\bibnamefont
  {Heidrich-Meisner}},\ and\ \bibinfo {author} {\bibfnamefont {A.~A.}\
  \bibnamefont {Aligia}},\ }\bibfield  {title} {\bibinfo {title} {Splitting of
  topological charge pumping in an interacting two-component fermionic
  {Rice}-{Mele} {Hubbard} model},\ }\href
  {https://doi.org/10.1103/PhysRevB.106.045141} {\bibfield  {journal} {\bibinfo
   {journal} {Phys. Rev. B}\ }\textbf {\bibinfo {volume} {106}},\ \bibinfo
  {pages} {045141} (\bibinfo {year} {2022})}\BibitemShut {NoStop}%
\bibitem [{\citenamefont {Stenzel}\ \emph {et~al.}(2019)\citenamefont
  {Stenzel}, \citenamefont {Hayward}, \citenamefont {Hubig}, \citenamefont
  {Schollw\"ock},\ and\ \citenamefont
  {Heidrich-Meisner}}]{stenzel_quantum_2019}%
  \BibitemOpen
  \bibfield  {author} {\bibinfo {author} {\bibfnamefont {L.}~\bibnamefont
  {Stenzel}}, \bibinfo {author} {\bibfnamefont {A.~L.~C.}\ \bibnamefont
  {Hayward}}, \bibinfo {author} {\bibfnamefont {C.}~\bibnamefont {Hubig}},
  \bibinfo {author} {\bibfnamefont {U.}~\bibnamefont {Schollw\"ock}},\ and\
  \bibinfo {author} {\bibfnamefont {F.}~\bibnamefont {Heidrich-Meisner}},\
  }\bibfield  {title} {\bibinfo {title} {Quantum phases and topological
  properties of interacting fermions in one-dimensional superlattices},\ }\href
  {https://doi.org/10.1103/PhysRevA.99.053614} {\bibfield  {journal} {\bibinfo
  {journal} {Phys. Rev. A}\ }\textbf {\bibinfo {volume} {99}},\ \bibinfo
  {pages} {053614} (\bibinfo {year} {2019})}\BibitemShut {NoStop}%
\bibitem [{\citenamefont {Berg}\ \emph {et~al.}(2011)\citenamefont {Berg},
  \citenamefont {Levin},\ and\ \citenamefont {Altman}}]{Berg_2011}%
  \BibitemOpen
  \bibfield  {author} {\bibinfo {author} {\bibfnamefont {E.}~\bibnamefont
  {Berg}}, \bibinfo {author} {\bibfnamefont {M.}~\bibnamefont {Levin}},\ and\
  \bibinfo {author} {\bibfnamefont {E.}~\bibnamefont {Altman}},\ }\bibfield
  {title} {\bibinfo {title} {Quantized pumping and topology of the phase
  diagram for a system of interacting bosons},\ }\href
  {https://doi.org/10.1103/PhysRevLett.106.110405} {\bibfield  {journal}
  {\bibinfo  {journal} {Phys. Rev. Lett.}\ }\textbf {\bibinfo {volume} {106}},\
  \bibinfo {pages} {110405} (\bibinfo {year} {2011})}\BibitemShut {NoStop}%
\bibitem [{\citenamefont {Qian}\ \emph {et~al.}(2011)\citenamefont {Qian},
  \citenamefont {Gong},\ and\ \citenamefont {Zhang}}]{qian_quantum_2011}%
  \BibitemOpen
  \bibfield  {author} {\bibinfo {author} {\bibfnamefont {Y.}~\bibnamefont
  {Qian}}, \bibinfo {author} {\bibfnamefont {M.}~\bibnamefont {Gong}},\ and\
  \bibinfo {author} {\bibfnamefont {C.}~\bibnamefont {Zhang}},\ }\bibfield
  {title} {\bibinfo {title} {Quantum transport of bosonic cold atoms in
  double-well optical lattices},\ }\href
  {https://doi.org/10.1103/PhysRevA.84.013608} {\bibfield  {journal} {\bibinfo
  {journal} {Phys. Rev. A}\ }\textbf {\bibinfo {volume} {84}},\ \bibinfo
  {pages} {013608} (\bibinfo {year} {2011})}\BibitemShut {NoStop}%
\bibitem [{\citenamefont {Grusdt}\ and\ \citenamefont
  {H\"oning}(2014)}]{grusdt_realization_2014}%
  \BibitemOpen
  \bibfield  {author} {\bibinfo {author} {\bibfnamefont {F.}~\bibnamefont
  {Grusdt}}\ and\ \bibinfo {author} {\bibfnamefont {M.}~\bibnamefont
  {H\"oning}},\ }\bibfield  {title} {\bibinfo {title} {Realization of
  fractional chern insulators in the thin-torus limit with ultracold bosons},\
  }\href {https://doi.org/10.1103/PhysRevA.90.053623} {\bibfield  {journal}
  {\bibinfo  {journal} {Phys. Rev. A}\ }\textbf {\bibinfo {volume} {90}},\
  \bibinfo {pages} {053623} (\bibinfo {year} {2014})}\BibitemShut {NoStop}%
\bibitem [{\citenamefont {Zeng}\ \emph {et~al.}(2016)\citenamefont {Zeng},
  \citenamefont {Zhu},\ and\ \citenamefont {Sheng}}]{zeng_fractional_2016}%
  \BibitemOpen
  \bibfield  {author} {\bibinfo {author} {\bibfnamefont {T.-S.}\ \bibnamefont
  {Zeng}}, \bibinfo {author} {\bibfnamefont {W.}~\bibnamefont {Zhu}},\ and\
  \bibinfo {author} {\bibfnamefont {D.~N.}\ \bibnamefont {Sheng}},\ }\bibfield
  {title} {\bibinfo {title} {Fractional charge pumping of interacting bosons in
  one-dimensional superlattice},\ }\href
  {https://doi.org/10.1103/PhysRevB.94.235139} {\bibfield  {journal} {\bibinfo
  {journal} {Phys. Rev. B}\ }\textbf {\bibinfo {volume} {94}},\ \bibinfo
  {pages} {235139} (\bibinfo {year} {2016})}\BibitemShut {NoStop}%
\bibitem [{\citenamefont {Greschner}\ \emph {et~al.}(2020)\citenamefont
  {Greschner}, \citenamefont {Mondal},\ and\ \citenamefont
  {Mishra}}]{greschner_topological_2020}%
  \BibitemOpen
  \bibfield  {author} {\bibinfo {author} {\bibfnamefont {S.}~\bibnamefont
  {Greschner}}, \bibinfo {author} {\bibfnamefont {S.}~\bibnamefont {Mondal}},\
  and\ \bibinfo {author} {\bibfnamefont {T.}~\bibnamefont {Mishra}},\
  }\bibfield  {title} {\bibinfo {title} {Topological charge pumping of bound
  bosonic pairs},\ }\href {https://doi.org/10.1103/PhysRevA.101.053630}
  {\bibfield  {journal} {\bibinfo  {journal} {Phys. Rev. A}\ }\textbf {\bibinfo
  {volume} {101}},\ \bibinfo {pages} {053630} (\bibinfo {year}
  {2020})}\BibitemShut {NoStop}%
\bibitem [{\citenamefont {Kohn}(1964)}]{kohn_1964}%
  \BibitemOpen
  \bibfield  {author} {\bibinfo {author} {\bibfnamefont {W.}~\bibnamefont
  {Kohn}},\ }\bibfield  {title} {\bibinfo {title} {Theory of the insulating
  state},\ }\href {https://doi.org/10.1103/PhysRev.133.A171} {\bibfield
  {journal} {\bibinfo  {journal} {Phys. Rev.}\ }\textbf {\bibinfo {volume}
  {133}},\ \bibinfo {pages} {A171} (\bibinfo {year} {1964})}\BibitemShut
  {NoStop}%
\bibitem [{\citenamefont {Thouless}\ \emph {et~al.}(1982)\citenamefont
  {Thouless}, \citenamefont {Kohmoto}, \citenamefont {Nightingale},\ and\
  \citenamefont {den Nijs}}]{Thouless_1982}%
  \BibitemOpen
  \bibfield  {author} {\bibinfo {author} {\bibfnamefont {D.~J.}\ \bibnamefont
  {Thouless}}, \bibinfo {author} {\bibfnamefont {M.}~\bibnamefont {Kohmoto}},
  \bibinfo {author} {\bibfnamefont {M.~P.}\ \bibnamefont {Nightingale}},\ and\
  \bibinfo {author} {\bibfnamefont {M.}~\bibnamefont {den Nijs}},\ }\bibfield
  {title} {\bibinfo {title} {Quantized hall conductance in a two-dimensional
  periodic potential},\ }\href {https://doi.org/10.1103/PhysRevLett.49.405}
  {\bibfield  {journal} {\bibinfo  {journal} {Phys. Rev. Lett.}\ }\textbf
  {\bibinfo {volume} {49}},\ \bibinfo {pages} {405} (\bibinfo {year}
  {1982})}\BibitemShut {NoStop}%
\bibitem [{\citenamefont {Harper}(1955{\natexlab{a}})}]{aah_model}%
  \BibitemOpen
  \bibfield  {author} {\bibinfo {author} {\bibfnamefont {P.~G.}\ \bibnamefont
  {Harper}},\ }\bibfield  {title} {\bibinfo {title} {Single band motion of
  conduction electrons in a uniform magnetic field},\ }\href
  {https://doi.org/10.1088/0370-1298/68/10/304} {\bibfield  {journal} {\bibinfo
   {journal} {Proc. Phys. Soc. A}\ }\textbf {\bibinfo {volume} {68}},\ \bibinfo
  {pages} {874} (\bibinfo {year} {1955}{\natexlab{a}})}\BibitemShut {NoStop}%
\bibitem [{\citenamefont {Mostaan}\ \emph {et~al.}(2022)\citenamefont
  {Mostaan}, \citenamefont {Grusdt},\ and\ \citenamefont
  {Goldman}}]{mostaan_quantized_2022}%
  \BibitemOpen
  \bibfield  {author} {\bibinfo {author} {\bibfnamefont {N.}~\bibnamefont
  {Mostaan}}, \bibinfo {author} {\bibfnamefont {F.}~\bibnamefont {Grusdt}},\
  and\ \bibinfo {author} {\bibfnamefont {N.}~\bibnamefont {Goldman}},\
  }\bibfield  {title} {\bibinfo {title} {Quantized transport of solitons in
  nonlinear {Thouless} pumps: {From} {Wannier} drags to ultracold topological
  mixtures},\ }\bibfield  {journal} {\bibinfo  {journal} {arXiv preprint}\
  }\href {https://doi.org/10.48550/arXiv.2110.13075}
  {10.48550/arXiv.2110.13075} (\bibinfo {year} {2022})\BibitemShut {NoStop}%
\bibitem [{\citenamefont {J{\"u}rgensen}\ and\ \citenamefont
  {Rechtsman}(2022)}]{jurgensen_chern_vs_soliton_2022}%
  \BibitemOpen
  \bibfield  {author} {\bibinfo {author} {\bibfnamefont {M.}~\bibnamefont
  {J{\"u}rgensen}}\ and\ \bibinfo {author} {\bibfnamefont {M.~C.}\ \bibnamefont
  {Rechtsman}},\ }\bibfield  {title} {\bibinfo {title} {Chern number governs
  soliton motion in nonlinear thouless pumps},\ }\href
  {https://doi.org/10.1103/PhysRevLett.128.113901} {\bibfield  {journal}
  {\bibinfo  {journal} {Phys. Rev. Lett.}\ }\textbf {\bibinfo {volume} {128}},\
  \bibinfo {pages} {113901} (\bibinfo {year} {2022})}\BibitemShut {NoStop}%
\bibitem [{\citenamefont {Fu}\ \emph {et~al.}(2022)\citenamefont {Fu},
  \citenamefont {Wang}, \citenamefont {Kartashov}, \citenamefont {Konotop},\
  and\ \citenamefont {Ye}}]{konotop_2022}%
  \BibitemOpen
  \bibfield  {author} {\bibinfo {author} {\bibfnamefont {Q.}~\bibnamefont
  {Fu}}, \bibinfo {author} {\bibfnamefont {P.}~\bibnamefont {Wang}}, \bibinfo
  {author} {\bibfnamefont {Y.~V.}\ \bibnamefont {Kartashov}}, \bibinfo {author}
  {\bibfnamefont {V.~V.}\ \bibnamefont {Konotop}},\ and\ \bibinfo {author}
  {\bibfnamefont {F.}~\bibnamefont {Ye}},\ }\bibfield  {title} {\bibinfo
  {title} {Nonlinear thouless pumping: Solitons and transport breakdown},\
  }\href {https://doi.org/10.1103/PhysRevLett.128.154101} {\bibfield  {journal}
  {\bibinfo  {journal} {Phys. Rev. Lett.}\ }\textbf {\bibinfo {volume} {128}},\
  \bibinfo {pages} {154101} (\bibinfo {year} {2022})}\BibitemShut {NoStop}%
\bibitem [{\citenamefont {Jürgensen}\ \emph {et~al.}(2022)\citenamefont
  {Jürgensen}, \citenamefont {Mukherjee}, \citenamefont {Jörg},\ and\
  \citenamefont {Rechtsman}}]{jurgensen_quantized_2022}%
  \BibitemOpen
  \bibfield  {author} {\bibinfo {author} {\bibfnamefont {M.}~\bibnamefont
  {Jürgensen}}, \bibinfo {author} {\bibfnamefont {S.}~\bibnamefont
  {Mukherjee}}, \bibinfo {author} {\bibfnamefont {C.}~\bibnamefont {Jörg}},\
  and\ \bibinfo {author} {\bibfnamefont {M.~C.}\ \bibnamefont {Rechtsman}},\
  }\bibfield  {title} {\bibinfo {title} {Quantized {Fractional} {Thouless}
  {Pumping} of {Solitons}},\ }\bibfield  {journal} {\bibinfo  {journal} {arXiv
  preprint}\ }\href {https://doi.org/10.48550/arXiv.2201.08258}
  {10.48550/arXiv.2201.08258} (\bibinfo {year} {2022})\BibitemShut {NoStop}%
\bibitem [{\citenamefont {Jung}\ \emph {et~al.}(2022)\citenamefont {Jung},
  \citenamefont {Parto}, \citenamefont {Pyrialakos}, \citenamefont {Nasari},
  \citenamefont {Rutkowska}, \citenamefont {Trippenbach}, \citenamefont
  {Khajavikhan}, \citenamefont {Krolikowski},\ and\ \citenamefont
  {Christodoulides}}]{nematic_liquid_crystals}%
  \BibitemOpen
  \bibfield  {author} {\bibinfo {author} {\bibfnamefont {P.~S.}\ \bibnamefont
  {Jung}}, \bibinfo {author} {\bibfnamefont {M.}~\bibnamefont {Parto}},
  \bibinfo {author} {\bibfnamefont {G.~G.}\ \bibnamefont {Pyrialakos}},
  \bibinfo {author} {\bibfnamefont {H.}~\bibnamefont {Nasari}}, \bibinfo
  {author} {\bibfnamefont {K.}~\bibnamefont {Rutkowska}}, \bibinfo {author}
  {\bibfnamefont {M.}~\bibnamefont {Trippenbach}}, \bibinfo {author}
  {\bibfnamefont {M.}~\bibnamefont {Khajavikhan}}, \bibinfo {author}
  {\bibfnamefont {W.}~\bibnamefont {Krolikowski}},\ and\ \bibinfo {author}
  {\bibfnamefont {D.~N.}\ \bibnamefont {Christodoulides}},\ }\bibfield  {title}
  {\bibinfo {title} {Optical thouless pumping transport and nonlinear switching
  in a topological low-dimensional discrete nematic liquid crystal array},\
  }\href {https://doi.org/10.1103/PhysRevA.105.013513} {\bibfield  {journal}
  {\bibinfo  {journal} {Phys. Rev. A}\ }\textbf {\bibinfo {volume} {105}},\
  \bibinfo {pages} {013513} (\bibinfo {year} {2022})}\BibitemShut {NoStop}%
\bibitem [{\citenamefont {Hayward}\ \emph {et~al.}(2018)\citenamefont
  {Hayward}, \citenamefont {Schweizer}, \citenamefont {Lohse}, \citenamefont
  {Aidelsburger},\ and\ \citenamefont
  {Heidrich-Meisner}}]{Lohse_interaction_2018}%
  \BibitemOpen
  \bibfield  {author} {\bibinfo {author} {\bibfnamefont {A.}~\bibnamefont
  {Hayward}}, \bibinfo {author} {\bibfnamefont {C.}~\bibnamefont {Schweizer}},
  \bibinfo {author} {\bibfnamefont {M.}~\bibnamefont {Lohse}}, \bibinfo
  {author} {\bibfnamefont {M.}~\bibnamefont {Aidelsburger}},\ and\ \bibinfo
  {author} {\bibfnamefont {F.}~\bibnamefont {Heidrich-Meisner}},\ }\bibfield
  {title} {\bibinfo {title} {Topological charge pumping in the interacting
  bosonic rice-mele model},\ }\href
  {https://doi.org/10.1103/PhysRevB.98.245148} {\bibfield  {journal} {\bibinfo
  {journal} {Phys. Rev. B}\ }\textbf {\bibinfo {volume} {98}},\ \bibinfo
  {pages} {245148} (\bibinfo {year} {2018})}\BibitemShut {NoStop}%
\bibitem [{\citenamefont {Schollwöck}(2011)}]{schollwocks_2011}%
  \BibitemOpen
  \bibfield  {author} {\bibinfo {author} {\bibfnamefont {U.}~\bibnamefont
  {Schollwöck}},\ }\bibfield  {title} {\bibinfo {title} {The density-matrix
  renormalization group in the age of matrix product states},\ }\href
  {https://doi.org/https://doi.org/10.1016/j.aop.2010.09.012} {\bibfield
  {journal} {\bibinfo  {journal} {Annals of Physics}\ }\textbf {\bibinfo
  {volume} {326}},\ \bibinfo {pages} {96} (\bibinfo {year} {2011})}\BibitemShut
  {NoStop}%
\bibitem [{\citenamefont {Lin}\ \emph {et~al.}(2020)\citenamefont {Lin},
  \citenamefont {Ke},\ and\ \citenamefont
  {Lee}}]{lin_interaction-induced_2020}%
  \BibitemOpen
  \bibfield  {author} {\bibinfo {author} {\bibfnamefont {L.}~\bibnamefont
  {Lin}}, \bibinfo {author} {\bibfnamefont {Y.}~\bibnamefont {Ke}},\ and\
  \bibinfo {author} {\bibfnamefont {C.}~\bibnamefont {Lee}},\ }\bibfield
  {title} {\bibinfo {title} {Interaction-induced topological bound states and
  {Thouless} pumping in a one-dimensional optical lattice},\ }\href
  {https://doi.org/10.1103/PhysRevA.101.023620} {\bibfield  {journal} {\bibinfo
   {journal} {Phys. Rev. A}\ }\textbf {\bibinfo {volume} {101}},\ \bibinfo
  {pages} {023620} (\bibinfo {year} {2020})}\BibitemShut {NoStop}%
\bibitem [{\citenamefont {Kuno}\ and\ \citenamefont
  {Hatsugai}(2020)}]{Yoshihito_interaction_induced_pump_2022}%
  \BibitemOpen
  \bibfield  {author} {\bibinfo {author} {\bibfnamefont {Y.}~\bibnamefont
  {Kuno}}\ and\ \bibinfo {author} {\bibfnamefont {Y.}~\bibnamefont
  {Hatsugai}},\ }\bibfield  {title} {\bibinfo {title} {Interaction-induced
  topological charge pump},\ }\href
  {https://doi.org/10.1103/PhysRevResearch.2.042024} {\bibfield  {journal}
  {\bibinfo  {journal} {Phys. Rev. Research}\ }\textbf {\bibinfo {volume}
  {2}},\ \bibinfo {pages} {042024} (\bibinfo {year} {2020})}\BibitemShut
  {NoStop}%
\bibitem [{\citenamefont {Esin}\ \emph {et~al.}(2022)\citenamefont {Esin},
  \citenamefont {Kuhlenkamp}, \citenamefont {Refael}, \citenamefont {Berg},
  \citenamefont {Rudner},\ and\ \citenamefont {Lindner}}]{esin_universal_2022}%
  \BibitemOpen
  \bibfield  {author} {\bibinfo {author} {\bibfnamefont {I.}~\bibnamefont
  {Esin}}, \bibinfo {author} {\bibfnamefont {C.}~\bibnamefont {Kuhlenkamp}},
  \bibinfo {author} {\bibfnamefont {G.}~\bibnamefont {Refael}}, \bibinfo
  {author} {\bibfnamefont {E.}~\bibnamefont {Berg}}, \bibinfo {author}
  {\bibfnamefont {M.~S.}\ \bibnamefont {Rudner}},\ and\ \bibinfo {author}
  {\bibfnamefont {N.~H.}\ \bibnamefont {Lindner}},\ }\bibfield  {title}
  {\bibinfo {title} {Universal transport in periodically driven systems without
  long-lived quasiparticles},\ }\bibfield  {journal} {\bibinfo  {journal}
  {arXiv preprint}\ }\href {https://doi.org/10.48550/arXiv.2203.01313}
  {10.48550/arXiv.2203.01313} (\bibinfo {year} {2022})\BibitemShut {NoStop}%
\bibitem [{\citenamefont {Ke}\ \emph {et~al.}(2017)\citenamefont {Ke},
  \citenamefont {Qin}, \citenamefont {Kivshar},\ and\ \citenamefont
  {Lee}}]{multiparticle_wannier}%
  \BibitemOpen
  \bibfield  {author} {\bibinfo {author} {\bibfnamefont {Y.}~\bibnamefont
  {Ke}}, \bibinfo {author} {\bibfnamefont {X.}~\bibnamefont {Qin}}, \bibinfo
  {author} {\bibfnamefont {Y.~S.}\ \bibnamefont {Kivshar}},\ and\ \bibinfo
  {author} {\bibfnamefont {C.}~\bibnamefont {Lee}},\ }\bibfield  {title}
  {\bibinfo {title} {Multiparticle wannier states and thouless pumping of
  interacting bosons},\ }\href {https://doi.org/10.1103/PhysRevA.95.063630}
  {\bibfield  {journal} {\bibinfo  {journal} {Phys. Rev. A}\ }\textbf {\bibinfo
  {volume} {95}},\ \bibinfo {pages} {063630} (\bibinfo {year}
  {2017})}\BibitemShut {NoStop}%
\bibitem [{\citenamefont {Haug}\ \emph {et~al.}(2020)\citenamefont {Haug},
  \citenamefont {Amico}, \citenamefont {Kwek}, \citenamefont {Munro},\ and\
  \citenamefont {Bastidas}}]{amico_2020}%
  \BibitemOpen
  \bibfield  {author} {\bibinfo {author} {\bibfnamefont {T.}~\bibnamefont
  {Haug}}, \bibinfo {author} {\bibfnamefont {L.}~\bibnamefont {Amico}},
  \bibinfo {author} {\bibfnamefont {L.-C.}\ \bibnamefont {Kwek}}, \bibinfo
  {author} {\bibfnamefont {W.~J.}\ \bibnamefont {Munro}},\ and\ \bibinfo
  {author} {\bibfnamefont {V.~M.}\ \bibnamefont {Bastidas}},\ }\bibfield
  {title} {\bibinfo {title} {Topological pumping of quantum correlations},\
  }\href {https://doi.org/10.1103/PhysRevResearch.2.013135} {\bibfield
  {journal} {\bibinfo  {journal} {Phys. Rev. Research}\ }\textbf {\bibinfo
  {volume} {2}},\ \bibinfo {pages} {013135} (\bibinfo {year}
  {2020})}\BibitemShut {NoStop}%
\bibitem [{\citenamefont {Ando}\ \emph {et~al.}(1975)\citenamefont {Ando},
  \citenamefont {Matsumoto},\ and\ \citenamefont {Uemura}}]{ando_1975_theory}%
  \BibitemOpen
  \bibfield  {author} {\bibinfo {author} {\bibfnamefont {T.}~\bibnamefont
  {Ando}}, \bibinfo {author} {\bibfnamefont {Y.}~\bibnamefont {Matsumoto}},\
  and\ \bibinfo {author} {\bibfnamefont {Y.}~\bibnamefont {Uemura}},\
  }\bibfield  {title} {\bibinfo {title} {Theory of hall effect in a
  two-dimensional electron system},\ }\href
  {https://doi.org/10.1143/JPSJ.39.279} {\bibfield  {journal} {\bibinfo
  {journal} {J. Phys. Soc. Jpn.}\ }\textbf {\bibinfo {volume} {39}},\ \bibinfo
  {pages} {279} (\bibinfo {year} {1975})}\BibitemShut {NoStop}%
\bibitem [{\citenamefont {Laughlin}(1981)}]{Laughlin_1981}%
  \BibitemOpen
  \bibfield  {author} {\bibinfo {author} {\bibfnamefont {R.~B.}\ \bibnamefont
  {Laughlin}},\ }\bibfield  {title} {\bibinfo {title} {Quantized hall
  conductivity in two dimensions},\ }\href
  {https://doi.org/10.1103/PhysRevB.23.5632} {\bibfield  {journal} {\bibinfo
  {journal} {Phys. Rev. B}\ }\textbf {\bibinfo {volume} {23}},\ \bibinfo
  {pages} {5632} (\bibinfo {year} {1981})}\BibitemShut {NoStop}%
\bibitem [{\citenamefont {Harper}(1955{\natexlab{b}})}]{harper_general_1955}%
  \BibitemOpen
  \bibfield  {author} {\bibinfo {author} {\bibfnamefont {P.~G.}\ \bibnamefont
  {Harper}},\ }\bibfield  {title} {\bibinfo {title} {The {General} {Motion} of
  {Conduction} {Electrons} in a {Uniform} {Magnetic} {Field}, with
  {Application} to the {Diamagnetism} of {Metals}},\ }\href
  {https://doi.org/10.1088/0370-1298/68/10/305} {\bibfield  {journal} {\bibinfo
   {journal} {Proc. Phys. Soc. A}\ }\textbf {\bibinfo {volume} {68}},\ \bibinfo
  {pages} {879} (\bibinfo {year} {1955}{\natexlab{b}})}\BibitemShut {NoStop}%
\bibitem [{\citenamefont {Azbel}(1964)}]{azbel_1964}%
  \BibitemOpen
  \bibfield  {author} {\bibinfo {author} {\bibfnamefont {M.~Y.}\ \bibnamefont
  {Azbel}},\ }\bibfield  {title} {\bibinfo {title} {Energy spectrum of a
  conduction electron in a magnetic field},\ }\href@noop {} {\bibfield
  {journal} {\bibinfo  {journal} {Zh. Eksp. Teor. Fiz.}\ }\textbf {\bibinfo
  {volume} {46}},\ \bibinfo {pages} {929} (\bibinfo {year} {1964})}\BibitemShut
  {NoStop}%
\bibitem [{\citenamefont {Hofstadter}(1976)}]{hofstadter_energy_1976}%
  \BibitemOpen
  \bibfield  {author} {\bibinfo {author} {\bibfnamefont {D.~R.}\ \bibnamefont
  {Hofstadter}},\ }\bibfield  {title} {\bibinfo {title} {Energy levels and wave
  functions of {Bloch} electrons in rational and irrational magnetic fields},\
  }\href {https://doi.org/10.1103/PhysRevB.14.2239} {\bibfield  {journal}
  {\bibinfo  {journal} {Phys. Rev. B}\ }\textbf {\bibinfo {volume} {14}},\
  \bibinfo {pages} {2239} (\bibinfo {year} {1976})}\BibitemShut {NoStop}%
\bibitem [{\citenamefont {Aubry}\ and\ \citenamefont
  {André}(1980)}]{aubry_1980}%
  \BibitemOpen
  \bibfield  {author} {\bibinfo {author} {\bibfnamefont {S.}~\bibnamefont
  {Aubry}}\ and\ \bibinfo {author} {\bibfnamefont {G.}~\bibnamefont {André}},\
  }\bibfield  {title} {\bibinfo {title} {Analyticity breaking and anderson
  localization in incommensurate lattices},\ }\href@noop {} {\bibfield
  {journal} {\bibinfo  {journal} {Ann. Israel Phys. Soc.}\ }\textbf {\bibinfo
  {volume} {3}},\ \bibinfo {pages} {133} (\bibinfo {year} {1980})}\BibitemShut
  {NoStop}%
\bibitem [{\citenamefont {Kraus}\ \emph
  {et~al.}(2012{\natexlab{b}})\citenamefont {Kraus}, \citenamefont {Lahini},
  \citenamefont {Ringel}, \citenamefont {Verbin},\ and\ \citenamefont
  {Zilberberg}}]{kraus_topological_2012}%
  \BibitemOpen
  \bibfield  {author} {\bibinfo {author} {\bibfnamefont {Y.~E.}\ \bibnamefont
  {Kraus}}, \bibinfo {author} {\bibfnamefont {Y.}~\bibnamefont {Lahini}},
  \bibinfo {author} {\bibfnamefont {Z.}~\bibnamefont {Ringel}}, \bibinfo
  {author} {\bibfnamefont {M.}~\bibnamefont {Verbin}},\ and\ \bibinfo {author}
  {\bibfnamefont {O.}~\bibnamefont {Zilberberg}},\ }\bibfield  {title}
  {\bibinfo {title} {Topological {States} and {Adiabatic} {Pumping} in
  {Quasicrystals}},\ }\href {https://doi.org/10.1103/PhysRevLett.109.106402}
  {\bibfield  {journal} {\bibinfo  {journal} {Phys. Rev. Lett.}\ }\textbf
  {\bibinfo {volume} {109}},\ \bibinfo {pages} {106402} (\bibinfo {year}
  {2012}{\natexlab{b}})}\BibitemShut {NoStop}%
\bibitem [{\citenamefont {Kraus}\ \emph {et~al.}(2013)\citenamefont {Kraus},
  \citenamefont {Ringel},\ and\ \citenamefont
  {Zilberberg}}]{kraus_four-dimensional_2013}%
  \BibitemOpen
  \bibfield  {author} {\bibinfo {author} {\bibfnamefont {Y.~E.}\ \bibnamefont
  {Kraus}}, \bibinfo {author} {\bibfnamefont {Z.}~\bibnamefont {Ringel}},\ and\
  \bibinfo {author} {\bibfnamefont {O.}~\bibnamefont {Zilberberg}},\ }\bibfield
   {title} {\bibinfo {title} {Four-{Dimensional} {Quantum} {Hall} {Effect} in a
  {Two}-{Dimensional} {Quasicrystal}},\ }\href
  {https://doi.org/10.1103/PhysRevLett.111.226401} {\bibfield  {journal}
  {\bibinfo  {journal} {Phys. Rev. Lett.}\ }\textbf {\bibinfo {volume} {111}},\
  \bibinfo {pages} {226401} (\bibinfo {year} {2013})}\BibitemShut {NoStop}%
\bibitem [{\citenamefont {Marra}\ \emph {et~al.}(2015)\citenamefont {Marra},
  \citenamefont {Citro},\ and\ \citenamefont {Ortix}}]{Marra_2015}%
  \BibitemOpen
  \bibfield  {author} {\bibinfo {author} {\bibfnamefont {P.}~\bibnamefont
  {Marra}}, \bibinfo {author} {\bibfnamefont {R.}~\bibnamefont {Citro}},\ and\
  \bibinfo {author} {\bibfnamefont {C.}~\bibnamefont {Ortix}},\ }\bibfield
  {title} {\bibinfo {title} {Fractional quantization of the topological charge
  pumping in a one-dimensional superlattice},\ }\href
  {https://doi.org/10.1103/PhysRevB.91.125411} {\bibfield  {journal} {\bibinfo
  {journal} {Phys. Rev. B}\ }\textbf {\bibinfo {volume} {91}},\ \bibinfo
  {pages} {125411} (\bibinfo {year} {2015})}\BibitemShut {NoStop}%
\bibitem [{\citenamefont {Wei}\ and\ \citenamefont
  {Mueller}(2015)}]{wei_anomalous_2015}%
  \BibitemOpen
  \bibfield  {author} {\bibinfo {author} {\bibfnamefont {R.}~\bibnamefont
  {Wei}}\ and\ \bibinfo {author} {\bibfnamefont {E.~J.}\ \bibnamefont
  {Mueller}},\ }\bibfield  {title} {\bibinfo {title} {Anomalous charge pumping
  in a one-dimensional optical superlattice},\ }\href
  {https://doi.org/10.1103/PhysRevA.92.013609} {\bibfield  {journal} {\bibinfo
  {journal} {Phys. Rev. A}\ }\textbf {\bibinfo {volume} {92}},\ \bibinfo
  {pages} {013609} (\bibinfo {year} {2015})}\BibitemShut {NoStop}%
\bibitem [{\citenamefont {Mancini}\ \emph {et~al.}(2015)\citenamefont
  {Mancini}, \citenamefont {Pagano}, \citenamefont {Cappellini}, \citenamefont
  {Livi}, \citenamefont {Rider}, \citenamefont {Catani}, \citenamefont {Sias},
  \citenamefont {Zoller}, \citenamefont {Inguscio}, \citenamefont {Dalmonte},\
  and\ \citenamefont {Fallani}}]{mancini_observation_2015}%
  \BibitemOpen
  \bibfield  {author} {\bibinfo {author} {\bibfnamefont {M.}~\bibnamefont
  {Mancini}}, \bibinfo {author} {\bibfnamefont {G.}~\bibnamefont {Pagano}},
  \bibinfo {author} {\bibfnamefont {G.}~\bibnamefont {Cappellini}}, \bibinfo
  {author} {\bibfnamefont {L.}~\bibnamefont {Livi}}, \bibinfo {author}
  {\bibfnamefont {M.}~\bibnamefont {Rider}}, \bibinfo {author} {\bibfnamefont
  {J.}~\bibnamefont {Catani}}, \bibinfo {author} {\bibfnamefont
  {C.}~\bibnamefont {Sias}}, \bibinfo {author} {\bibfnamefont {P.}~\bibnamefont
  {Zoller}}, \bibinfo {author} {\bibfnamefont {M.}~\bibnamefont {Inguscio}},
  \bibinfo {author} {\bibfnamefont {M.}~\bibnamefont {Dalmonte}},\ and\
  \bibinfo {author} {\bibfnamefont {L.}~\bibnamefont {Fallani}},\ }\bibfield
  {title} {\bibinfo {title} {Observation of chiral edge states with neutral
  fermions in synthetic {Hall} ribbons},\ }\href
  {https://doi.org/10.1126/science.aaa8736} {\bibfield  {journal} {\bibinfo
  {journal} {Science}\ }\textbf {\bibinfo {volume} {349}},\ \bibinfo {pages}
  {1510} (\bibinfo {year} {2015})}\BibitemShut {NoStop}%
\bibitem [{\citenamefont {Stuhl}\ \emph {et~al.}(2015)\citenamefont {Stuhl},
  \citenamefont {Lu}, \citenamefont {Aycock}, \citenamefont {Genkina},\ and\
  \citenamefont {Spielman}}]{stuhl_visualizing_2015}%
  \BibitemOpen
  \bibfield  {author} {\bibinfo {author} {\bibfnamefont {B.~K.}\ \bibnamefont
  {Stuhl}}, \bibinfo {author} {\bibfnamefont {H.-I.}\ \bibnamefont {Lu}},
  \bibinfo {author} {\bibfnamefont {L.~M.}\ \bibnamefont {Aycock}}, \bibinfo
  {author} {\bibfnamefont {D.}~\bibnamefont {Genkina}},\ and\ \bibinfo {author}
  {\bibfnamefont {I.~B.}\ \bibnamefont {Spielman}},\ }\bibfield  {title}
  {\bibinfo {title} {Visualizing edge states with an atomic {Bose} gas in the
  quantum {Hall} regime},\ }\href {https://doi.org/10.1126/science.aaa8515}
  {\bibfield  {journal} {\bibinfo  {journal} {Science}\ }\textbf {\bibinfo
  {volume} {349}},\ \bibinfo {pages} {1514} (\bibinfo {year}
  {2015})}\BibitemShut {NoStop}%
\bibitem [{\citenamefont {Fabre}\ \emph {et~al.}(2022)\citenamefont {Fabre},
  \citenamefont {Bouhiron}, \citenamefont {Satoor}, \citenamefont {Lopes},\
  and\ \citenamefont {Nascimbene}}]{fabre_laughlins_2022}%
  \BibitemOpen
  \bibfield  {author} {\bibinfo {author} {\bibfnamefont {A.}~\bibnamefont
  {Fabre}}, \bibinfo {author} {\bibfnamefont {J.-B.}\ \bibnamefont {Bouhiron}},
  \bibinfo {author} {\bibfnamefont {T.}~\bibnamefont {Satoor}}, \bibinfo
  {author} {\bibfnamefont {R.}~\bibnamefont {Lopes}},\ and\ \bibinfo {author}
  {\bibfnamefont {S.}~\bibnamefont {Nascimbene}},\ }\bibfield  {title}
  {\bibinfo {title} {Laughlin's {Topological} {Charge} {Pump} in an {Atomic}
  {Hall} {Cylinder}},\ }\href {https://doi.org/10.1103/PhysRevLett.128.173202}
  {\bibfield  {journal} {\bibinfo  {journal} {Phys. Rev. Lett.}\ }\textbf
  {\bibinfo {volume} {128}},\ \bibinfo {pages} {173202} (\bibinfo {year}
  {2022})}\BibitemShut {NoStop}%
\bibitem [{\citenamefont {Celi}\ \emph {et~al.}(2014)\citenamefont {Celi},
  \citenamefont {Massignan}, \citenamefont {Ruseckas}, \citenamefont {Goldman},
  \citenamefont {Spielman}, \citenamefont {Juzeliūnas},\ and\ \citenamefont
  {Lewenstein}}]{celi_synthetic_2014}%
  \BibitemOpen
  \bibfield  {author} {\bibinfo {author} {\bibfnamefont {A.}~\bibnamefont
  {Celi}}, \bibinfo {author} {\bibfnamefont {P.}~\bibnamefont {Massignan}},
  \bibinfo {author} {\bibfnamefont {J.}~\bibnamefont {Ruseckas}}, \bibinfo
  {author} {\bibfnamefont {N.}~\bibnamefont {Goldman}}, \bibinfo {author}
  {\bibfnamefont {I.}~\bibnamefont {Spielman}}, \bibinfo {author}
  {\bibfnamefont {G.}~\bibnamefont {Juzeliūnas}},\ and\ \bibinfo {author}
  {\bibfnamefont {M.}~\bibnamefont {Lewenstein}},\ }\bibfield  {title}
  {\bibinfo {title} {Synthetic {Gauge} {Fields} in {Synthetic} {Dimensions}},\
  }\href {https://doi.org/10.1103/PhysRevLett.112.043001} {\bibfield  {journal}
  {\bibinfo  {journal} {Phys. Rev. Lett.}\ }\textbf {\bibinfo {volume} {112}},\
  \bibinfo {pages} {043001} (\bibinfo {year} {2014})}\BibitemShut {NoStop}%
\bibitem [{\citenamefont {Livi}\ \emph {et~al.}(2016)\citenamefont {Livi},
  \citenamefont {Cappellini}, \citenamefont {Diem}, \citenamefont {Franchi},
  \citenamefont {Clivati}, \citenamefont {Frittelli}, \citenamefont {Levi},
  \citenamefont {Calonico}, \citenamefont {Catani}, \citenamefont {Inguscio},\
  and\ \citenamefont {Fallani}}]{livi_synthetic_2016}%
  \BibitemOpen
  \bibfield  {author} {\bibinfo {author} {\bibfnamefont {L.}~\bibnamefont
  {Livi}}, \bibinfo {author} {\bibfnamefont {G.}~\bibnamefont {Cappellini}},
  \bibinfo {author} {\bibfnamefont {M.}~\bibnamefont {Diem}}, \bibinfo {author}
  {\bibfnamefont {L.}~\bibnamefont {Franchi}}, \bibinfo {author} {\bibfnamefont
  {C.}~\bibnamefont {Clivati}}, \bibinfo {author} {\bibfnamefont
  {M.}~\bibnamefont {Frittelli}}, \bibinfo {author} {\bibfnamefont
  {F.}~\bibnamefont {Levi}}, \bibinfo {author} {\bibfnamefont {D.}~\bibnamefont
  {Calonico}}, \bibinfo {author} {\bibfnamefont {J.}~\bibnamefont {Catani}},
  \bibinfo {author} {\bibfnamefont {M.}~\bibnamefont {Inguscio}},\ and\
  \bibinfo {author} {\bibfnamefont {L.}~\bibnamefont {Fallani}},\ }\bibfield
  {title} {\bibinfo {title} {Synthetic {Dimensions} and {Spin}-{Orbit}
  {Coupling} with an {Optical} {Clock} {Transition}},\ }\href
  {https://doi.org/10.1103/PhysRevLett.117.220401} {\bibfield  {journal}
  {\bibinfo  {journal} {Phys. Rev. Lett.}\ }\textbf {\bibinfo {volume} {117}},\
  \bibinfo {pages} {220401} (\bibinfo {year} {2016})}\BibitemShut {NoStop}%
\bibitem [{\citenamefont {Kolkowitz}\ \emph {et~al.}(2017)\citenamefont
  {Kolkowitz}, \citenamefont {Bromley}, \citenamefont {Bothwell}, \citenamefont
  {Wall}, \citenamefont {Marti}, \citenamefont {Koller}, \citenamefont {Zhang},
  \citenamefont {Rey},\ and\ \citenamefont
  {Ye}}]{kolkowitz_spinorbit-coupled_2017}%
  \BibitemOpen
  \bibfield  {author} {\bibinfo {author} {\bibfnamefont {S.}~\bibnamefont
  {Kolkowitz}}, \bibinfo {author} {\bibfnamefont {S.~L.}\ \bibnamefont
  {Bromley}}, \bibinfo {author} {\bibfnamefont {T.}~\bibnamefont {Bothwell}},
  \bibinfo {author} {\bibfnamefont {M.~L.}\ \bibnamefont {Wall}}, \bibinfo
  {author} {\bibfnamefont {G.~E.}\ \bibnamefont {Marti}}, \bibinfo {author}
  {\bibfnamefont {A.~P.}\ \bibnamefont {Koller}}, \bibinfo {author}
  {\bibfnamefont {X.}~\bibnamefont {Zhang}}, \bibinfo {author} {\bibfnamefont
  {A.~M.}\ \bibnamefont {Rey}},\ and\ \bibinfo {author} {\bibfnamefont
  {J.}~\bibnamefont {Ye}},\ }\bibfield  {title} {\bibinfo {title}
  {Spin–orbit-coupled fermions in an optical lattice clock},\ }\href
  {https://doi.org/10.1038/nature20811} {\bibfield  {journal} {\bibinfo
  {journal} {Nature}\ }\textbf {\bibinfo {volume} {542}},\ \bibinfo {pages}
  {66} (\bibinfo {year} {2017})}\BibitemShut {NoStop}%
\bibitem [{\citenamefont {An}\ \emph {et~al.}(2017)\citenamefont {An},
  \citenamefont {Meier},\ and\ \citenamefont {Gadway}}]{an_direct_2017}%
  \BibitemOpen
  \bibfield  {author} {\bibinfo {author} {\bibfnamefont {F.~A.}\ \bibnamefont
  {An}}, \bibinfo {author} {\bibfnamefont {E.~J.}\ \bibnamefont {Meier}},\ and\
  \bibinfo {author} {\bibfnamefont {B.}~\bibnamefont {Gadway}},\ }\bibfield
  {title} {\bibinfo {title} {Direct observation of chiral currents and magnetic
  reflection in atomic flux lattices},\ }\href
  {https://doi.org/10.1126/sciadv.1602685} {\bibfield  {journal} {\bibinfo
  {journal} {Science Advances}\ }\textbf {\bibinfo {volume} {3}},\ \bibinfo
  {pages} {e1602685} (\bibinfo {year} {2017})}\BibitemShut {NoStop}%
\bibitem [{\citenamefont {Lustig}\ \emph {et~al.}(2019)\citenamefont {Lustig},
  \citenamefont {Weimann}, \citenamefont {Plotnik}, \citenamefont {Lumer},
  \citenamefont {Bandres}, \citenamefont {Szameit},\ and\ \citenamefont
  {Segev}}]{lustig_photonic_2019}%
  \BibitemOpen
  \bibfield  {author} {\bibinfo {author} {\bibfnamefont {E.}~\bibnamefont
  {Lustig}}, \bibinfo {author} {\bibfnamefont {S.}~\bibnamefont {Weimann}},
  \bibinfo {author} {\bibfnamefont {Y.}~\bibnamefont {Plotnik}}, \bibinfo
  {author} {\bibfnamefont {Y.}~\bibnamefont {Lumer}}, \bibinfo {author}
  {\bibfnamefont {M.~A.}\ \bibnamefont {Bandres}}, \bibinfo {author}
  {\bibfnamefont {A.}~\bibnamefont {Szameit}},\ and\ \bibinfo {author}
  {\bibfnamefont {M.}~\bibnamefont {Segev}},\ }\bibfield  {title} {\bibinfo
  {title} {Photonic topological insulator in synthetic dimensions},\ }\href
  {https://doi.org/10.1038/s41586-019-0943-7} {\bibfield  {journal} {\bibinfo
  {journal} {Nature}\ }\textbf {\bibinfo {volume} {567}},\ \bibinfo {pages}
  {356} (\bibinfo {year} {2019})}\BibitemShut {NoStop}%
\bibitem [{\citenamefont {Ozawa}\ \emph {et~al.}(2019)\citenamefont {Ozawa},
  \citenamefont {Price}, \citenamefont {Amo}, \citenamefont {Goldman},
  \citenamefont {Hafezi}, \citenamefont {Lu}, \citenamefont {Rechtsman},
  \citenamefont {Schuster}, \citenamefont {Simon}, \citenamefont {Zilberberg},\
  and\ \citenamefont {Carusotto}}]{ozawa_topological_2019}%
  \BibitemOpen
  \bibfield  {author} {\bibinfo {author} {\bibfnamefont {T.}~\bibnamefont
  {Ozawa}}, \bibinfo {author} {\bibfnamefont {H.~M.}\ \bibnamefont {Price}},
  \bibinfo {author} {\bibfnamefont {A.}~\bibnamefont {Amo}}, \bibinfo {author}
  {\bibfnamefont {N.}~\bibnamefont {Goldman}}, \bibinfo {author} {\bibfnamefont
  {M.}~\bibnamefont {Hafezi}}, \bibinfo {author} {\bibfnamefont
  {L.}~\bibnamefont {Lu}}, \bibinfo {author} {\bibfnamefont {M.~C.}\
  \bibnamefont {Rechtsman}}, \bibinfo {author} {\bibfnamefont {D.}~\bibnamefont
  {Schuster}}, \bibinfo {author} {\bibfnamefont {J.}~\bibnamefont {Simon}},
  \bibinfo {author} {\bibfnamefont {O.}~\bibnamefont {Zilberberg}},\ and\
  \bibinfo {author} {\bibfnamefont {I.}~\bibnamefont {Carusotto}},\ }\bibfield
  {title} {\bibinfo {title} {Topological photonics},\ }\href
  {https://doi.org/10.1103/RevModPhys.91.015006} {\bibfield  {journal}
  {\bibinfo  {journal} {Rev. Mod. Phys.}\ }\textbf {\bibinfo {volume} {91}},\
  \bibinfo {pages} {015006} (\bibinfo {year} {2019})}\BibitemShut {NoStop}%
\bibitem [{\citenamefont {Meier}\ \emph {et~al.}(2018)\citenamefont {Meier},
  \citenamefont {An}, \citenamefont {Dauphin}, \citenamefont {Maffei},
  \citenamefont {Massignan}, \citenamefont {Hughes},\ and\ \citenamefont
  {Gadway}}]{meier_2018}%
  \BibitemOpen
  \bibfield  {author} {\bibinfo {author} {\bibfnamefont {E.~J.}\ \bibnamefont
  {Meier}}, \bibinfo {author} {\bibfnamefont {F.~A.}\ \bibnamefont {An}},
  \bibinfo {author} {\bibfnamefont {A.}~\bibnamefont {Dauphin}}, \bibinfo
  {author} {\bibfnamefont {M.}~\bibnamefont {Maffei}}, \bibinfo {author}
  {\bibfnamefont {P.}~\bibnamefont {Massignan}}, \bibinfo {author}
  {\bibfnamefont {T.~L.}\ \bibnamefont {Hughes}},\ and\ \bibinfo {author}
  {\bibfnamefont {B.}~\bibnamefont {Gadway}},\ }\bibfield  {title} {\bibinfo
  {title} {Observation of the topological anderson insulator in disordered
  atomic wires},\ }\href {https://doi.org/10.1126/science.aat34} {\bibfield
  {journal} {\bibinfo  {journal} {Science}\ }\textbf {\bibinfo {volume}
  {362}},\ \bibinfo {pages} {929} (\bibinfo {year} {2018})}\BibitemShut
  {NoStop}%
\bibitem [{\citenamefont {Han}\ \emph {et~al.}(2019)\citenamefont {Han},
  \citenamefont {Kang},\ and\ \citenamefont {Shin}}]{han_band_2019}%
  \BibitemOpen
  \bibfield  {author} {\bibinfo {author} {\bibfnamefont {J.~H.}\ \bibnamefont
  {Han}}, \bibinfo {author} {\bibfnamefont {J.~H.}\ \bibnamefont {Kang}},\ and\
  \bibinfo {author} {\bibfnamefont {Y.}~\bibnamefont {Shin}},\ }\bibfield
  {title} {\bibinfo {title} {Band {Gap} {Closing} in a {Synthetic} {Hall}
  {Tube} of {Neutral} {Fermions}},\ }\href
  {https://doi.org/10.1103/PhysRevLett.122.065303} {\bibfield  {journal}
  {\bibinfo  {journal} {Phys. Rev. Lett.}\ }\textbf {\bibinfo {volume} {122}},\
  \bibinfo {pages} {065303} (\bibinfo {year} {2019})}\BibitemShut {NoStop}%
\bibitem [{\citenamefont {Liang}\ \emph {et~al.}(2021)\citenamefont {Liang},
  \citenamefont {Trypogeorgos}, \citenamefont {Valdés-Curiel}, \citenamefont
  {Tao}, \citenamefont {Zhao},\ and\ \citenamefont
  {Spielman}}]{liang_coherence_2021}%
  \BibitemOpen
  \bibfield  {author} {\bibinfo {author} {\bibfnamefont {Q.-Y.}\ \bibnamefont
  {Liang}}, \bibinfo {author} {\bibfnamefont {D.}~\bibnamefont {Trypogeorgos}},
  \bibinfo {author} {\bibfnamefont {A.}~\bibnamefont {Valdés-Curiel}},
  \bibinfo {author} {\bibfnamefont {J.}~\bibnamefont {Tao}}, \bibinfo {author}
  {\bibfnamefont {M.}~\bibnamefont {Zhao}},\ and\ \bibinfo {author}
  {\bibfnamefont {I.~B.}\ \bibnamefont {Spielman}},\ }\bibfield  {title}
  {\bibinfo {title} {Coherence and decoherence in the {Harper}-{Hofstadter}
  model},\ }\href {https://doi.org/10.1103/PhysRevResearch.3.023058} {\bibfield
   {journal} {\bibinfo  {journal} {Phys. Rev. Research}\ }\textbf {\bibinfo
  {volume} {3}},\ \bibinfo {pages} {023058} (\bibinfo {year}
  {2021})}\BibitemShut {NoStop}%
\bibitem [{\citenamefont {Li}\ \emph {et~al.}(2022)\citenamefont {Li},
  \citenamefont {Yan}, \citenamefont {Feng}, \citenamefont {Choudhury},
  \citenamefont {Blasing}, \citenamefont {Zhou},\ and\ \citenamefont
  {Chen}}]{li_bose-einstein_2022}%
  \BibitemOpen
  \bibfield  {author} {\bibinfo {author} {\bibfnamefont {C.-H.}\ \bibnamefont
  {Li}}, \bibinfo {author} {\bibfnamefont {Y.}~\bibnamefont {Yan}}, \bibinfo
  {author} {\bibfnamefont {S.-W.}\ \bibnamefont {Feng}}, \bibinfo {author}
  {\bibfnamefont {S.}~\bibnamefont {Choudhury}}, \bibinfo {author}
  {\bibfnamefont {D.~B.}\ \bibnamefont {Blasing}}, \bibinfo {author}
  {\bibfnamefont {Q.}~\bibnamefont {Zhou}},\ and\ \bibinfo {author}
  {\bibfnamefont {Y.~P.}\ \bibnamefont {Chen}},\ }\bibfield  {title} {\bibinfo
  {title} {Bose-{Einstein} {Condensate} on a {Synthetic} {Topological} {Hall}
  {Cylinder}},\ }\href {https://doi.org/10.1103/PRXQuantum.3.010316} {\bibfield
   {journal} {\bibinfo  {journal} {PRX Quantum}\ }\textbf {\bibinfo {volume}
  {3}},\ \bibinfo {pages} {010316} (\bibinfo {year} {2022})}\BibitemShut
  {NoStop}%
\bibitem [{\citenamefont {Price}\ \emph {et~al.}(2015)\citenamefont {Price},
  \citenamefont {Zilberberg}, \citenamefont {Ozawa}, \citenamefont
  {Carusotto},\ and\ \citenamefont {Goldman}}]{price_four-dimensional_2015}%
  \BibitemOpen
  \bibfield  {author} {\bibinfo {author} {\bibfnamefont {H.}~\bibnamefont
  {Price}}, \bibinfo {author} {\bibfnamefont {O.}~\bibnamefont {Zilberberg}},
  \bibinfo {author} {\bibfnamefont {T.}~\bibnamefont {Ozawa}}, \bibinfo
  {author} {\bibfnamefont {I.}~\bibnamefont {Carusotto}},\ and\ \bibinfo
  {author} {\bibfnamefont {N.}~\bibnamefont {Goldman}},\ }\bibfield  {title}
  {\bibinfo {title} {Four-{Dimensional} {Quantum} {Hall} {Effect} with
  {Ultracold} {Atoms}},\ }\href
  {https://doi.org/10.1103/PhysRevLett.115.195303} {\bibfield  {journal}
  {\bibinfo  {journal} {Phys. Rev. Lett.}\ }\textbf {\bibinfo {volume} {115}},\
  \bibinfo {pages} {195303} (\bibinfo {year} {2015})}\BibitemShut {NoStop}%
\bibitem [{\citenamefont {Ozawa}\ \emph {et~al.}(2016)\citenamefont {Ozawa},
  \citenamefont {Price}, \citenamefont {Goldman}, \citenamefont {Zilberberg},\
  and\ \citenamefont {Carusotto}}]{ozawa_synthetic_2016}%
  \BibitemOpen
  \bibfield  {author} {\bibinfo {author} {\bibfnamefont {T.}~\bibnamefont
  {Ozawa}}, \bibinfo {author} {\bibfnamefont {H.~M.}\ \bibnamefont {Price}},
  \bibinfo {author} {\bibfnamefont {N.}~\bibnamefont {Goldman}}, \bibinfo
  {author} {\bibfnamefont {O.}~\bibnamefont {Zilberberg}},\ and\ \bibinfo
  {author} {\bibfnamefont {I.}~\bibnamefont {Carusotto}},\ }\bibfield  {title}
  {\bibinfo {title} {Synthetic dimensions in integrated photonics: {From}
  optical isolation to four-dimensional quantum {Hall} physics},\ }\href
  {https://doi.org/10.1103/PhysRevA.93.043827} {\bibfield  {journal} {\bibinfo
  {journal} {Phys. Rev. A}\ }\textbf {\bibinfo {volume} {93}},\ \bibinfo
  {pages} {043827} (\bibinfo {year} {2016})}\BibitemShut {NoStop}%
\bibitem [{\citenamefont {Lu}\ \emph {et~al.}(2018)\citenamefont {Lu},
  \citenamefont {Gao},\ and\ \citenamefont {Wang}}]{lu_topological_2018}%
  \BibitemOpen
  \bibfield  {author} {\bibinfo {author} {\bibfnamefont {L.}~\bibnamefont
  {Lu}}, \bibinfo {author} {\bibfnamefont {H.}~\bibnamefont {Gao}},\ and\
  \bibinfo {author} {\bibfnamefont {Z.}~\bibnamefont {Wang}},\ }\bibfield
  {title} {\bibinfo {title} {Topological one-way fiber of second {Chern}
  number},\ }\href {https://doi.org/10.1038/s41467-018-07817-3} {\bibfield
  {journal} {\bibinfo  {journal} {Nat. Commun.}\ }\textbf {\bibinfo {volume}
  {9}},\ \bibinfo {pages} {5384} (\bibinfo {year} {2018})}\BibitemShut
  {NoStop}%
\bibitem [{\citenamefont {Kolodrubetz}(2016)}]{kolodrubetz_measuring_2016}%
  \BibitemOpen
  \bibfield  {author} {\bibinfo {author} {\bibfnamefont {M.}~\bibnamefont
  {Kolodrubetz}},\ }\bibfield  {title} {\bibinfo {title} {Measuring the
  {Second} {Chern} {Number} from {Nonadiabatic} {Effects}},\ }\href
  {https://doi.org/10.1103/PhysRevLett.117.015301} {\bibfield  {journal}
  {\bibinfo  {journal} {Phys. Rev. Lett.}\ }\textbf {\bibinfo {volume} {117}},\
  \bibinfo {pages} {015301} (\bibinfo {year} {2016})}\BibitemShut {NoStop}%
\bibitem [{\citenamefont {Lohse}\ \emph {et~al.}(2018)\citenamefont {Lohse},
  \citenamefont {Schweizer}, \citenamefont {Price}, \citenamefont
  {Zilberberg},\ and\ \citenamefont {Bloch}}]{lohse2018}%
  \BibitemOpen
  \bibfield  {author} {\bibinfo {author} {\bibfnamefont {M.}~\bibnamefont
  {Lohse}}, \bibinfo {author} {\bibfnamefont {C.}~\bibnamefont {Schweizer}},
  \bibinfo {author} {\bibfnamefont {H.~M.}\ \bibnamefont {Price}}, \bibinfo
  {author} {\bibfnamefont {O.}~\bibnamefont {Zilberberg}},\ and\ \bibinfo
  {author} {\bibfnamefont {I.}~\bibnamefont {Bloch}},\ }\bibfield  {title}
  {\bibinfo {title} {Exploring {4D} quantum {Hall} physics with a {2D}
  topological charge pump},\ }\href {https://doi.org/10.1038/nature25000}
  {\bibfield  {journal} {\bibinfo  {journal} {Nature}\ }\textbf {\bibinfo
  {volume} {553}},\ \bibinfo {pages} {55} (\bibinfo {year} {2018})}\BibitemShut
  {NoStop}%
\bibitem [{\citenamefont {Zilberberg}\ \emph {et~al.}(2018)\citenamefont
  {Zilberberg}, \citenamefont {Huang}, \citenamefont {Guglielmon},
  \citenamefont {Wang}, \citenamefont {Chen}, \citenamefont {Kraus},\ and\
  \citenamefont {Rechtsman}}]{zilberberg_photonic_2018}%
  \BibitemOpen
  \bibfield  {author} {\bibinfo {author} {\bibfnamefont {O.}~\bibnamefont
  {Zilberberg}}, \bibinfo {author} {\bibfnamefont {S.}~\bibnamefont {Huang}},
  \bibinfo {author} {\bibfnamefont {J.}~\bibnamefont {Guglielmon}}, \bibinfo
  {author} {\bibfnamefont {M.}~\bibnamefont {Wang}}, \bibinfo {author}
  {\bibfnamefont {K.~P.}\ \bibnamefont {Chen}}, \bibinfo {author}
  {\bibfnamefont {Y.~E.}\ \bibnamefont {Kraus}},\ and\ \bibinfo {author}
  {\bibfnamefont {M.~C.}\ \bibnamefont {Rechtsman}},\ }\bibfield  {title}
  {\bibinfo {title} {Photonic topological boundary pumping as a probe of {4D}
  quantum {Hall} physics},\ }\href {https://doi.org/10.1038/nature25011}
  {\bibfield  {journal} {\bibinfo  {journal} {Nature}\ }\textbf {\bibinfo
  {volume} {553}},\ \bibinfo {pages} {59} (\bibinfo {year} {2018})}\BibitemShut
  {NoStop}%
\bibitem [{\citenamefont {Sugawa}\ \emph {et~al.}(2018)\citenamefont {Sugawa},
  \citenamefont {Salces-Carcoba}, \citenamefont {Perry}, \citenamefont {Yue},\
  and\ \citenamefont {Spielman}}]{sugawa_second_2018}%
  \BibitemOpen
  \bibfield  {author} {\bibinfo {author} {\bibfnamefont {S.}~\bibnamefont
  {Sugawa}}, \bibinfo {author} {\bibfnamefont {F.}~\bibnamefont
  {Salces-Carcoba}}, \bibinfo {author} {\bibfnamefont {A.~R.}\ \bibnamefont
  {Perry}}, \bibinfo {author} {\bibfnamefont {Y.}~\bibnamefont {Yue}},\ and\
  \bibinfo {author} {\bibfnamefont {I.~B.}\ \bibnamefont {Spielman}},\
  }\bibfield  {title} {\bibinfo {title} {Second {Chern} number of a
  quantum-simulated non-{Abelian} {Yang} monopole},\ }\href
  {https://doi.org/10.1126/science.aam9031} {\bibfield  {journal} {\bibinfo
  {journal} {Science}\ }\textbf {\bibinfo {volume} {360}},\ \bibinfo {pages}
  {1429} (\bibinfo {year} {2018})}\BibitemShut {NoStop}%
\bibitem [{\citenamefont {Benalcazar}\ \emph
  {et~al.}(2017{\natexlab{a}})\citenamefont {Benalcazar}, \citenamefont
  {Bernevig},\ and\ \citenamefont {Hughes}}]{Benalcazar2017}%
  \BibitemOpen
  \bibfield  {author} {\bibinfo {author} {\bibfnamefont {W.~A.}\ \bibnamefont
  {Benalcazar}}, \bibinfo {author} {\bibfnamefont {B.~A.}\ \bibnamefont
  {Bernevig}},\ and\ \bibinfo {author} {\bibfnamefont {T.~L.}\ \bibnamefont
  {Hughes}},\ }\bibfield  {title} {\bibinfo {title} {Quantized electric
  multipole insulators},\ }\href {https://doi.org/10.1126/science.aah6442}
  {\bibfield  {journal} {\bibinfo  {journal} {Science}\ }\textbf {\bibinfo
  {volume} {357}},\ \bibinfo {pages} {61} (\bibinfo {year}
  {2017}{\natexlab{a}})}\BibitemShut {NoStop}%
\bibitem [{\citenamefont {Benalcazar}\ \emph
  {et~al.}(2017{\natexlab{b}})\citenamefont {Benalcazar}, \citenamefont
  {Bernevig},\ and\ \citenamefont {Hughes}}]{Benalcazar2017a}%
  \BibitemOpen
  \bibfield  {author} {\bibinfo {author} {\bibfnamefont {W.~A.}\ \bibnamefont
  {Benalcazar}}, \bibinfo {author} {\bibfnamefont {B.~A.}\ \bibnamefont
  {Bernevig}},\ and\ \bibinfo {author} {\bibfnamefont {T.~L.}\ \bibnamefont
  {Hughes}},\ }\bibfield  {title} {\bibinfo {title} {Electric multipole
  moments, topological multipole moment pumping, and chiral hinge states in
  crystalline insulators},\ }\href {https://doi.org/10.1103/PhysRevB.96.245115}
  {\bibfield  {journal} {\bibinfo  {journal} {Phys. Rev. B}\ }\textbf {\bibinfo
  {volume} {96}},\ \bibinfo {pages} {245115} (\bibinfo {year}
  {2017}{\natexlab{b}})}\BibitemShut {NoStop}%
\bibitem [{\citenamefont {Benalcazar}\ \emph {et~al.}(2022)\citenamefont
  {Benalcazar}, \citenamefont {Noh}, \citenamefont {Wang}, \citenamefont
  {Huang}, \citenamefont {Chen},\ and\ \citenamefont
  {Rechtsman}}]{Benalcazar2020}%
  \BibitemOpen
  \bibfield  {author} {\bibinfo {author} {\bibfnamefont {W.~A.}\ \bibnamefont
  {Benalcazar}}, \bibinfo {author} {\bibfnamefont {J.}~\bibnamefont {Noh}},
  \bibinfo {author} {\bibfnamefont {M.}~\bibnamefont {Wang}}, \bibinfo {author}
  {\bibfnamefont {S.}~\bibnamefont {Huang}}, \bibinfo {author} {\bibfnamefont
  {K.~P.}\ \bibnamefont {Chen}},\ and\ \bibinfo {author} {\bibfnamefont
  {M.~C.}\ \bibnamefont {Rechtsman}},\ }\bibfield  {title} {\bibinfo {title}
  {Higher-order topological pumping},\ }\href
  {https://doi.org/10.1103/PhysRevB.105.195129} {\bibfield  {journal} {\bibinfo
   {journal} {Phys. Rev. B}\ }\textbf {\bibinfo {volume} {105}},\ \bibinfo
  {pages} {195129} (\bibinfo {year} {2022})}\BibitemShut {NoStop}%
\bibitem [{\citenamefont {Noguchi}\ \emph {et~al.}(2021)\citenamefont
  {Noguchi}, \citenamefont {Kobayashi}, \citenamefont {Jiang}, \citenamefont
  {Kuroda}, \citenamefont {Takahashi}, \citenamefont {Xu}, \citenamefont {Lee},
  \citenamefont {Hirayama}, \citenamefont {Ochi}, \citenamefont {Shirasawa},
  \citenamefont {Zhang}, \citenamefont {Lin}, \citenamefont {Bareille},
  \citenamefont {Sakuragi}, \citenamefont {Tanaka}, \citenamefont {Kunisada},
  \citenamefont {Kurokawa}, \citenamefont {Yaji}, \citenamefont {Harasawa},
  \citenamefont {Kandyba}, \citenamefont {Giampietri}, \citenamefont {Barinov},
  \citenamefont {Kim}, \citenamefont {Cacho}, \citenamefont {Hashimoto},
  \citenamefont {Lu}, \citenamefont {Shin}, \citenamefont {Arita},
  \citenamefont {Lai}, \citenamefont {Sasagawa},\ and\ \citenamefont
  {Kondo}}]{Noguchi2021}%
  \BibitemOpen
  \bibfield  {author} {\bibinfo {author} {\bibfnamefont {R.}~\bibnamefont
  {Noguchi}}, \bibinfo {author} {\bibfnamefont {M.}~\bibnamefont {Kobayashi}},
  \bibinfo {author} {\bibfnamefont {Z.}~\bibnamefont {Jiang}}, \bibinfo
  {author} {\bibfnamefont {K.}~\bibnamefont {Kuroda}}, \bibinfo {author}
  {\bibfnamefont {T.}~\bibnamefont {Takahashi}}, \bibinfo {author}
  {\bibfnamefont {Z.}~\bibnamefont {Xu}}, \bibinfo {author} {\bibfnamefont
  {D.}~\bibnamefont {Lee}}, \bibinfo {author} {\bibfnamefont {M.}~\bibnamefont
  {Hirayama}}, \bibinfo {author} {\bibfnamefont {M.}~\bibnamefont {Ochi}},
  \bibinfo {author} {\bibfnamefont {T.}~\bibnamefont {Shirasawa}}, \bibinfo
  {author} {\bibfnamefont {P.}~\bibnamefont {Zhang}}, \bibinfo {author}
  {\bibfnamefont {C.}~\bibnamefont {Lin}}, \bibinfo {author} {\bibfnamefont
  {C.}~\bibnamefont {Bareille}}, \bibinfo {author} {\bibfnamefont
  {S.}~\bibnamefont {Sakuragi}}, \bibinfo {author} {\bibfnamefont
  {H.}~\bibnamefont {Tanaka}}, \bibinfo {author} {\bibfnamefont
  {S.}~\bibnamefont {Kunisada}}, \bibinfo {author} {\bibfnamefont
  {K.}~\bibnamefont {Kurokawa}}, \bibinfo {author} {\bibfnamefont
  {K.}~\bibnamefont {Yaji}}, \bibinfo {author} {\bibfnamefont {A.}~\bibnamefont
  {Harasawa}}, \bibinfo {author} {\bibfnamefont {V.}~\bibnamefont {Kandyba}},
  \bibinfo {author} {\bibfnamefont {A.}~\bibnamefont {Giampietri}}, \bibinfo
  {author} {\bibfnamefont {A.}~\bibnamefont {Barinov}}, \bibinfo {author}
  {\bibfnamefont {T.~K.}\ \bibnamefont {Kim}}, \bibinfo {author} {\bibfnamefont
  {C.}~\bibnamefont {Cacho}}, \bibinfo {author} {\bibfnamefont
  {M.}~\bibnamefont {Hashimoto}}, \bibinfo {author} {\bibfnamefont
  {D.}~\bibnamefont {Lu}}, \bibinfo {author} {\bibfnamefont {S.}~\bibnamefont
  {Shin}}, \bibinfo {author} {\bibfnamefont {R.}~\bibnamefont {Arita}},
  \bibinfo {author} {\bibfnamefont {K.}~\bibnamefont {Lai}}, \bibinfo {author}
  {\bibfnamefont {T.}~\bibnamefont {Sasagawa}},\ and\ \bibinfo {author}
  {\bibfnamefont {T.}~\bibnamefont {Kondo}},\ }\bibfield  {title} {\bibinfo
  {title} {Evidence for a higher-order topological insulator in a
  three-dimensional material built from van der {Waals} stacking of
  bismuth-halide chains},\ }\href {https://doi.org/10.1038/s41563-020-00871-7}
  {\bibfield  {journal} {\bibinfo  {journal} {Nature Materials}\ }\textbf
  {\bibinfo {volume} {20}},\ \bibinfo {pages} {473} (\bibinfo {year}
  {2021})}\BibitemShut {NoStop}%
\bibitem [{\citenamefont {Peterson}\ \emph {et~al.}(2018)\citenamefont
  {Peterson}, \citenamefont {Benalcazar}, \citenamefont {Hughes},\ and\
  \citenamefont {Bahl}}]{Peterson2018}%
  \BibitemOpen
  \bibfield  {author} {\bibinfo {author} {\bibfnamefont {C.~W.}\ \bibnamefont
  {Peterson}}, \bibinfo {author} {\bibfnamefont {W.~A.}\ \bibnamefont
  {Benalcazar}}, \bibinfo {author} {\bibfnamefont {T.~L.}\ \bibnamefont
  {Hughes}},\ and\ \bibinfo {author} {\bibfnamefont {G.}~\bibnamefont {Bahl}},\
  }\bibfield  {title} {\bibinfo {title} {A quantized microwave quadrupole
  insulator with topologically protected corner states},\ }\href
  {https://doi.org/10.1038/nature25777} {\bibfield  {journal} {\bibinfo
  {journal} {Nature}\ }\textbf {\bibinfo {volume} {555}},\ \bibinfo {pages}
  {346} (\bibinfo {year} {2018})}\BibitemShut {NoStop}%
\bibitem [{\citenamefont {Imhof}\ \emph {et~al.}(2018)\citenamefont {Imhof},
  \citenamefont {Berger}, \citenamefont {Bayer}, \citenamefont {Brehm},
  \citenamefont {Molenkamp}, \citenamefont {Kiessling}, \citenamefont
  {Schindler}, \citenamefont {Lee}, \citenamefont {Greiter}, \citenamefont
  {Neupert},\ and\ \citenamefont {Thomale}}]{Imhof2018}%
  \BibitemOpen
  \bibfield  {author} {\bibinfo {author} {\bibfnamefont {S.}~\bibnamefont
  {Imhof}}, \bibinfo {author} {\bibfnamefont {C.}~\bibnamefont {Berger}},
  \bibinfo {author} {\bibfnamefont {F.}~\bibnamefont {Bayer}}, \bibinfo
  {author} {\bibfnamefont {J.}~\bibnamefont {Brehm}}, \bibinfo {author}
  {\bibfnamefont {L.~W.}\ \bibnamefont {Molenkamp}}, \bibinfo {author}
  {\bibfnamefont {T.}~\bibnamefont {Kiessling}}, \bibinfo {author}
  {\bibfnamefont {F.}~\bibnamefont {Schindler}}, \bibinfo {author}
  {\bibfnamefont {C.~H.}\ \bibnamefont {Lee}}, \bibinfo {author} {\bibfnamefont
  {M.}~\bibnamefont {Greiter}}, \bibinfo {author} {\bibfnamefont
  {T.}~\bibnamefont {Neupert}},\ and\ \bibinfo {author} {\bibfnamefont
  {R.}~\bibnamefont {Thomale}},\ }\bibfield  {title} {\bibinfo {title}
  {Topolectrical-circuit realization of topological corner modes},\ }\href
  {https://doi.org/10.1038/s41567-018-0246-1} {\bibfield  {journal} {\bibinfo
  {journal} {Nature Physics}\ }\textbf {\bibinfo {volume} {14}},\ \bibinfo
  {pages} {925} (\bibinfo {year} {2018})}\BibitemShut {NoStop}%
\bibitem [{\citenamefont {Serra-Garcia}\ \emph {et~al.}(2018)\citenamefont
  {Serra-Garcia}, \citenamefont {Peri}, \citenamefont {Süsstrunk},
  \citenamefont {Bilal}, \citenamefont {Larsen}, \citenamefont {Villanueva},\
  and\ \citenamefont {Huber}}]{SerraGarcia2018}%
  \BibitemOpen
  \bibfield  {author} {\bibinfo {author} {\bibfnamefont {M.}~\bibnamefont
  {Serra-Garcia}}, \bibinfo {author} {\bibfnamefont {V.}~\bibnamefont {Peri}},
  \bibinfo {author} {\bibfnamefont {R.}~\bibnamefont {Süsstrunk}}, \bibinfo
  {author} {\bibfnamefont {O.~R.}\ \bibnamefont {Bilal}}, \bibinfo {author}
  {\bibfnamefont {T.}~\bibnamefont {Larsen}}, \bibinfo {author} {\bibfnamefont
  {L.~G.}\ \bibnamefont {Villanueva}},\ and\ \bibinfo {author} {\bibfnamefont
  {S.~D.}\ \bibnamefont {Huber}},\ }\bibfield  {title} {\bibinfo {title}
  {Observation of a phononic quadrupole topological insulator},\ }\href
  {https://doi.org/10.1038/nature25156} {\bibfield  {journal} {\bibinfo
  {journal} {Nature}\ }\textbf {\bibinfo {volume} {555}},\ \bibinfo {pages}
  {342} (\bibinfo {year} {2018})}\BibitemShut {NoStop}%
\bibitem [{\citenamefont {Bao}\ \emph {et~al.}(2019)\citenamefont {Bao},
  \citenamefont {Zou}, \citenamefont {Zhang}, \citenamefont {He}, \citenamefont
  {Sun},\ and\ \citenamefont {Zhang}}]{Bao2019}%
  \BibitemOpen
  \bibfield  {author} {\bibinfo {author} {\bibfnamefont {J.}~\bibnamefont
  {Bao}}, \bibinfo {author} {\bibfnamefont {D.}~\bibnamefont {Zou}}, \bibinfo
  {author} {\bibfnamefont {W.}~\bibnamefont {Zhang}}, \bibinfo {author}
  {\bibfnamefont {W.}~\bibnamefont {He}}, \bibinfo {author} {\bibfnamefont
  {H.}~\bibnamefont {Sun}},\ and\ \bibinfo {author} {\bibfnamefont
  {X.}~\bibnamefont {Zhang}},\ }\bibfield  {title} {\bibinfo {title}
  {Topoelectrical circuit octupole insulator with topologically protected
  corner states},\ }\href {https://doi.org/10.1103/PhysRevB.100.201406}
  {\bibfield  {journal} {\bibinfo  {journal} {Phys. Rev. B}\ }\textbf {\bibinfo
  {volume} {100}},\ \bibinfo {pages} {201406} (\bibinfo {year}
  {2019})}\BibitemShut {NoStop}%
\bibitem [{\citenamefont {Mittal}\ \emph {et~al.}(2019)\citenamefont {Mittal},
  \citenamefont {Orre}, \citenamefont {Zhu}, \citenamefont {Gorlach},
  \citenamefont {Poddubny},\ and\ \citenamefont {Hafezi}}]{Mittal2019}%
  \BibitemOpen
  \bibfield  {author} {\bibinfo {author} {\bibfnamefont {S.}~\bibnamefont
  {Mittal}}, \bibinfo {author} {\bibfnamefont {V.~V.}\ \bibnamefont {Orre}},
  \bibinfo {author} {\bibfnamefont {G.}~\bibnamefont {Zhu}}, \bibinfo {author}
  {\bibfnamefont {M.~A.}\ \bibnamefont {Gorlach}}, \bibinfo {author}
  {\bibfnamefont {A.}~\bibnamefont {Poddubny}},\ and\ \bibinfo {author}
  {\bibfnamefont {M.}~\bibnamefont {Hafezi}},\ }\bibfield  {title} {\bibinfo
  {title} {Photonic quadrupole topological phases},\ }\href
  {https://doi.org/10.1038/s41566-019-0452-0} {\bibfield  {journal} {\bibinfo
  {journal} {Nature Photonics}\ }\textbf {\bibinfo {volume} {13}},\ \bibinfo
  {pages} {692} (\bibinfo {year} {2019})}\BibitemShut {NoStop}%
\bibitem [{\citenamefont {Ni}\ \emph {et~al.}(2019)\citenamefont {Ni},
  \citenamefont {Weiner}, \citenamefont {Alù},\ and\ \citenamefont
  {Khanikaev}}]{Ni2019}%
  \BibitemOpen
  \bibfield  {author} {\bibinfo {author} {\bibfnamefont {X.}~\bibnamefont
  {Ni}}, \bibinfo {author} {\bibfnamefont {M.}~\bibnamefont {Weiner}}, \bibinfo
  {author} {\bibfnamefont {A.}~\bibnamefont {Alù}},\ and\ \bibinfo {author}
  {\bibfnamefont {A.~B.}\ \bibnamefont {Khanikaev}},\ }\bibfield  {title}
  {\bibinfo {title} {Observation of higher-order topological acoustic states
  protected by generalized chiral symmetry},\ }\href
  {https://doi.org/10.1038/s41563-018-0252-9} {\bibfield  {journal} {\bibinfo
  {journal} {Nature Materials}\ }\textbf {\bibinfo {volume} {18}},\ \bibinfo
  {pages} {113} (\bibinfo {year} {2019})}\BibitemShut {NoStop}%
\bibitem [{\citenamefont {Ni}\ \emph {et~al.}(2020)\citenamefont {Ni},
  \citenamefont {Li}, \citenamefont {Weiner}, \citenamefont {Alù},\ and\
  \citenamefont {Khanikaev}}]{Ni2020}%
  \BibitemOpen
  \bibfield  {author} {\bibinfo {author} {\bibfnamefont {X.}~\bibnamefont
  {Ni}}, \bibinfo {author} {\bibfnamefont {M.}~\bibnamefont {Li}}, \bibinfo
  {author} {\bibfnamefont {M.}~\bibnamefont {Weiner}}, \bibinfo {author}
  {\bibfnamefont {A.}~\bibnamefont {Alù}},\ and\ \bibinfo {author}
  {\bibfnamefont {A.~B.}\ \bibnamefont {Khanikaev}},\ }\bibfield  {title}
  {\bibinfo {title} {Demonstration of a quantized acoustic octupole topological
  insulator},\ }\href {https://doi.org/10.1038/s41467-020-15705-y} {\bibfield
  {journal} {\bibinfo  {journal} {Nat. Commun.}\ }\textbf {\bibinfo {volume}
  {11}},\ \bibinfo {pages} {2108} (\bibinfo {year} {2020})}\BibitemShut
  {NoStop}%
\bibitem [{\citenamefont {Xue}\ \emph {et~al.}(2018)\citenamefont {Xue},
  \citenamefont {Yang}, \citenamefont {Gao}, \citenamefont {Chong},\ and\
  \citenamefont {Zhang}}]{Xue2018}%
  \BibitemOpen
  \bibfield  {author} {\bibinfo {author} {\bibfnamefont {H.}~\bibnamefont
  {Xue}}, \bibinfo {author} {\bibfnamefont {Y.}~\bibnamefont {Yang}}, \bibinfo
  {author} {\bibfnamefont {F.}~\bibnamefont {Gao}}, \bibinfo {author}
  {\bibfnamefont {Y.}~\bibnamefont {Chong}},\ and\ \bibinfo {author}
  {\bibfnamefont {B.}~\bibnamefont {Zhang}},\ }\bibfield  {title} {\bibinfo
  {title} {Acoustic higher-order topological insulator on a kagome lattice},\
  }\href {https://doi.org/10.1038/s41563-018-0251-x} {\bibfield  {journal}
  {\bibinfo  {journal} {Nature Materials}\ }\textbf {\bibinfo {volume} {18}},\
  \bibinfo {pages} {108} (\bibinfo {year} {2018})}\BibitemShut {NoStop}%
\bibitem [{\citenamefont {Dutt}\ \emph {et~al.}(2020)\citenamefont {Dutt},
  \citenamefont {Minkov}, \citenamefont {Williamson},\ and\ \citenamefont
  {Fan}}]{Dutt2020}%
  \BibitemOpen
  \bibfield  {author} {\bibinfo {author} {\bibfnamefont {A.}~\bibnamefont
  {Dutt}}, \bibinfo {author} {\bibfnamefont {M.}~\bibnamefont {Minkov}},
  \bibinfo {author} {\bibfnamefont {I.~A.~D.}\ \bibnamefont {Williamson}},\
  and\ \bibinfo {author} {\bibfnamefont {S.}~\bibnamefont {Fan}},\ }\bibfield
  {title} {\bibinfo {title} {Higher-order topological insulators in synthetic
  dimensions},\ }\href {https://doi.org/10.1038/s41377-020-0334-8} {\bibfield
  {journal} {\bibinfo  {journal} {Light Sci. Appl.}\ }\textbf {\bibinfo
  {volume} {9}},\ \bibinfo {pages} {131} (\bibinfo {year} {2020})}\BibitemShut
  {NoStop}%
\bibitem [{\citenamefont {Kang}\ \emph {et~al.}(2019)\citenamefont {Kang},
  \citenamefont {Shiozaki},\ and\ \citenamefont {Cho}}]{Kang2019}%
  \BibitemOpen
  \bibfield  {author} {\bibinfo {author} {\bibfnamefont {B.}~\bibnamefont
  {Kang}}, \bibinfo {author} {\bibfnamefont {K.}~\bibnamefont {Shiozaki}},\
  and\ \bibinfo {author} {\bibfnamefont {G.~Y.}\ \bibnamefont {Cho}},\
  }\bibfield  {title} {\bibinfo {title} {Many-body order parameters for
  multipoles in solids},\ }\href {https://doi.org/10.1103/PhysRevB.100.245134}
  {\bibfield  {journal} {\bibinfo  {journal} {Phys. Rev. B}\ }\textbf {\bibinfo
  {volume} {100}},\ \bibinfo {pages} {245134} (\bibinfo {year}
  {2019})}\BibitemShut {NoStop}%
\bibitem [{\citenamefont {Petrides}\ and\ \citenamefont
  {Zilberberg}(2020)}]{Petrides2020}%
  \BibitemOpen
  \bibfield  {author} {\bibinfo {author} {\bibfnamefont {I.}~\bibnamefont
  {Petrides}}\ and\ \bibinfo {author} {\bibfnamefont {O.}~\bibnamefont
  {Zilberberg}},\ }\bibfield  {title} {\bibinfo {title} {Higher-order
  topological insulators, topological pumps and the quantum {Hall} effect in
  high dimensions},\ }\href {https://doi.org/10.1103/PhysRevResearch.2.022049}
  {\bibfield  {journal} {\bibinfo  {journal} {Physical Review Research}\
  }\textbf {\bibinfo {volume} {2}},\ \bibinfo {pages} {022049} (\bibinfo {year}
  {2020})}\BibitemShut {NoStop}%
\bibitem [{\citenamefont {Kang}\ \emph {et~al.}(2021)\citenamefont {Kang},
  \citenamefont {Lee},\ and\ \citenamefont {Cho}}]{Kang2021}%
  \BibitemOpen
  \bibfield  {author} {\bibinfo {author} {\bibfnamefont {B.}~\bibnamefont
  {Kang}}, \bibinfo {author} {\bibfnamefont {W.}~\bibnamefont {Lee}},\ and\
  \bibinfo {author} {\bibfnamefont {G.~Y.}\ \bibnamefont {Cho}},\ }\bibfield
  {title} {\bibinfo {title} {Many-{Body} {Invariants} for {Chern} and {Chiral}
  {Hinge} {Insulators}},\ }\href
  {https://doi.org/10.1103/PhysRevLett.126.016402} {\bibfield  {journal}
  {\bibinfo  {journal} {Phys. Rev. Lett.}\ }\textbf {\bibinfo {volume} {126}},\
  \bibinfo {pages} {016402} (\bibinfo {year} {2021})}\BibitemShut {NoStop}%
\bibitem [{\citenamefont {Wienand}\ \emph {et~al.}(2022)\citenamefont
  {Wienand}, \citenamefont {Horn}, \citenamefont {Aidelsburger}, \citenamefont
  {Bibo},\ and\ \citenamefont {Grusdt}}]{wienand_2021}%
  \BibitemOpen
  \bibfield  {author} {\bibinfo {author} {\bibfnamefont {J.~F.}\ \bibnamefont
  {Wienand}}, \bibinfo {author} {\bibfnamefont {F.}~\bibnamefont {Horn}},
  \bibinfo {author} {\bibfnamefont {M.}~\bibnamefont {Aidelsburger}}, \bibinfo
  {author} {\bibfnamefont {J.}~\bibnamefont {Bibo}},\ and\ \bibinfo {author}
  {\bibfnamefont {F.}~\bibnamefont {Grusdt}},\ }\bibfield  {title} {\bibinfo
  {title} {Thouless pumps and bulk-boundary correspondence in higher-order
  symmetry-protected topological phases},\ }\href
  {https://doi.org/10.1103/PhysRevLett.128.246602} {\bibfield  {journal}
  {\bibinfo  {journal} {Phys. Rev. Lett.}\ }\textbf {\bibinfo {volume} {128}},\
  \bibinfo {pages} {246602} (\bibinfo {year} {2022})}\BibitemShut {NoStop}%
\bibitem [{\citenamefont {Resta}(1998)}]{Resta1998}%
  \BibitemOpen
  \bibfield  {author} {\bibinfo {author} {\bibfnamefont {R.}~\bibnamefont
  {Resta}},\ }\bibfield  {title} {\bibinfo {title} {Quantum-{Mechanical}
  {Position} {Operator} in {Extended} {Systems}},\ }\href
  {https://doi.org/10.1103/PhysRevLett.80.1800} {\bibfield  {journal} {\bibinfo
   {journal} {Phys. Rev. Lett.}\ }\textbf {\bibinfo {volume} {80}},\ \bibinfo
  {pages} {1800} (\bibinfo {year} {1998})}\BibitemShut {NoStop}%
\bibitem [{\citenamefont {Araki}\ \emph {et~al.}(2020)\citenamefont {Araki},
  \citenamefont {Mizoguchi},\ and\ \citenamefont
  {Hatsugai}}]{araki_berry_2020}%
  \BibitemOpen
  \bibfield  {author} {\bibinfo {author} {\bibfnamefont {H.}~\bibnamefont
  {Araki}}, \bibinfo {author} {\bibfnamefont {T.}~\bibnamefont {Mizoguchi}},\
  and\ \bibinfo {author} {\bibfnamefont {Y.}~\bibnamefont {Hatsugai}},\
  }\bibfield  {title} {\bibinfo {title} {{$\mathbb{Z}_{Q}$} berry phase for
  higher-order symmetry-protected topological phases},\ }\href
  {https://doi.org/10.1103/PhysRevResearch.2.012009} {\bibfield  {journal}
  {\bibinfo  {journal} {Phys. Rev. Research}\ }\textbf {\bibinfo {volume}
  {2}},\ \bibinfo {pages} {012009} (\bibinfo {year} {2020})}\BibitemShut
  {NoStop}%
\bibitem [{\citenamefont {V.}(1984)}]{Berry_1984}%
  \BibitemOpen
  \bibfield  {author} {\bibinfo {author} {\bibfnamefont {B.~M.}\ \bibnamefont
  {V.}},\ }\bibfield  {title} {\bibinfo {title} {Berry phase},\ }\href@noop {}
  {\bibfield  {journal} {\bibinfo  {journal} {Proc. R. Soc. London Ser. A}\
  }\textbf {\bibinfo {volume} {392}},\ \bibinfo {pages} {45} (\bibinfo {year}
  {1984})}\BibitemShut {NoStop}%
\bibitem [{\citenamefont {Zak}(1989)}]{zak_1989}%
  \BibitemOpen
  \bibfield  {author} {\bibinfo {author} {\bibfnamefont {J.}~\bibnamefont
  {Zak}},\ }\bibfield  {title} {\bibinfo {title} {Berry's phase for energy
  bands in solids},\ }\href {https://doi.org/10.1103/PhysRevLett.62.2747}
  {\bibfield  {journal} {\bibinfo  {journal} {Phys. Rev. Lett.}\ }\textbf
  {\bibinfo {volume} {62}},\ \bibinfo {pages} {2747} (\bibinfo {year}
  {1989})}\BibitemShut {NoStop}%
\bibitem [{\citenamefont {Brosco}\ \emph {et~al.}(2021)\citenamefont {Brosco},
  \citenamefont {Pilozzi}, \citenamefont {Fazio},\ and\ \citenamefont
  {Conti}}]{brosco_2021}%
  \BibitemOpen
  \bibfield  {author} {\bibinfo {author} {\bibfnamefont {V.}~\bibnamefont
  {Brosco}}, \bibinfo {author} {\bibfnamefont {L.}~\bibnamefont {Pilozzi}},
  \bibinfo {author} {\bibfnamefont {R.}~\bibnamefont {Fazio}},\ and\ \bibinfo
  {author} {\bibfnamefont {C.}~\bibnamefont {Conti}},\ }\bibfield  {title}
  {\bibinfo {title} {Non-abelian thouless pumping in a photonic lattice},\
  }\href {https://doi.org/10.1103/PhysRevA.103.063518} {\bibfield  {journal}
  {\bibinfo  {journal} {Phys. Rev. A}\ }\textbf {\bibinfo {volume} {103}},\
  \bibinfo {pages} {063518} (\bibinfo {year} {2021})}\BibitemShut {NoStop}%
\bibitem [{\citenamefont {Resta}(1994)}]{resta_1994}%
  \BibitemOpen
  \bibfield  {author} {\bibinfo {author} {\bibfnamefont {R.}~\bibnamefont
  {Resta}},\ }\bibfield  {title} {\bibinfo {title} {Macroscopic polarization in
  crystalline dielectrics: the geometric phase approach},\ }\href
  {https://doi.org/10.1103/RevModPhys.66.899} {\bibfield  {journal} {\bibinfo
  {journal} {Rev. Mod. Phys.}\ }\textbf {\bibinfo {volume} {66}},\ \bibinfo
  {pages} {899} (\bibinfo {year} {1994})}\BibitemShut {NoStop}%
\bibitem [{\citenamefont {Chang}\ \emph {et~al.}(2013)\citenamefont {Chang},
  \citenamefont {Zhang}, \citenamefont {Feng}, \citenamefont {Shen},
  \citenamefont {Zhang}, \citenamefont {Guo}, \citenamefont {Li}, \citenamefont
  {Ou}, \citenamefont {Wei}, \citenamefont {Wang} \emph {et~al.}}]{chang_2013}%
  \BibitemOpen
  \bibfield  {author} {\bibinfo {author} {\bibfnamefont {C.-Z.}\ \bibnamefont
  {Chang}}, \bibinfo {author} {\bibfnamefont {J.}~\bibnamefont {Zhang}},
  \bibinfo {author} {\bibfnamefont {X.}~\bibnamefont {Feng}}, \bibinfo {author}
  {\bibfnamefont {J.}~\bibnamefont {Shen}}, \bibinfo {author} {\bibfnamefont
  {Z.}~\bibnamefont {Zhang}}, \bibinfo {author} {\bibfnamefont
  {M.}~\bibnamefont {Guo}}, \bibinfo {author} {\bibfnamefont {K.}~\bibnamefont
  {Li}}, \bibinfo {author} {\bibfnamefont {Y.}~\bibnamefont {Ou}}, \bibinfo
  {author} {\bibfnamefont {P.}~\bibnamefont {Wei}}, \bibinfo {author}
  {\bibfnamefont {L.-L.}\ \bibnamefont {Wang}}, \emph {et~al.},\ }\bibfield
  {title} {\bibinfo {title} {Experimental observation of the quantum anomalous
  hall effect in a magnetic topological insulator},\ }\href
  {https://doi.org/10.1126/science.1234414} {\bibfield  {journal} {\bibinfo
  {journal} {Science}\ }\textbf {\bibinfo {volume} {340}},\ \bibinfo {pages}
  {167} (\bibinfo {year} {2013})}\BibitemShut {NoStop}%
\bibitem [{\citenamefont {Sambe}(1973)}]{Sambe_1973}%
  \BibitemOpen
  \bibfield  {author} {\bibinfo {author} {\bibfnamefont {H.}~\bibnamefont
  {Sambe}},\ }\bibfield  {title} {\bibinfo {title} {Steady states and
  quasienergies of a quantum-mechanical system in an oscillating field},\
  }\href {https://doi.org/10.1103/PhysRevA.7.2203} {\bibfield  {journal}
  {\bibinfo  {journal} {Phys. Rev. A}\ }\textbf {\bibinfo {volume} {7}},\
  \bibinfo {pages} {2203} (\bibinfo {year} {1973})}\BibitemShut {NoStop}%
\bibitem [{\citenamefont {Shirley}(1965)}]{Shirley_1979}%
  \BibitemOpen
  \bibfield  {author} {\bibinfo {author} {\bibfnamefont {J.~H.}\ \bibnamefont
  {Shirley}},\ }\bibfield  {title} {\bibinfo {title} {Solution of the
  schr\"odinger equation with a hamiltonian periodic in time},\ }\href
  {https://doi.org/10.1103/PhysRev.138.B979} {\bibfield  {journal} {\bibinfo
  {journal} {Phys. Rev.}\ }\textbf {\bibinfo {volume} {138}},\ \bibinfo {pages}
  {B979} (\bibinfo {year} {1965})}\BibitemShut {NoStop}%
\bibitem [{\citenamefont {Resta}(1992)}]{resta_1992}%
  \BibitemOpen
  \bibfield  {author} {\bibinfo {author} {\bibfnamefont {R.}~\bibnamefont
  {Resta}},\ }\bibfield  {title} {\bibinfo {title} {Theory of the electric
  polarization in crystals},\ }\href
  {https://doi.org/10.1080/00150199208016065} {\bibfield  {journal} {\bibinfo
  {journal} {Ferroelectrics}\ }\textbf {\bibinfo {volume} {136}},\ \bibinfo
  {pages} {51} (\bibinfo {year} {1992})}\BibitemShut {NoStop}%
\bibitem [{\citenamefont {King-Smith}\ and\ \citenamefont
  {Vanderbilt}(1993)}]{king_1993}%
  \BibitemOpen
  \bibfield  {author} {\bibinfo {author} {\bibfnamefont {R.~D.}\ \bibnamefont
  {King-Smith}}\ and\ \bibinfo {author} {\bibfnamefont {D.}~\bibnamefont
  {Vanderbilt}},\ }\bibfield  {title} {\bibinfo {title} {Theory of polarization
  of crystalline solids},\ }\href {https://doi.org/10.1103/PhysRevB.47.1651}
  {\bibfield  {journal} {\bibinfo  {journal} {Phys. Rev. B}\ }\textbf {\bibinfo
  {volume} {47}},\ \bibinfo {pages} {1651} (\bibinfo {year}
  {1993})}\BibitemShut {NoStop}%
\end{thebibliography}%




\newpage

\appendix

\section*{Supplementary Information}

\section{The Chern number}
\label{sec:box1}
Quantized pumping can be understood as the result of geometric phases. The most well-known example is the Aharonov-Bohm phase: In brief, charged particles in the presence of a magnetic field, that evolve along a closed trajectory, acquire a phase that is equal to the magnetic flux piercing the area that is enclosed by the trajectory in units of $h/e$, where $h$ is Planck's constant and $e$ the charge of an electron. This phase can be measured directly and manifests itself, e.g., in the anomalous magnetoresistance~\cite{chang_2013}.

Berry demonstrated~\cite{Berry_1984} that the Aharonov-Bohm phase is an example of a more general phenomenon. Suppose the Hamiltonian of a closed quantum system depends on some parameters, which are changed adiabatically in time, such that after a period $T$ the system returns to the original values, thereby encircling a closed path ${\cal C}$ in the parameter space. Simultaneously, the wave function of the system may acquire an additional geometric phase, so-called Berry phase, which depends on ${\cal C}$. Similar to the Aharonov-Bohm phase, the Berry phase can be thought of as the result of a flux of some effective ``magnetic field'', known as Berry curvature in the mathematical literature, through the contour ${\cal C}$. The difference is that both the contour and the ``magnetic field'' exist in an abstract parameter space rather than in real space. The single-valuedness of the wave function requires that the Berry curvature integrated over a closed manifold is quantized in units of $2\pi$. For a 1D time-periodic system with Bloch bands, the pumped charge results from a fictitious ``magnetic flux'' in an abstract (1+1)-D parameter space formed by quasimomentum $k$ and time $t$, which are both periodic. The charge pumped after one cycle, can be expressed as an integral of the effective ``magnetic field'' over the area enclosed by the pump path ${\cal C}$. For a uniformly filled Bloch band the pumped charge after one period $T$ is quantized according to the Chern number $\nu$, which is an integer.

\section{Floquet theory}
\label{sec:box2}
Due to discrete time-translation invariance of an Hamiltonian with periodicity $T$, there exists a basis of solutions of the time-dependent Schr\"{o}dinger equation, that are periodic up to a phase, the so called Floquet states $\ket{\psi_\alpha (t)} = \nep^{-i\varepsilon_{\alpha} t/\hbar} \ket{\phi_\alpha (t)} $~\cite{Sambe_1973, Shirley_1979}.
The $T$-periodic states $\ket{\phi_\alpha (t)}$ are the so-called Floquet modes and $\varepsilon_\alpha$ are the corresponding quasienergies:
they are defined modulo an integer number of $\hbar\omega=2\pi\,\hbar/T$, so that it is possible to restrict them to the first Floquet Brillouin zone (FBZ)
$[-\hbar\omega/2,\hbar\omega/2)$. In the case of periodic boundary conditions and of a translationally-invariant system Floquet-Bloch bands are formed $\varepsilon_{\alpha,k}$ where $k$ is a wavevector of the first Brillouin zone (BZ).

\section{Polarization theory}
\label{sec:box3}

The modern theory of polarization associates $P$ with the mean position of the
charge distribution per unit cell in the case of translationally invariant states~\cite{resta_1994}.
For a system of length $La$ with periodic boundary conditions,
one can define the polarization for a many-body wave function $\ket{\Psi(t)}$ via~\cite{Resta1998}:
\begin{equation}
  \label{eq:pol-MB1}
  P(t)=\frac{qa}{2\pi}\operatorname{Im}\ln\left\langle\Psi(t) \middle|
  e^{\frac{i2\pi}{La}\hat{X}} \middle| \Psi(t)\right\rangle\quad(\bmod \,qa),
\end{equation}
where $\hat{X} =\sum_{x=0}^{L-1} a \left(x-x_{0}\right)\hat{n}_{x}$ is the
position operator, $\hat n_x$ is the local density operator,   $q$ is the charge per particle ($q=1$ for neutral atoms), $a$
is the length of the unit cell, $L$ is the number of sites, and $x_0$ is the unit-cell center.
Let us note that, $P$ is only defined modulo $qa$ and it has the units of a dipole moment.

It is insightful to see that for the case of non-interacting fermions for a filled band, the expression for the many-body polarization indeed reduces to the usual form~\cite{resta_1992,king_1993,resta_1994}:
\begin{equation}
\label{eq:pol-fermi}
P_{\mathrm{NI}}(t) = \frac{-iq}{2\pi}\int_{-\pi/a}^{\pi/a} \text{d}k \bra{u(k,t)}\partial_k \ket{u(k,t)},
\end{equation}
where $\ket{u(k,t)}$ are single-particle momentum-eigenstates of the instantaneous Hamiltonian and denotes the quasimomentum.
In fact one can see that the polarization is determined by the Zak phase associated with a homogeneously filled band,
which is defined modulo $2\pi$~\cite{zak_1989}.

\end{document}